\documentclass[onecolumn]{emulateapj}
\usepackage{graphicx}
\begin{document}

\newcommand{\ftd}{F_{\rm td}}
\newcommand{\ratio}{(R_{\rm app}/D)^2}
\newcommand{\fortyeight}{SAX~J1748.9$-$2021}
\newcommand{\twenty}{4U~1820$-$30}
\newcommand{\fortyfive}{EXO~1745$-$248}
\newcommand{\thirtyone}{KS~1731$-$260}
\newcommand{\twentyfour}{4U~1724$-$207}
\newcommand{\oeight}{4U~1608$-$52}

\title{The Dense Matter Equation of State from Neutron Star Radius and Mass 
Measurements}

\author{%
  Feryal \"Ozel\altaffilmark{1},
  Dimitrios Psaltis\altaffilmark{1},
  Tolga G\"uver\altaffilmark{2},
  Gordon Baym\altaffilmark{3},\\
  Craig Heinke\altaffilmark{4,5}, and
  Sebastien Guillot\altaffilmark{6}%
}

\email{E-mail: fozel@email.arizona.edu}

\altaffiltext{1}{Department of Astronomy, University of Arizona, 933
  N. Cherry Ave., Tucson, AZ 85721}

\altaffiltext{2}{Istanbul University, Science Faculty, Department of
  Astronomy and Space Sciences, Beyazit, 34119, Istanbul, Turkey}

\altaffiltext{3}{Department of Physics, University of Illinois at Urbana-Champaign,
1110 W. Green Street, Urbana, Illinois 61801}

\altaffiltext{4}{Department of Physics, University of Alberta, CCIS 4-183, Edmonton, AB T6G 2E1, Canada}

\altaffiltext{5} {Alexander von Humboldt Fellow, at Max-Planck-Institut f\"ur Radioastronomie, Auf dem H\"ugel 69, 
D-53121 Bonn, Germany}

\altaffiltext{6}{Instituto de Astrof\'{i}sica, Facultad de F\'{i}sica, Pontificia Universidad Cat\'{o}lica de 
Chile, Av. Vicu\~{n}a Mackenna 4860, 782-0436 Macul, Santiago, Chile}

\begin{abstract}
We present a comprehensive study of spectroscopic radius measurements
of twelve neutron stars obtained during thermonuclear bursts or in
quiescence. We incorporate, for the first time, a large number of
systematic uncertainties in the measurement of the apparent angular
sizes, Eddington fluxes, and distances, in the composition of the
interstellar medium, and in the flux calibration of X-ray
detectors. We also take into account the results of recent theoretical
calculations of rotational effects on neutron star radii, of
atmospheric effects on surface spectra, and of relativistic
corrections to the Eddington critical flux. We employ Bayesian
statistical frameworks to obtain neutron star radii from the
spectroscopic measurements as well as to infer the equation of state
from the radius measurements. Combining these with the results of
experiments in the vicinity of nuclear saturation density and the
observations of $\sim 2\ M_\odot$ neutron stars, we place strong and
quantitative constraints on the properties of the equation of state
between $\approx 2-8$ times the nuclear saturation density. We find
that around $M=1.5~M_\odot$, the preferred equation of state predicts
radii between $10.1 - 11.1$~km. When interpreting the pressure
constraints in the context of high density equations of state based on
interacting nucleons, our results suggest a relatively weak
contribution of the three-body interaction potential.
\end{abstract}


\section{Introduction}
The densest matter in the universe at low temperatures and at finite
baryon density is found in the cores of neutron stars. Such conditions
are not accessible to current laboratory experiments. Measuring the
macroscopic properties of neutron stars, and in particular, their
radii, offers the most direct and powerful probe of the composition of
and interactions in cold, ultradense matter.

There has been a lot of recent progress in measuring neutron star
radii with a variety of techniques and using them to constrain the
equation of state (see \"Ozel 2013 for a recent review). Spectroscopic
observations of thermonuclear bursts from accreting neutron stars with
the last generation of X-ray telescopes have provided measurements of
both the radii and the masses of several sources with weakly
correlated uncertainties (e.g., \"Ozel et al.\ 2009, 2012; G\"uver et
al.\ 2010a,b; G\"uver \& \"Ozel 2013).  Observations of similar
neutron stars during quiescence have yielded measurements of their
apparent angular sizes, which leads to a determination of the radii
with correlated uncertainties with the neutron star mass (Heinke et
al.\ 2006, 2014; Webb \& Barret 2007; Guillot et al.\ 2011, 2013).

A parallel avenue of progress has taken place in understanding the
mapping from neutron star masses and radii to the equation of state.
Several parametric representations of the equation of state allow
radius measurements to be used for a direct inference of the pressure
at several fiducial densities above the nuclear saturation density
($\rho_{\rm ns}$; Lindblom 1992; Lattimer \& Prakash 2001; \"Ozel \&
Psaltis 2009; Read et al.\ 2009). \"Ozel et al. (2010) used one of
these mapping techniques on the radius measurements obtained from
thermonuclear bursters to place the first constraints on the neutron
star equation of state at high densities. They found that the
relatively small observed radii point to lower pressures at and above
$2\rho_{\rm ns}$ than those predicted by purely nucleonic equations of
state. Similarly small radii were found by Guillot et al.\ (2013) in
their analysis of quiescent neutron stars, which also point to softer
equations of state than expected for nucleonic compositions.  Steiner
et al.\ (2010) and Lattimer \& Steiner (2014a) found that substantial
reinterpretation of the observed astrophysical phenomena and/or
choosing extreme values for some of the measurements was needed to
make the observed radii larger and reduce the tension with the
predictions of some nucleonic equations of state.

Since these first studies, our understanding of systematics in the
spectroscopic measurements has been substantially improved and
theoretical work has identified a number of small, albeit important,
corrections that need to be applied to the inference of neutron star
radii. In this paper, we incorporate these corrections and sources of
uncertainty in the analysis of the spectroscopic data and infer the
parameters of the neutron star matter equation of state that are
consistent with all astrophysical data as well as with laboratory
experiments at low densities. 

Specifically, comprehensive studies of a large sample of sources
allowed for a quantitative assessment of the systematic uncertainties
in the spectroscopic measurements. For thermonuclear bursters, a
Bayesian mixture technique with an outlier detection scheme was
applied on a very large Rossi X-ray Timing Explorer (RXTE) sample
consisting of 13,095 burst spectra from 12 sources. This resulted in a
data-driven measurement of the intrinsic scatter in the apparent
angular sizes during the cooling tails of bursts (G\"uver et
al.\ 2012b) and of the scatter in the touchdown fluxes during
photosperic radius-expansion bursts in individual sources (G\"uver et
al.\ 2012a). Moreover, a comparison of the simultaneous observations
of bursts with multiple X-ray instruments, and specifically between
the PCA and {\it Chandra} ACIS) significantly constrained any biases
in the RXTE burst fluxes due to calibration to $\lesssim 1\%$ (G\"uver
et al.\ 2015). For quiescent neutron stars, long observations with
{\it Chandra} and XMM-{\it Newton} resulted in high signal-to-noise
spectra, which, in combination with explorations of uncertainties in
the atmospheric composition and the amount of interstellar extinction
in the soft X-rays, improved the inferences of their apparent angular
sizes (Guillot et al.\ 2013; Guillot \& Rutledge 2014; Heinke et
al.\ 2014).

Numerous developments have also taken place on the theoretical models
that are employed to interpret these observational data. Since most of
the X-ray burst sources spin at moderately high rates, general
relativistic corrections to the apparent angular sizes that go beyond
the Schwarzschild approximation and incorporate effects that are of
second order in spin have been calculated (Baub\"ock et al.\ 2012,
2015). Radiative equilibrium models of neutron star atmospheres have
also been improved by including the angle and energy dependence of the
scattering kernels (Suleimanov et al.\ 2011, 2012). These resulted in
more accurate relations between spectral and effective temperatures as
well as in notable temperature corrections to the Eddington
(touchdown) fluxes. Different statistical tools for inferring the
uncertainties in the neutron star radii and masses from burst data
have been assessed. This led to the identification of a Bayesian
method that does not introduce biases in the radius measurements, in
contrast to the frequentist method used in earlier studies (see the
companion paper \"Ozel \& Psaltis 2015). Finally, Bayesian methods
have also been developed to infer the properties of the dense matter
equation of state from radius and mass measurements (Steiner et
al.\ 2010).

There are additional astrophysical data and laboratory experiments
that can be used in conjunction with the neutron star radius
measurements to further constrain the equation of state of dense
matter. In this paper, we use the results of nucleon-nucleon
interaction calculations that are firmly based on scattering data
below 350~MeV and the properties of light nuclei (Akmal et al.\ 1998;
Peiper et al.\ 2001; Gandolfi et al.\ 2012) to place a lower limit on
the pressure of neutron-rich matter at densities $\sim 2 \rho_{\rm
  ns}$, where $\rho_{\rm ns}$ is the nuclear matter saturation
density, $\sim 2.7 \times 10^{14}~{\rm g~cm}^{-3}$ (or $n_{\rm ns}
\sim 0.16~{\rm fm} ^{-3}$). We also use the observations of the two
neutron stars with the highest measured masses (Demorest et al.\ 2010;
Antoniadis et al.\ 2013) to exclude the regions of the pressure
parameter space that do not produce $\sim 2~M_\odot$ neutron stars.

In Section 2, we introduce the improved theoretical models for the
rotational corrections to the apparent angular sizes, the temperature
corrections to the inferred Eddington limit, and the Bayesian
framework for combining these two measurements. In Section 3, we
present the spectroscopic data as well as the inferred radii for each
of the twelve sources included in this study. In Section 4, we obtain
constraints on the neutron star radius and a mono-parametric equation
of state by combining all of the individual radius measurements.  In
Section 5, we use a Bayesian technique to map the measured radii and
masses into the pressures at three fiducial densities taking into
account the additional astrophysical and laboratory constraints.
Finally, in Section 6, we discuss the implications of our findings for
neutron star astrophysics and nuclear models.

\begin{figure}
\centering
   \includegraphics[scale=0.45]{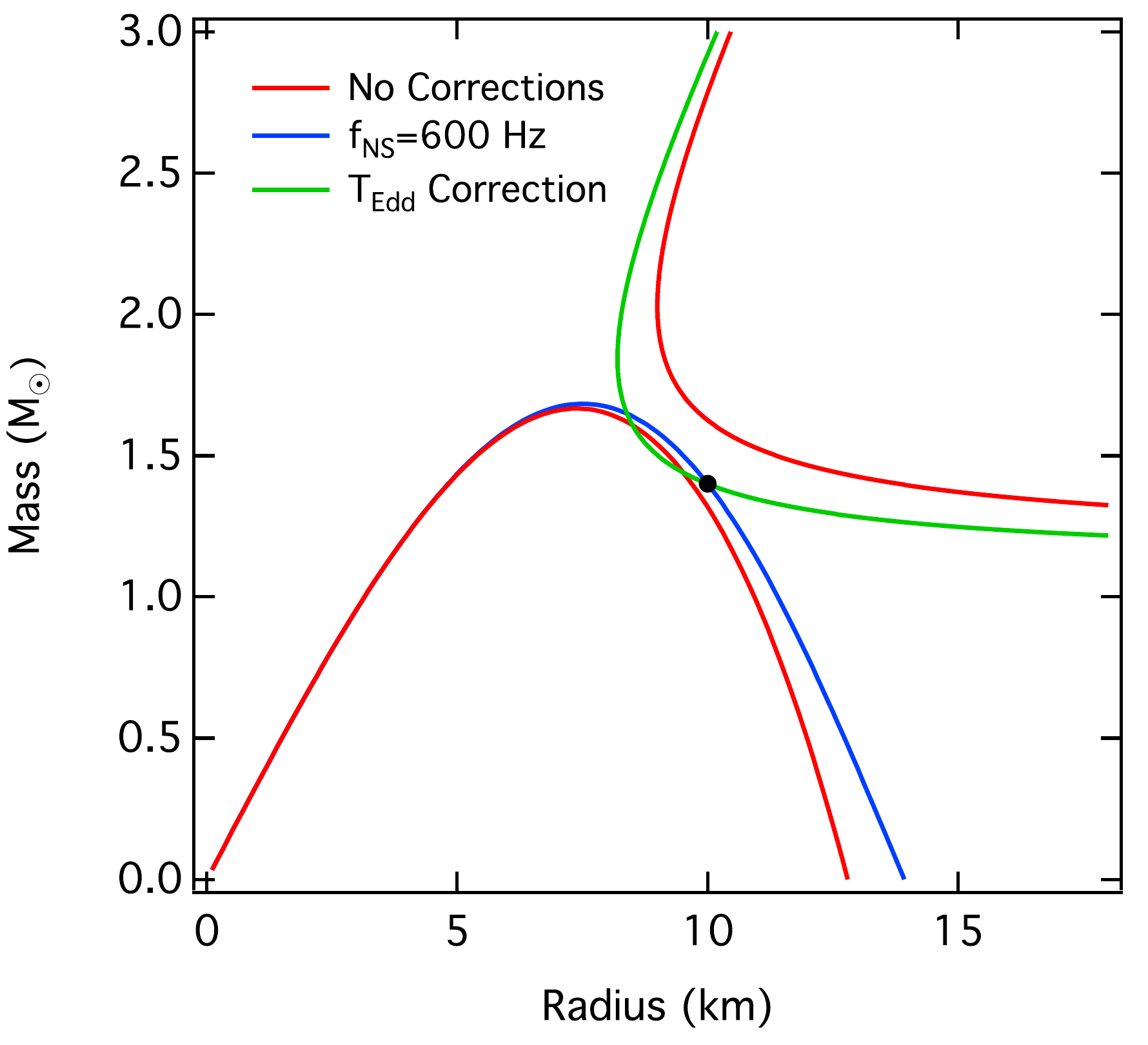}
\caption{Contours of constant apparent angular size (blue) and
  touchdown flux (green) for a $M=1.4~M_\odot$ and $R=10$~km neutron
  star spinning at 600~Hz and a spectral temperature during the
  touchdown moment of Eddington limited bursts calculated using
  equation~(\ref{eq:Tedd}) for this mass and radius. These curves include
  the corrections to the apparent area due to neutron star spin and
  the temperature correction to the Eddington limit, respectively. The
  red curves are the corresponding contours when these corrections are
  not taken into account and would have led to no solutions for the
  neutron star mass and radius.}  \mbox{}
\label{fig:rot_corr} 
\end{figure}

\section{Theoretical Framework for Spectroscopic Neutron Star Radius Measurements}

\subsection{Rotational Corrections to the Apparent Angular Size}

For neutron stars that show thermonuclear bursts or are in quiescence,
the measurements of the radii rely on the detection of thermal
emission from their surfaces. For such compact sources, the ratio
$F/\sigma_B T_{\rm c}^4$, where $F$ is the observed thermal flux,
$\sigma_B$ is the Stefan-Boltzmann constant, and $T_{\rm c}$ is the
temperature measured from the spectrum, yields the apparent angular
size $A_\infty$ of the source. For slowly spinning neutron stars, this
relates to the stellar radius $R$ via
\begin{equation}
A_\infty = \frac{R^2}{D^2 f_c^4}\left(1-\frac{2GM}{Rc^2} \right)^{-1},
\end{equation}
where $M$ is the mass of the neutron star, $D$ is its distance, $f_c$
is the color correction factor that takes into account the distortions
in the spectrum due to the stellar atmosphere, $G$ is the
gravitational constant, and $c$ is the speed of light.

All neutron stars with measured spin frequencies in our sample spin at
frequencies that are larger than 400~Hz. At such frequencies, the
Doppler broadening of the spectrum as well as distortions related to
the oblateness and the quadrupole moment of the neutron star become
important. Baub\"ock et al.\ (2015) explored these effects in detail
and devised an approximate formula for the spin corrections to the
apparent surface area. In the absence of any additional information
on the inclination of the source, the appropriately weighted angular
size for a spin frequency $f_{\rm NS}$ becomes
\begin{eqnarray}
A_\infty = && \frac{R^2}{D^2 f_c^4}\left(1-\frac{2GM}{Rc^2}
\right)^{-1} \times \nonumber\\ &&
\left\{1+\left[\left(0.108-0.096\frac{M}{M_\odot}\right)
  +\left(-0.061+0.114\frac{M}{M_\odot}\right)\frac{R}{10~{\rm km}} -
  0.128\left(\frac{R}{10~{\rm km}}\right)^2 \right] \left(\frac{f_{\rm
    NS}}{1000~{\rm Hz}}\right)^2 \right \}^2.
\label{eq:Arot}
\end{eqnarray}
In Figure~\ref{fig:rot_corr}, we show in blue the contour of constant
apparent angular size on the mass-radius plane for a $M=1.4~M_\odot$ 
and $R=10$~km neutron star spinning at 600~Hz. To highlight
the effect of the rotational corrections, we also plot in red the
corresponding contour obtained under the Schwarzschild approximation
used in the previous studies. As discussed in Baub\"ock et
al.\ (2015), the rotational effects lead to larger angular sizes for
the same neutron star mass and radius.

\subsection{Temperature Corrections to the Eddington Limit}

The atmospheres of neutron stars during thermonuclear bursts are
dominated by electron scattering. During strong bursts, the radiation
forces lift the photosphere above the neutron star surface and allow
for a measurement of the Eddington critical luminosity. When this
luminosity is measured at the touchdown point, i.e., when the
photosphere has returned to the neutron star surface, it is related to
the neutron star mass and radius via
\begin{equation}
\ftd\ = \frac{GMc}{k_{\rm es} D^2}\left(1-\frac{2GM}{Rc^2}
\right)^{1/2}, 
\end{equation}
where 
\begin{equation}
k_{\rm es} \equiv 0.2 (1+X)~{\rm cm}^2~{\rm g}^{-1} 
\end{equation}
is the electron scattering opacity and $X$ is the hydrogen mass
fraction of the atmosphere.

Because of the energy dependence in the Klein-Nishina cross section
and the fact that the photons exchange energy with the electrons at
each scattering, the Eddington flux depends on the temperature of the
atmosphere. Paczynski (1983) derived an approximation for this
temperature correction, which was further refined by Suleimanov et
al.\ (2012) by taking into account the angular dependence of the
scattering processes. With the approximate formula given in the latter
study, the Eddington flux becomes
\begin{equation}
\ftd\ = \frac{GMc}{k_{\rm es} D^2}\left(1-\frac{2GM}{Rc^2}
\right)^{1/2} \left[1+\left(\frac{kT_c}{38.8~{\rm keV}}\right)^{a_g}
\left(1-\frac{2GM}{Rc^2}\right)^{-a_g/2}\right], 
\label{eq:Tcorr}
\end{equation}
where
\begin{equation}
a_g = 1.01 + 0.067 \left( \frac{g_{\rm eff}}{10^{14}~{\rm cm~s}^{-2}}\right)
\end{equation}
and 
\begin{equation}
g_{\rm eff} = \frac{GM}{R^2} \left(1-\frac{2GM}{Rc^2} \right)^{-1/2}.
\end{equation}
In equation~(\ref{eq:Tcorr}), the correction to the Eddington flux depends 
on the color temperature when the atmosphere reaches that limit, and this 
color temperature, in turn, depends on the mass and radius of the neutron 
star and the composition of the atmosphere via
\begin{equation}
T_{\rm Edd, c} = f_{\rm c} T_{\rm Edd, eff} = f_{\rm c} 
\left(\frac{g_{\rm eff} c}{\sigma_{\rm B} k_{\rm es}}\right)^{1/4} 
= f_{\rm c} \left(\frac{GMc}{\sigma_{\rm B} k_{\rm es} R^2}\right)^{1/4}
 \left(1-\frac{2GM}{Rc^2} \right)^{-1/8}, 
\label{eq:Tedd}
\end{equation}
where $\sigma_{\rm B}$ is the Boltzmann constant. 

In Figure~\ref{fig:rot_corr}, we show in green the contour of constant
touchdown flux in the mass-radius plane for a $M=1.4~M_\odot$ and
$R=10$~km neutron star with hydrogen mass fraction $X=0$ and a color
correction factor at touchdown of $f_c=1.9$. We also plot in red the
corresponding contour obtained when the temperature correction to the
Eddington flux is not taken into account, as was done in the previous
studies. Because at high temperature the scaterring cross section
decreases, a measured Eddington flux corresponds to a smaller mass
(and radius).

\subsection{Statistics of Combining Observables to Infer Neutron Star Radii}

It is clear from Equations~\ref{eq:Arot} and \ref{eq:Tcorr} that
measurements of the apparent angular size and the Eddington flux of a
neutron star can be combined to determine its mass when the distance
to the source is known.  In earlier studies, this inference was
performed and the uncertainties were assessed in a frequentist
approach. We showed in a companion paper that this approach suffers
from significant biases for the range of masses and radii expected for
neutron stars (see \"Ozel \& Psaltis 2015). To alleviate these
shortcomings, we devised a Bayesian framework that more faithfully
reconstructs masses and radii from synthetic data and we utilize it in
the present analysis.

Using Bayes' theorem, we write the likelihood $P(M,R \; \vert \; {\rm
  data})$ that a neutron star has a given mass and radius given a set
of spectroscopic observables as
\begin{equation}
P(M,R \; |\; {\rm data}) = C P(\rm{data} \; | \; M, R) P_{\rm pri}(M) P_{\rm pri} (R),  
\label{eq:MR_burst}
\end{equation}
where $P_{\rm pri} (M)$ and $P_{\rm pri} (R)$ are the priors over the
mass and radius and $C$ is an appropriate normalization constant. 
Given that $A$ and $\ftd$ are ideally uncorrelated measurements, we
can write 
\begin{eqnarray}
P(\rm{data} \; | \; M, R) &&= \int P(D) \; dD \int P(f_c) \; df_c \int P(X) 
\; dX \int P(f_{\rm NS}) \; df_{\rm NS} \nonumber \\ && \times P[F_{\rm td}(M, R, D, X)] \;
P[A(M, R, D, f_{\rm NS}, f_c)]. 
\label{eq:bayes}
\end{eqnarray}
Here, $P(D)$, $P(\ftd)$, and $P(A)$ are the posterior likelihoods of
the measurements over the distance, the touchdown flux, and the
apparent angular size; $P(X)$, $P(f_c)$, and $P(f_{\rm NS})$ are the
priors over the hydrogen mass fraction, the color correction factor,
and the spin frequency of the neutron star, respectively. Hereafter,
when we use this expression, we assume flat priors over the mass
between 0.6 and 3.5~$M_\odot$ and over the radius between the
Schwarzschild radius that corresponds to each mass and 20~km. These
ranges are chosen to be large enough such that their precise values do
not affect the posterior likelihoods.

When the spin frequency of a source is previously measured, we take as
$P(f_{\rm NS})$ a delta function at the known frequency. When the spin
frequency is not known, we assume a flat prior between 250 and 650~Hz,
which is consistent with the range of spins observed in thermonuclear
bursters, and apply spin corrections to the apparent angular size with
this prior.


\begin{deluxetable}{cccccc}
\tabletypesize{\scriptsize}
\tablewidth{500pt}
\tablenum{1}
\tablecaption{Properties of Neutron Star Burst Sources}
\tablehead{
 \colhead{Source} &
 \colhead{App. Angular Size} &
 \colhead{Touchdown Flux\tablenotemark{a}} &
 \colhead{Spin Freq.\tablenotemark{b}} &
 \colhead{Distance\tablenotemark{b}} &
 \colhead{Radius\tablenotemark{c}} 
\cr
 \colhead{} &
 \colhead{(km/10~kpc)$^2$} &
 \colhead{($10^{-8}$~erg~s$^{-1}$~cm$^{-2}$)} &
 \colhead{(Hz)} &
 \colhead{(kpc)} &
 \colhead{(km)}
}
\startdata
\twenty\ & 89.9$\pm$15.9 & 5.98$\pm$0.66 & \nodata & $7.6\pm0.4$\tablenotemark{4} or $8.4\pm0.6$\tablenotemark{5,6} & 11.1 $\pm$ 1.8\\
\fortyeight\ & 89.7$\pm$9.6 & 4.03$\pm$0.54 & 410\tablenotemark{1} & $8.2 \pm 0.6$\tablenotemark{4,5,7} & 11.7 $\pm$ 1.7 \\
\fortyfive\ & 117.8$\pm$19.9 & 6.69$\pm$0.74 & \nodata & 6.3; $\Delta D=0.63$\tablenotemark{8,9} & 10.5 $\pm$ 1.6 \\
\thirtyone\ & 96.0$\pm$7.9 & 4.71$\pm$0.52 & 524\tablenotemark{2} & $\sim 7-9$\tablenotemark{10} & 10.0 $\pm$ 2.2 \\
\twentyfour\ & 113.8$\pm$15.4 & 5.29$\pm$0.58 & \nodata & 7.4$\pm$0.5 & 12.2 $\pm$ 1.4 \\
\oeight\ & 314$\pm$44.3 & 18.5$\pm$2.0 & 620\tablenotemark{3} & see Appendix & 9.8 $\pm$ 1.8 
\enddata
\scriptsize
\tablenotetext{a} {A minimum systematic uncertainty of 11\% has been assigned in accordance 
with G\"uver et al.\ (2012a).}
\tablenotetext{b} {References:
1.~Altamirano et al.\ 2008;
2.~Smith et al.\ 1997;
3.~Hartman et al.\ 2003;
4.~Kuulkers et al.\ 2003;
5.~Valenti et al.\ 2007;
6.~G\"uver et al.\ 2010b;
7.~G\"uver \& \"Ozel 2013;
8.~Ortolani et al.\ 2007;
9.~\"Ozel et al.\ 2009;
10.~\"Ozel et al.\ 2012a} 
\tablenotetext{c} {The radius and its 68\% uncertainty obtained by marginalizing the mass-radius likelihood 
of each source over the observed mass distribution, as in Figure~\ref{fig:combined_R}.}
\mbox{} 
\end{deluxetable}

\subsection{The Spectra of Thermonuclear Bursters}

\begin{figure}
\centering
   \includegraphics[scale=0.45]{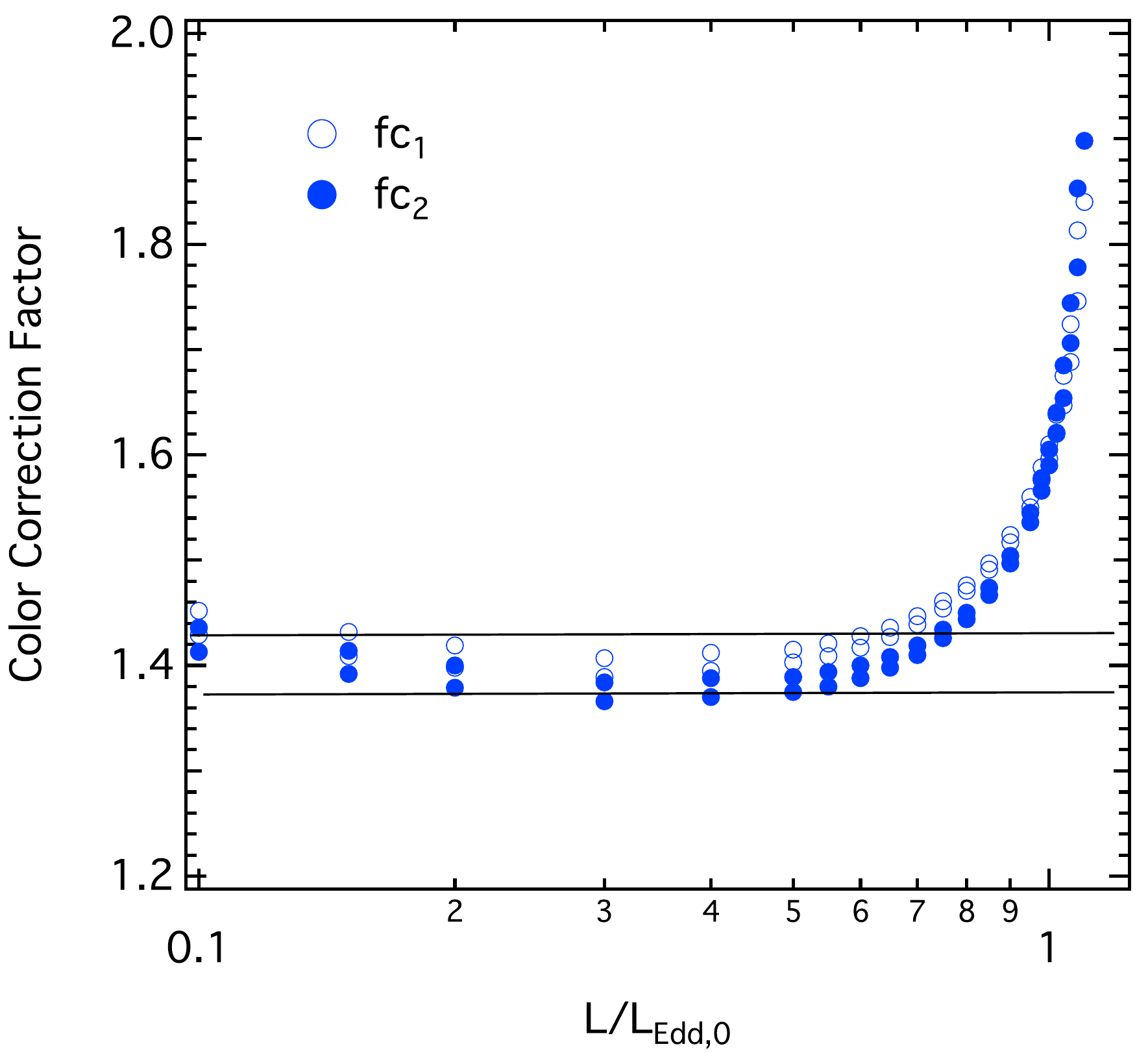}
\caption{Color correction factors from the models of He-rich
  neutron-star atmospheres by Suleimanov et al.\ (2012) for effective
  gravitational accelerations in the range $\log g_{\rm eff} =
  14.3-14.6$. The open and filled circles correspond to two different
  definitions of the color-correction factor explored by Suleimanov et
  al. (2012). The two horizontal lines show the range of values we use
  in this paper, which accurately reproduces the model results for
  fluxes less than about 0.7 of the critical Eddington flux.}
\label{fig:color_cor} 
\end{figure}

In order to obtain the mass-radius contours of a neutron star from the
measurement of its apparent angular size via equation~(\ref{eq:Arot}),
we need to employ models of the neutron star atmosphere that enter
through the color correction factor $f_c$. Since the initial work of
London \& Taam (1986), increasingly more sophisticated calculations
have been performed for a variety of compositions and effective
gravitational accelerations (Madej et al.\ 2004; Majczyna et
al.\ 2005; Suleimanov et al.\ 2011).  Most recently, Suleimanov et
al.\ (2012) calculated a large set of models taking into account the
full angular and energy dependence of the scattering process. They
found that, for fluxes between $\approx 0.1-0.7$ of the Eddington
critical flux, which is the range that we consider here, the color
correction factors depend very weakly on the effective gravitational
acceleration or on the particular definition of the color correction
factor and are approximately constant in the range $1.4 \pm 0.05$ (see
Figure~\ref{fig:color_cor}). Note that the color correction factors do
evolve significantly at very high and very low flux levels, which we
exclude when measuring neutron star apparent angular sizes during the
cooling tails of thermonuclear bursts (see G\"uver et al.\ 2012b). In
the remainder of this paper, we consider a flat prior distribution of
the color correction factor in the above range (see
equation~\ref{eq:bayes}).

\section{Observations and Radius Measurements of Individual Sources}

\subsection{Thermonuclear Bursters}

There are five sources for which thermonuclear burst data have been
previously used to measure neutron star radii using their apparent
angular sizes, touchdown fluxes, and distances (see Table~1).

We also include in the present analysis 4U~1724$-$307, for which
Suleimanov et al.\ (2011) reported a radius measurement based on the
spectral evolution during the cooling tail of one long burst observed
from this source. As discussed in G\"uver et al.\ (2012b), the spectra
from that long burst used in the Suleimanov et al.\ (2011) study are
significantly different from blackbodies and from model atmosphere
spectra, resulting in $\chi^2/$d.o.f. in the range 1-8 in the spectral
fits (see also in't Zand \& Weinberg 2010). This indicates significant
contamination of the surface emission, either by the accretion flow or
by atomic lines from the ashes of the burst that have been brought up
to the photosphere, which makes the results unreliable. Instead, we
make use of the cooling tails of the two normal bursts observed from
\twentyfour\ to determine the apparent angular size (see G\"uver et
al.\ 2012b). The spectra from these two bursts show the expected
thermal shape and result in acceptable values for $\chi^2$/d.o.f. We
also make use of the touchdown flux measured from these bursts
(G\"uver et al.\ 2012a) when determining the neutron star radius.

Since the earliest measurements, G\"uver et al.\ (2012a,b) conducted
studies on the entire RXTE burst dataset and found $\sim 10\%$
systematic uncertainties in the apparent angular sizes and the
touchdown fluxes in the most prolific bursters.  In addition, G\"uver
et al.\ (2015) placed an upper limit of $\sim 1\%$ on the systematic
differences in the flux calibration between RXTE and {\it Chandra},
which, in principle, can affect the measured burst fluxes.

We reanalyze the data on these six sources uniformly, following the
procedures used and described in G\"uver et al.\ (2012
a,b). Specifically, {\it (i)} we apply the appropriate deadtime
correction to the observed countrates, which leads to a small increase
in the angular sizes and touchdown fluxes for the brightest sources.
{\it (ii)} We employ a Bayesian Gaussian-mixture model to quantify the
intrinsic scatter in the measurements of the angular sizes and the
touchdown fluxes. This is typically larger than the formal
uncertainties in the measurements and increases the uncertainties in
the inferred radii. {\it(iii)} When the number of Eddington-limited
bursts of an individual source is too small to assess the scatter in
the touchdown flux, we take an 11\% systematic uncertainty in this
quantity following the analysis of G\"uver et al.\ (2012a) on the
sample of sources with limited number of bursts. {\it (iv)} We add an
uncertainty of 1\% in the apparent angular sizes and the touchdown
fluxes to account for the flux calibration uncertainties. We summarize
all the measurements in Table~1 and discuss the additional details of
the source distances and atmospheric compositions for individual
sources below. Note that the uncertainties in Table~1 do not include
the 1\% flux calibration uncertainties, which we add in quadrature
when inferring the radii. We also list the ID numbers of the bursts
used in this study in Table~A2 of the Appendix, following the
numbering system used in Galloway et al.\ (2008a).

\begin{figure}
\centering
   \includegraphics[scale=0.4]{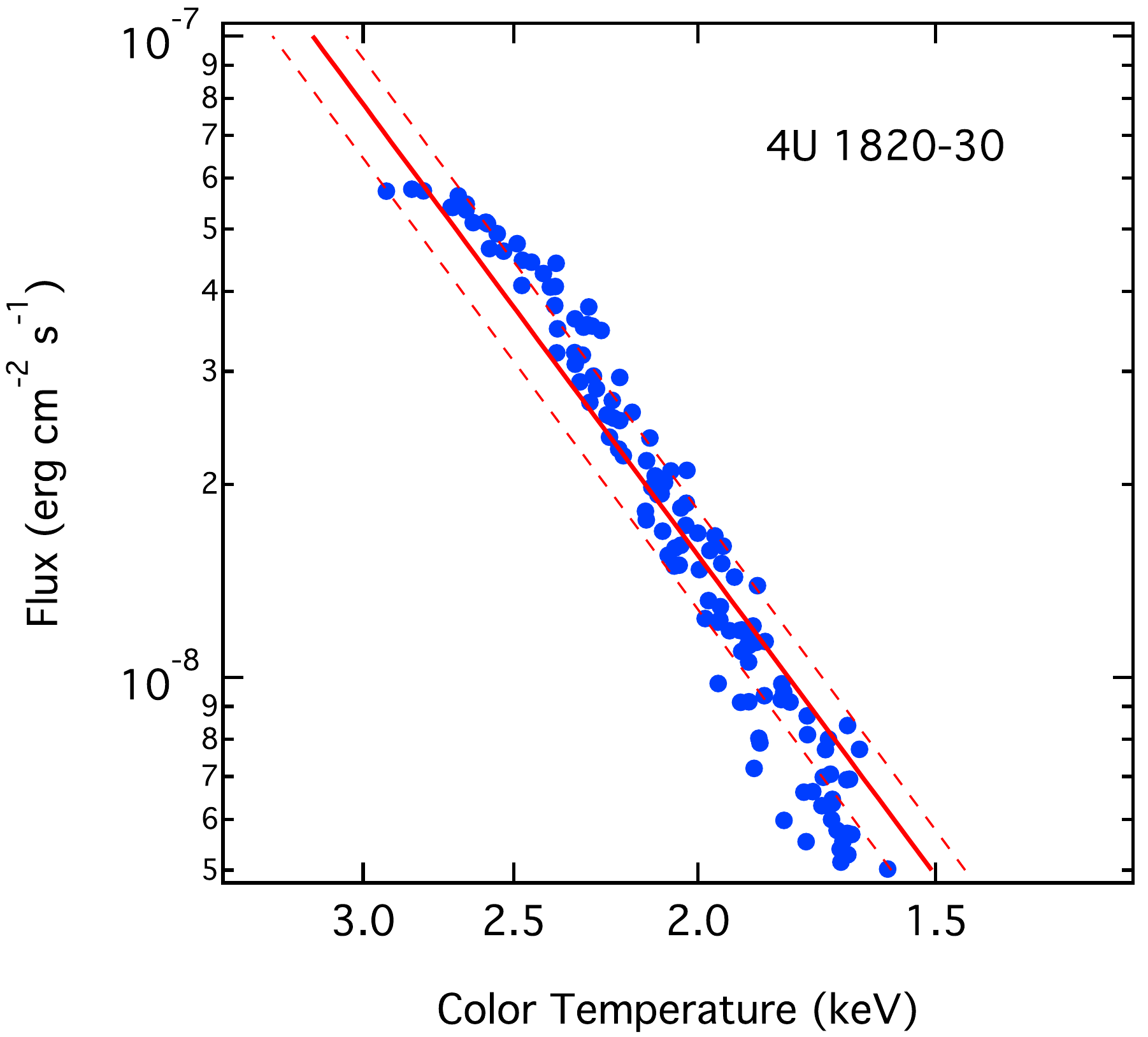}
   \includegraphics[scale=0.4]{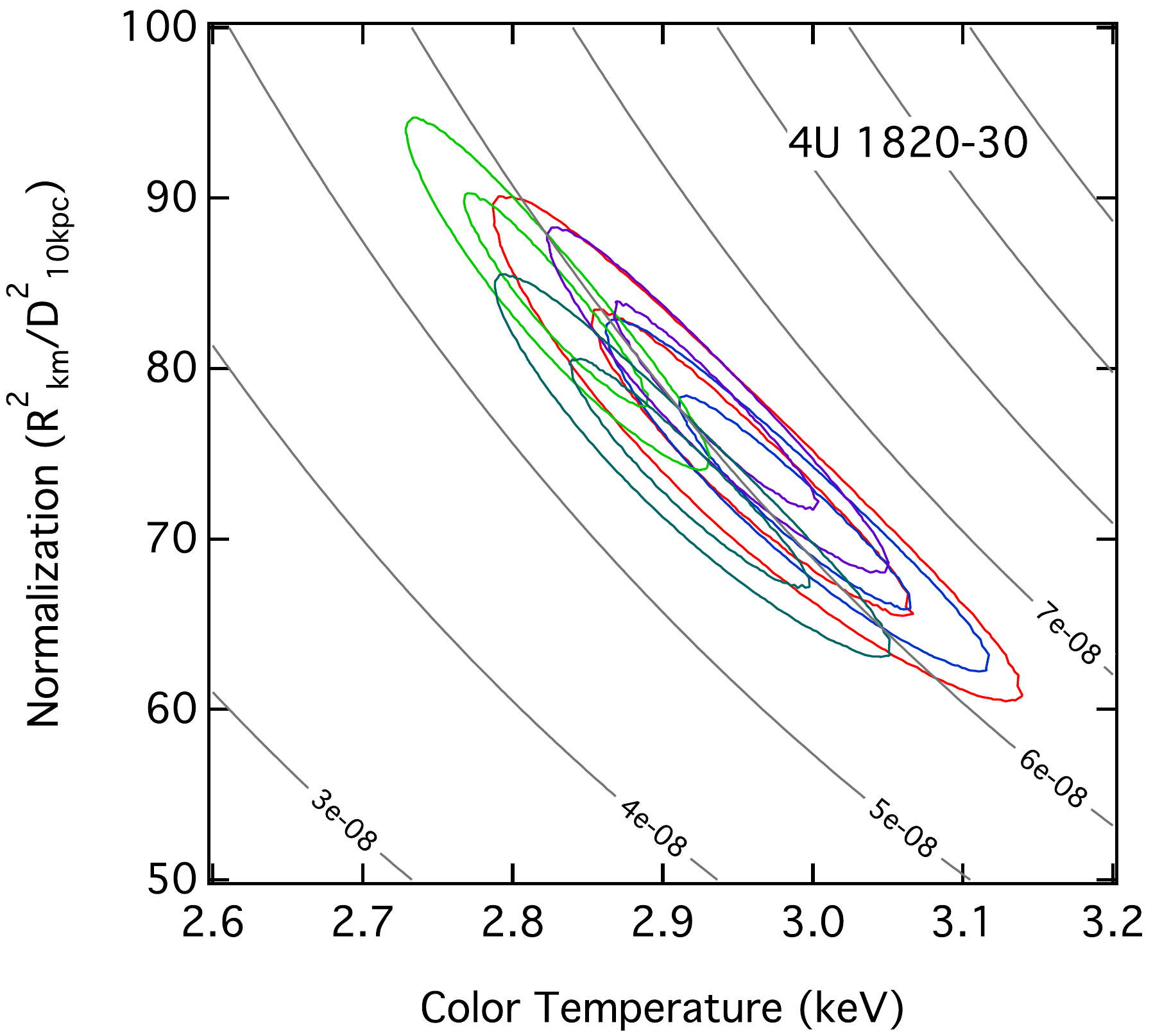}
   \includegraphics[scale=0.4]{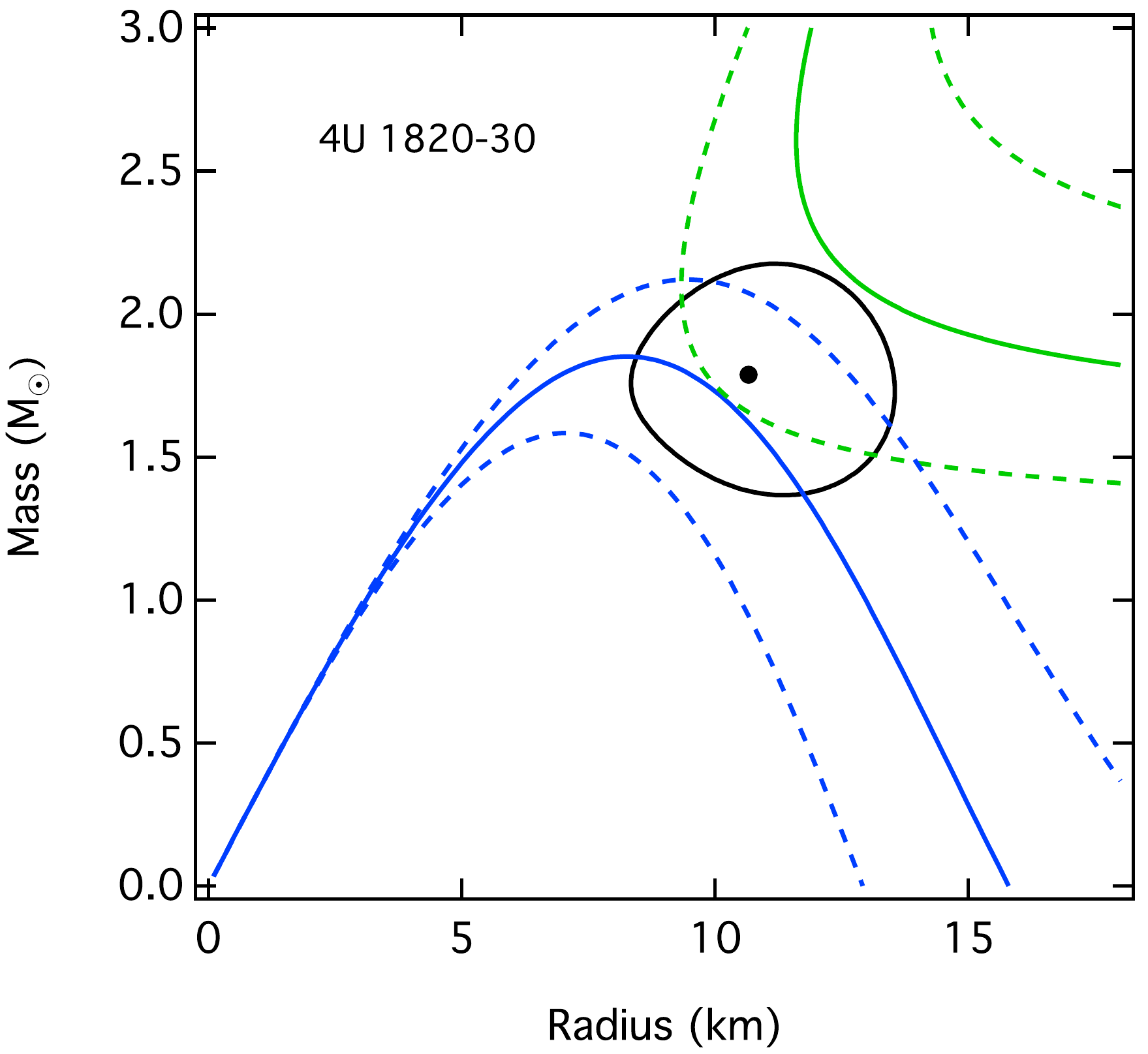}
   \includegraphics[scale=0.4]{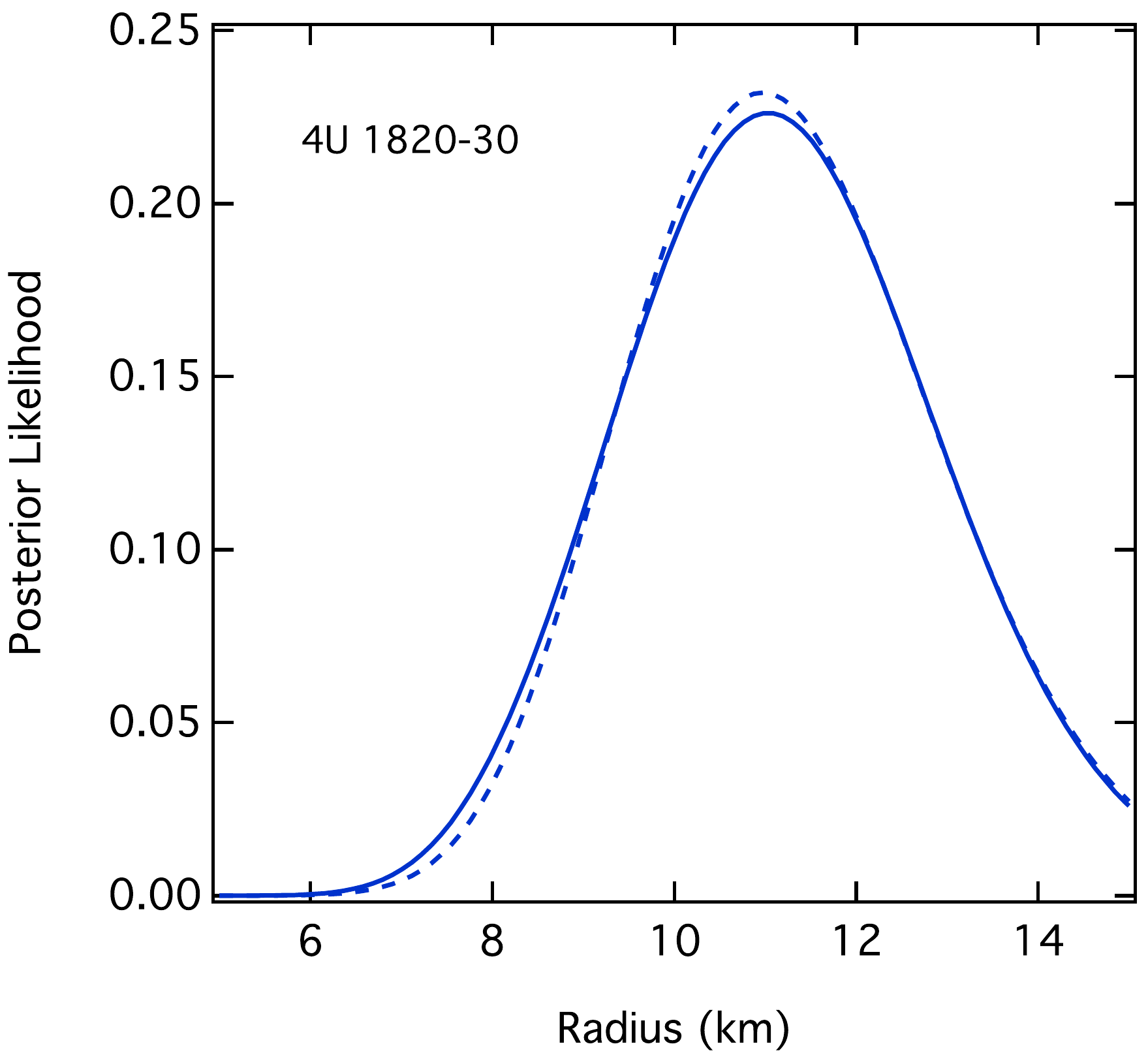}
\caption{{\em (Top Left)} The evolution of the flux and temperature
  measured for all the spectra in the cooling tails of bursts from
  \twenty. The diagonal red lines show the best-fit blackbody
  normalization and its 1$\sigma$ uncertainty. {\em (Top Right)} The
  1- and 2$\sigma$ confidence contours over the blackbody
  normalization and temperature measured at the touchdown moment in
  the PRE bursts. The black diagonal lines correspond to contours of
  constant touchdown flux. {\em (Bottom Left)} The black solid curve
  shows the 68\% confidence contour over the mass and radius of
  \twenty\ obtained by combining all the measurements and priors; the
  filled circle marks the location of the highest likelihood.  The
  solid blue and green lines denote the mass-radius curves obtained
  from the most likely values of the apparent angular size and
  Eddington flux, respectively, while the dashed curves denote the
  1$\sigma$ uncertainties of these measurements. {\em (Bottom right)}
  The posterior likelihood over the neutron star radius after the
  two-dimensional likelihoods are marginalized over mass. The dashed
  line assumes a flat prior over mass while the solid line assumes as
  a prior the observationally inferred mass distribution of recycled
  pulsars, as discussed in the text. }
\label{fig:data_1820} 
\end{figure}

\begin{figure}
\centering
   \includegraphics[scale=0.4]{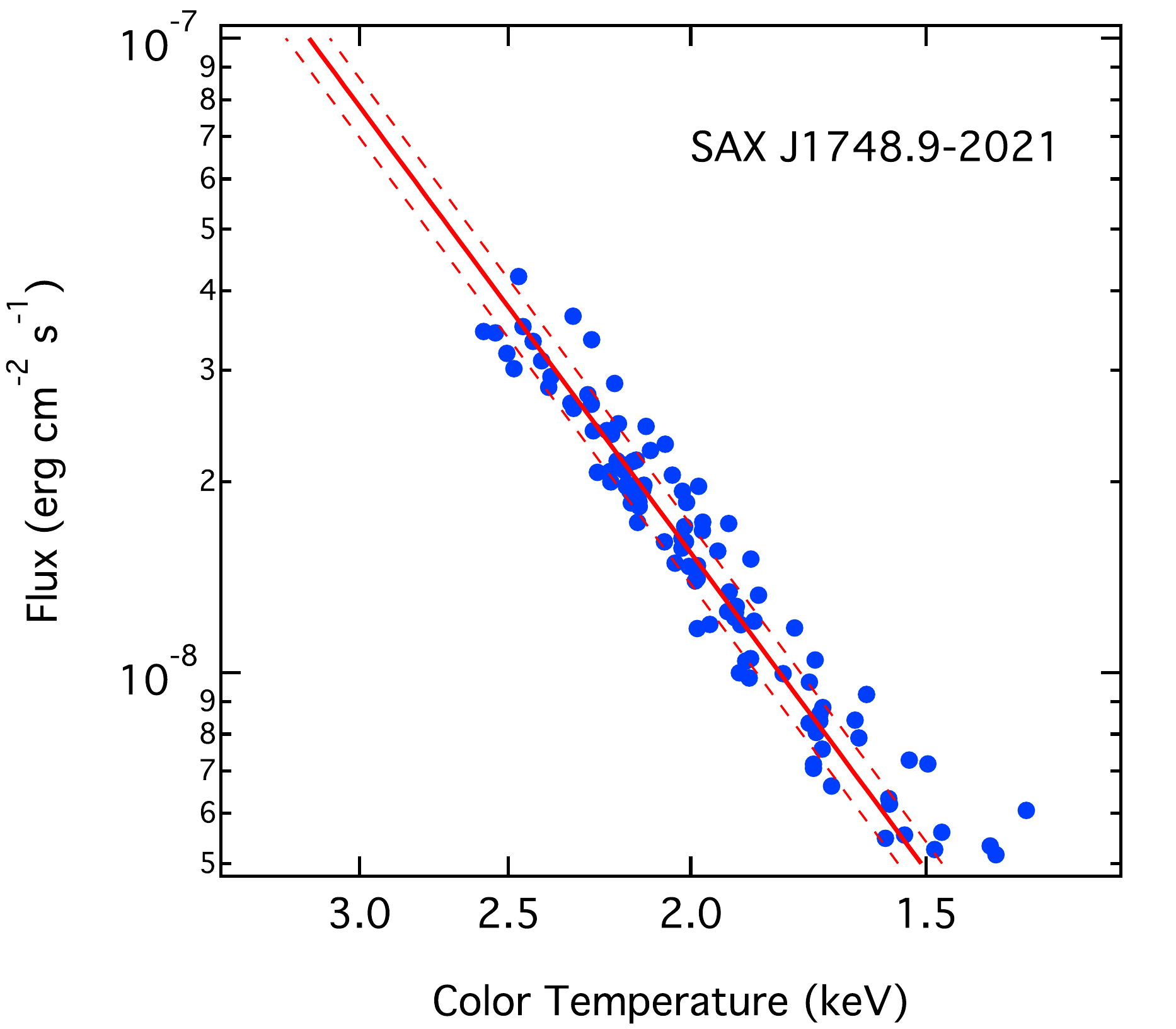}
   \includegraphics[scale=0.4]{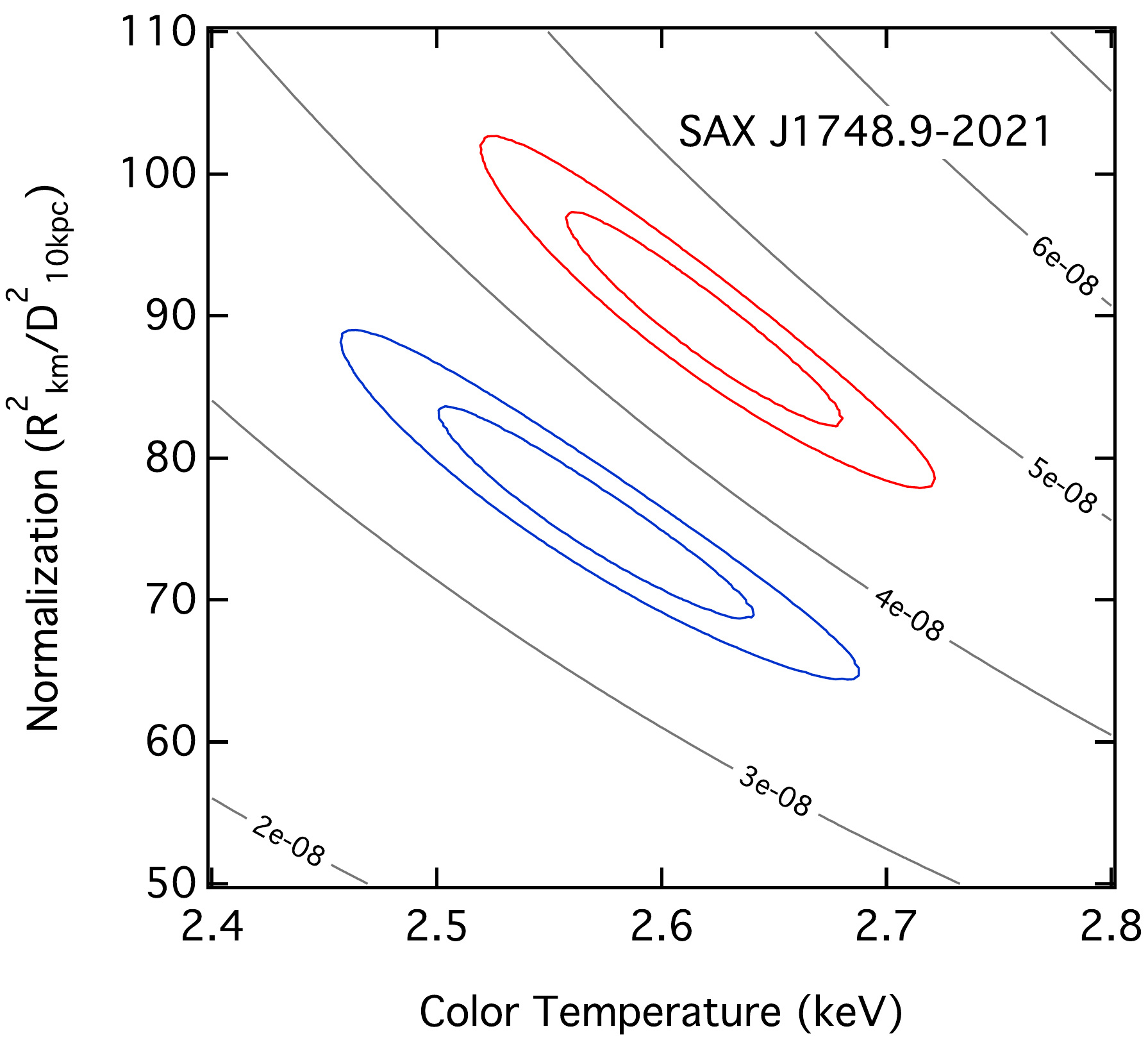}
   \includegraphics[scale=0.4]{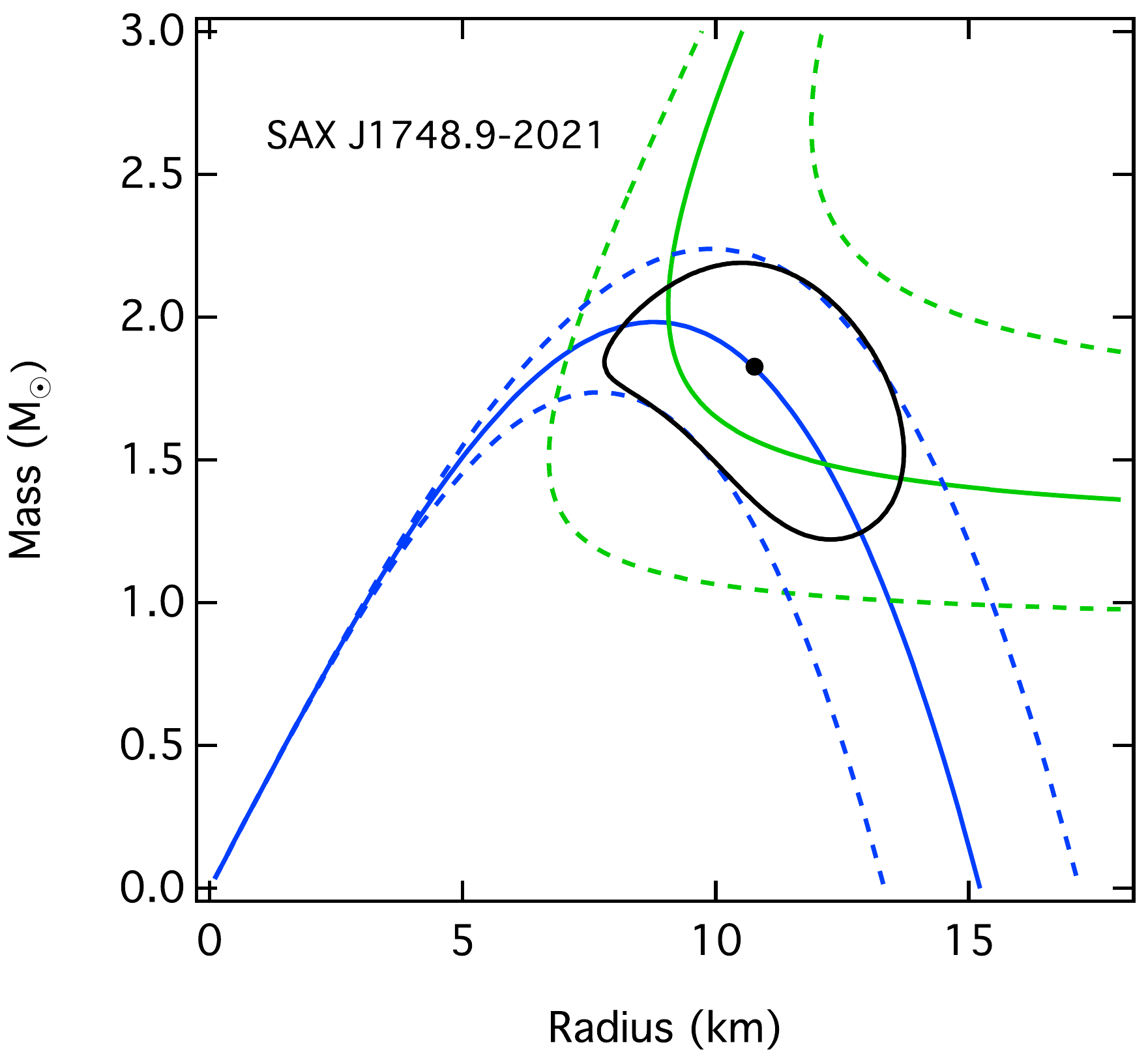}
   \includegraphics[scale=0.4]{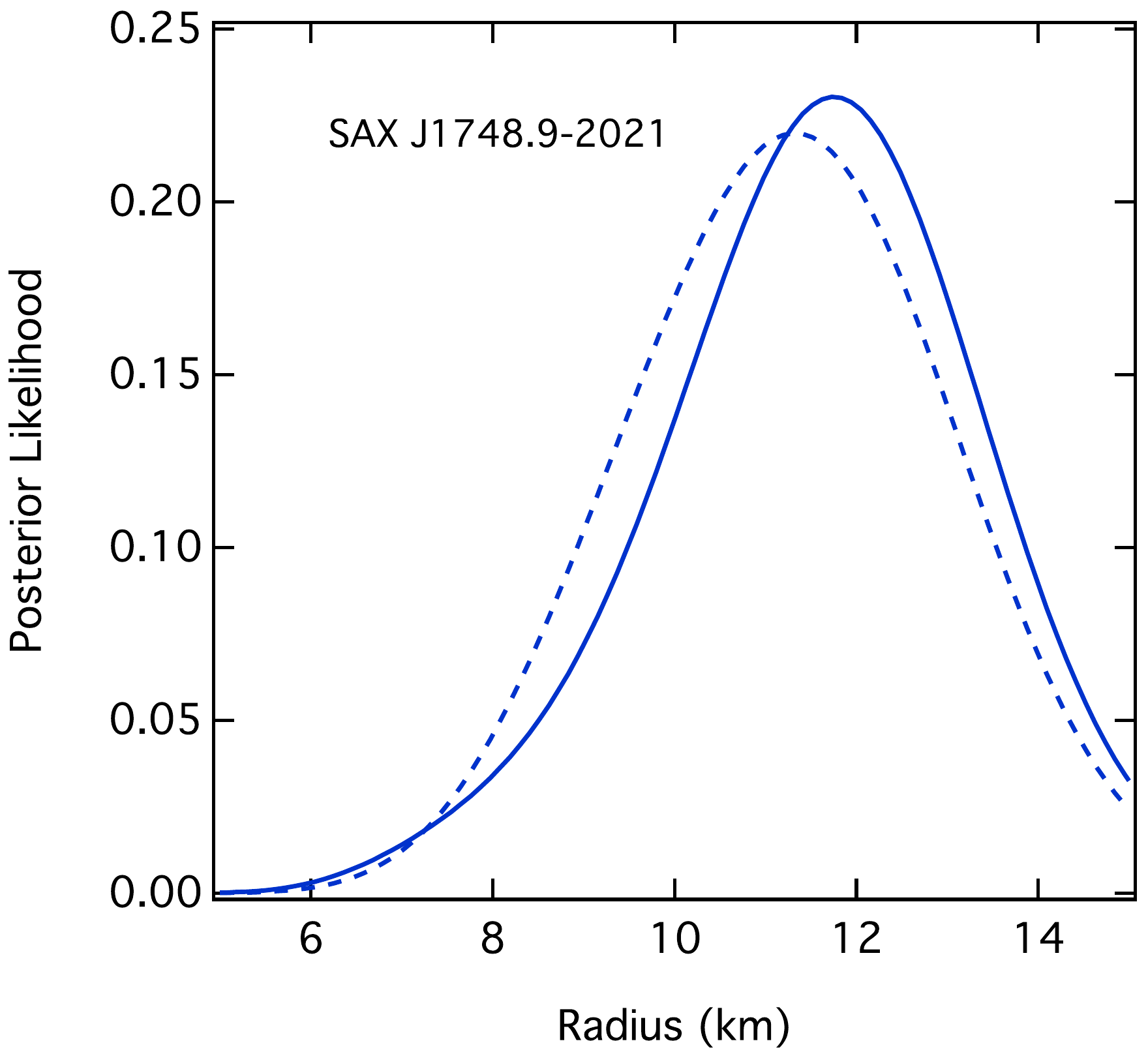}
\caption{Same as in Figure~\ref{fig:data_1820} but for \fortyeight.}
\label{fig:data_1748} 
\end{figure}

\subsubsection{\twenty}

\twenty\ is an ultracompact binary in the metal-rich globular cluster
NGC~6624. G\"uver et al.\ (2010) discussed two distance measurements
performed in the optical (Kuulkers et al.\ 2003) and in the near-IR
bands (Valenti et al.\ 2007). The first gives a distance of
$7.6\pm0.4$~kpc and the second gives $8.4\pm0.6$~kpc.  Harris et
al. (1996; 2010 revision) find a compatible distance estimate of
7.9~kpc in the optical band, with an uncertainty of
0.4~kpc\footnote[1]{see
  http://physwww.physics.mcmaster.ca/$\sim$harris/mwgc.ref}. Without any
further information to choose between the optical and the near-IR
measurements, G\"uver et al.\ (2010b) combined them into a single
boxcar likelihood between 6.8 to 9.6~kpc, which placed more than
warranted likelihoods at the shortest and intermediate
distances. Here, we instead opt to use a double Gaussian likelihood
with means and standard deviations that reflect the individual
measurements of Kuulkers et al.\ (2003) and Valenti et al.\ (2007) and
give equal integrated likelihood to each.

The fact that the neutron star is in a 11.4 minute binary (Stella et
al.\ 1987) requires that it is fed by a degenerate dwarf companion
that is free of hydrogen. For this reason, when inferring the radii,
we set the hydrogen abundance to $X=0$. No burst oscillations or
persistent pulsations have been observed from this source. Because of 
this, when applying spin corrections to the apparent angular size,
we assume a flat prior in spin between 250 and 650~Hz, as discussed in
Section 2.3.

The top left panel of Figure~\ref{fig:data_1820} shows the evolution
of the flux and temperature during the cooling tails of five bursts
observed from \twenty, while the top right panel shows the 68\% and
95\% confidence contours in the measurement of the blackbody
normalization vs. temperature during the touchdown phases in the five
Eddington-limited bursts. Because the intrinsic scatter in the
touchdown flux of this source is very small, we assign an 11\%
uncertainty to this measurement as discussed above. 

The lower left panel shows the 68\% and 95\% confidence contours over
the mass and radius of \twenty\ inferred within the Bayesian framework
discussed in Section 2.3, along with the contours of constant apparent
angular size (blue) and touchdown flux (green) obtained for this
source. The lower right panel shows the likelihood over the radius
when we marginalized the two-dimensional likelihood over mass. We do
this for a flat prior on mass between 0 and 3~$M_\odot$ as well as for
the observed mass distribution of fast radio pulsars, which are the
descendants of the low-mass X-ray binaries that make up our sample. As
discussed in \"Ozel et al.\ (2012b), the latter mass distribution can
be represented by a Gaussian with a mean of $1.46~M_\odot$ and a
dispersion of $0.21~M_\odot$. The difference in the result between
using the two different priors over the mass is minor. (Note that we
use the full two-dimensional likelihoods without these observational
priors on mass when inferring the parameters of the equation of
state.)

\begin{figure}
\centering
   \includegraphics[scale=0.4]{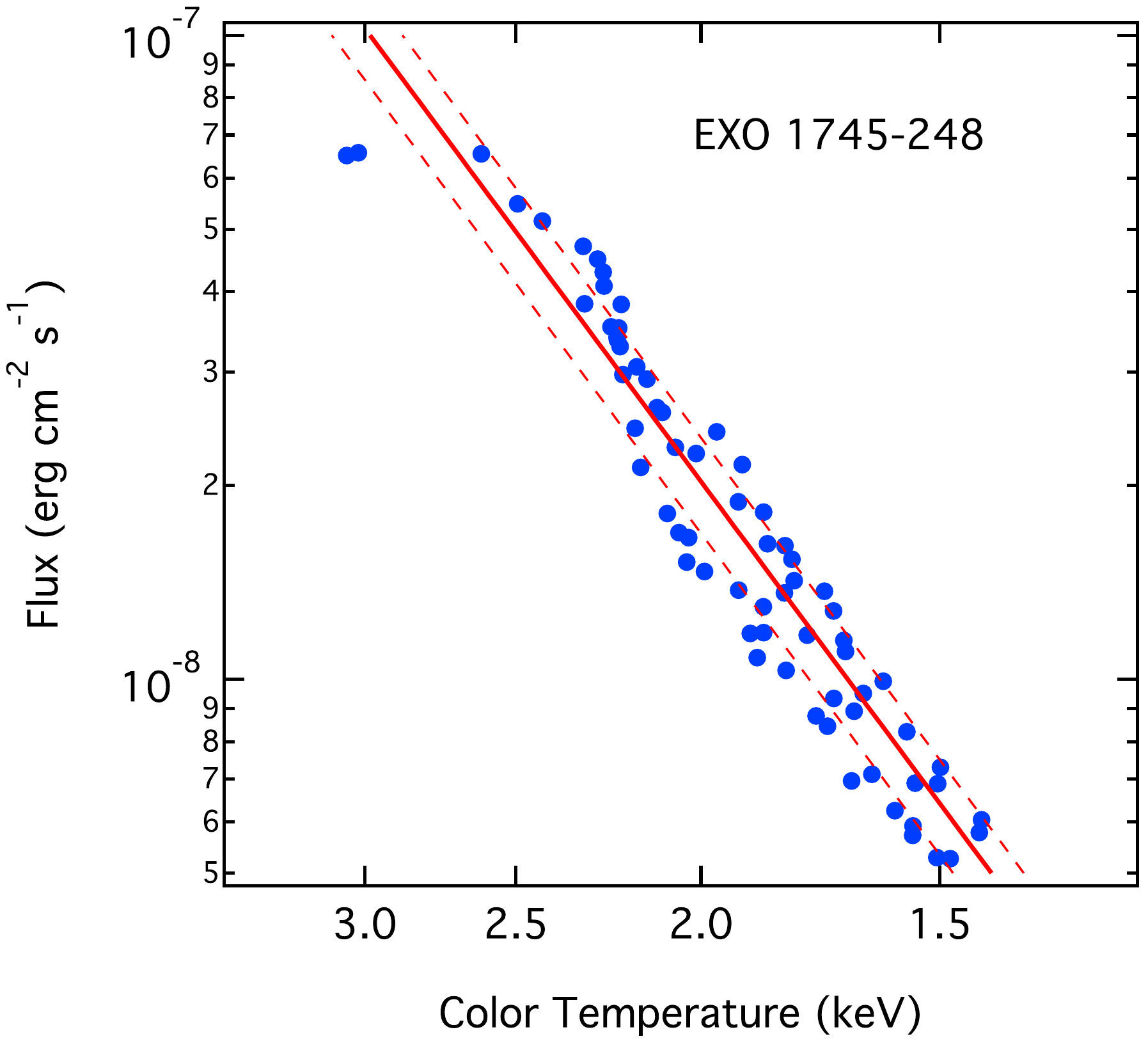}
   \includegraphics[scale=0.4]{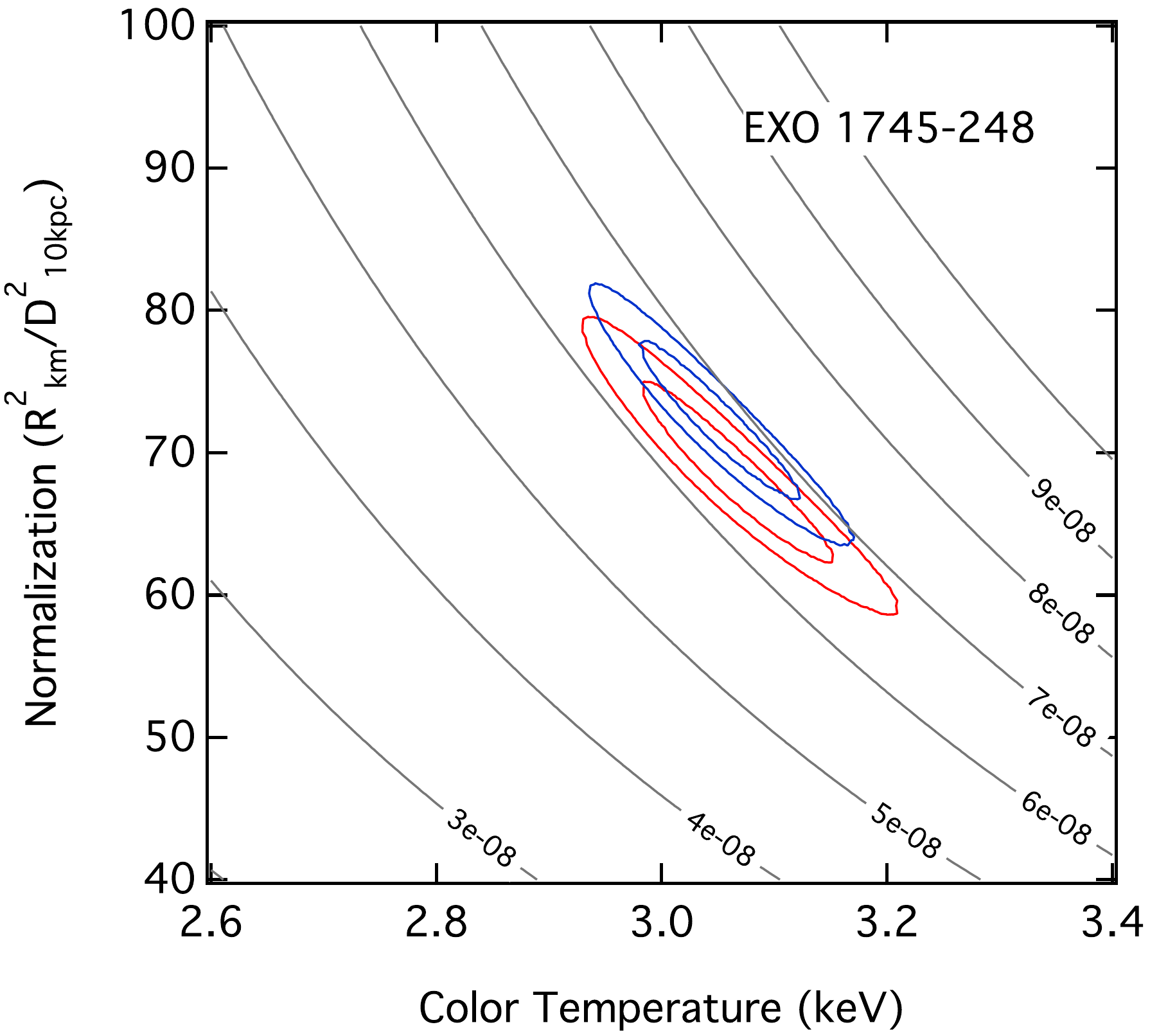}
   \includegraphics[scale=0.4]{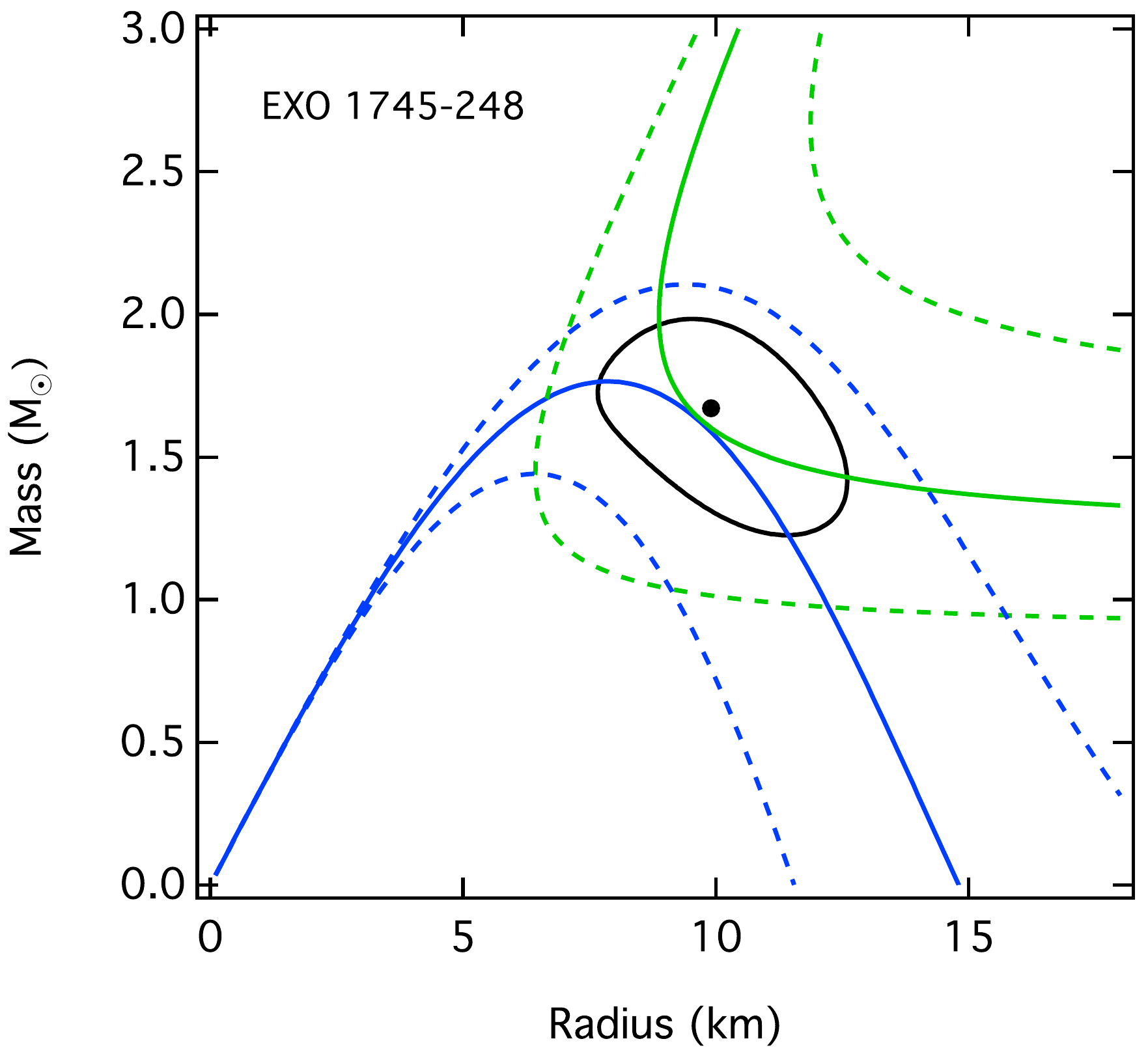}
   \includegraphics[scale=0.4]{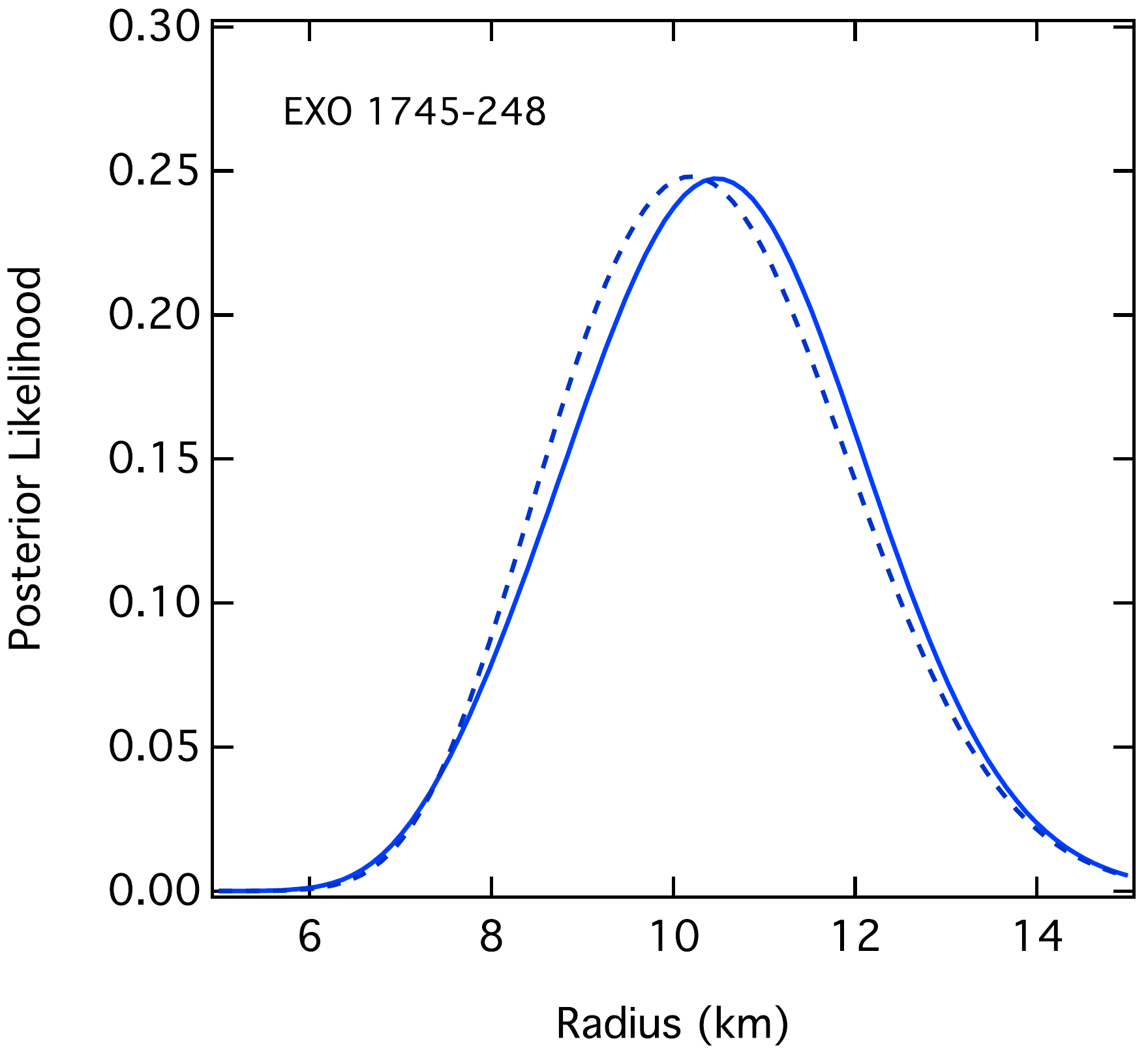}
\caption{Same as in Figure~\ref{fig:data_1820} but for \fortyfive.}
\label{fig:data_1745} 
\end{figure}

\subsubsection{\fortyeight}

The transient neutron star X-ray binary \fortyeight\ is located
in the globular cluster NGC~6440, which is a massive and old cluster
in the Galactic bulge. Two optical and one near-IR studies give
consistent and well-constrained distances to NGC~6440: Kuulkers et
al.\ (2003) reported $8.4^{+1.5}_{-1.3}$~kpc, Harris et al.\ (2010)
found 8.5~kpc, while Valenti et al.\ (2007) found $8.2 \pm 0.6$~kpc
using near-IR data. In this last study, the distance uncertainty is
improved and takes into account the systematic errors introduced by
the method of comparing the properties of NGC~6440, including its
metallicity and age, to the reference cluster. Because the central
values of the two measurements differ by less than the 1$\sigma$
uncertainty of either, we adopt here the latter distance and its
uncertainty.

\fortyeight\ has a spin frequency of 420~Hz, detected during
intermittent pulsations observed in the persistent emission
(Altamirano et al.\ 2008). The same study also found a binary orbital
period of 8.7~hr. Because there is no specific information about the
evolutionary state of the donor, we take a flat prior in the hydrogen
mass fraction between 0 and 0.7.

The top left panel of Figure~\ref{fig:data_1748} shows the evolution
of the flux and temperature during the cooling tails of four bursts
observed from \fortyeight, while the top right panel shows the 68\%
and 95\% confidence contours in the measurement of the blackbody
normalization vs. temperature during the touchdown phases in its two
Eddington-limited bursts. 

The lower panels of Figure~\ref{fig:data_1748} show the 68\%
confidence contour in mass and radius as well as the posterior
likelihood marginalized over mass using the Bayesian framework and
priors discussed above.

\begin{figure}
\centering
   \includegraphics[scale=0.4]{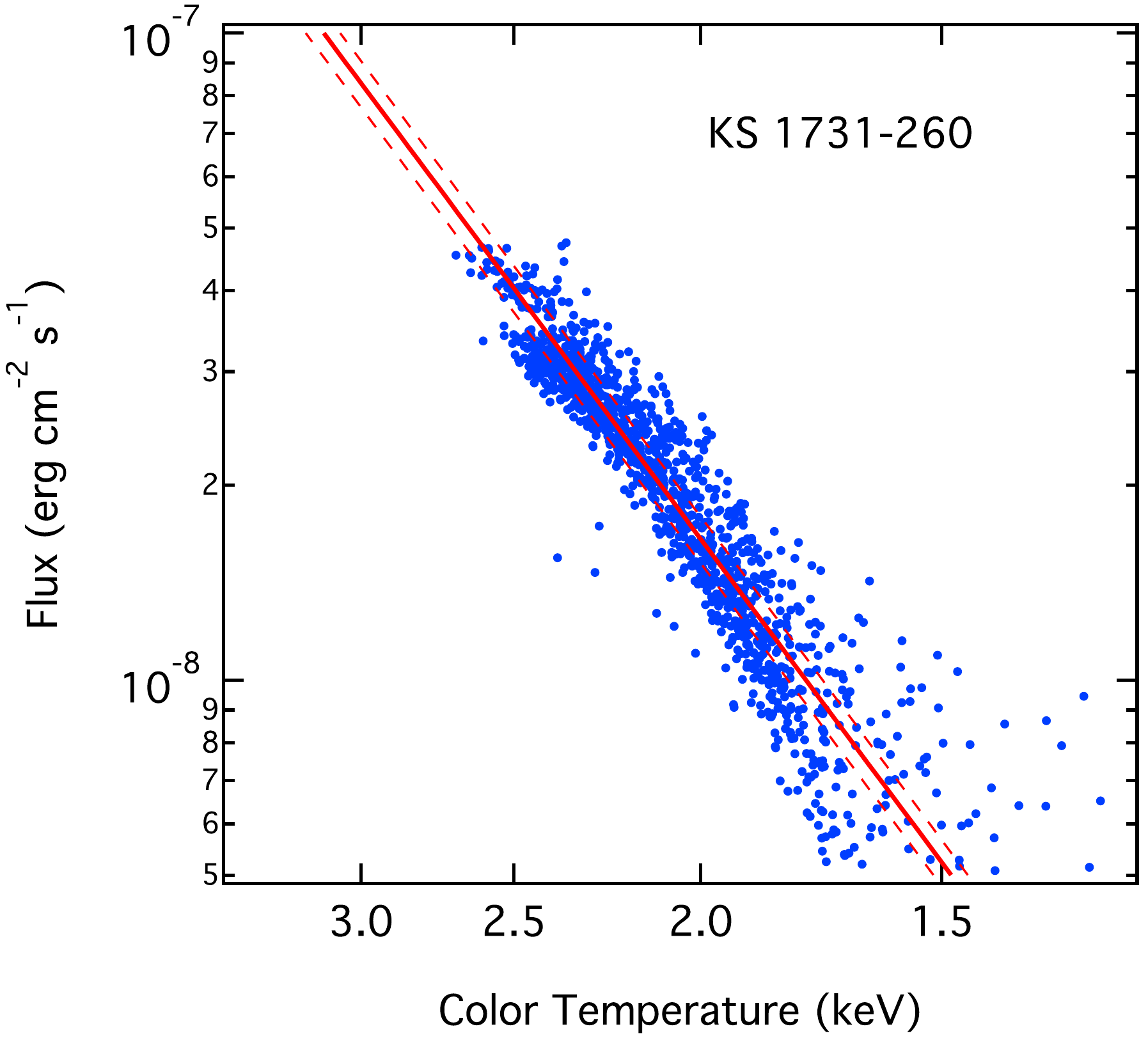}
   \includegraphics[scale=0.4]{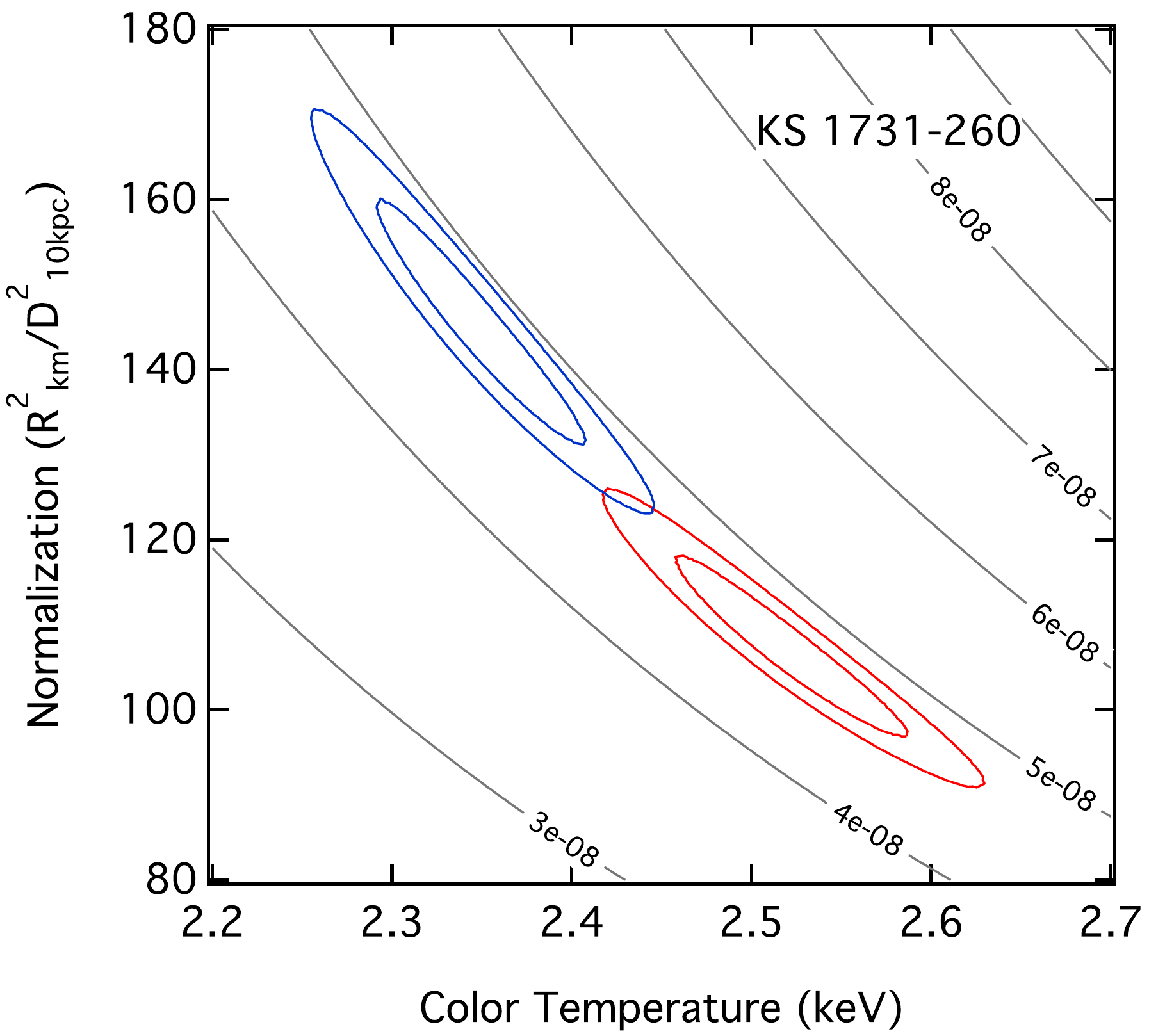}
   \includegraphics[scale=0.4]{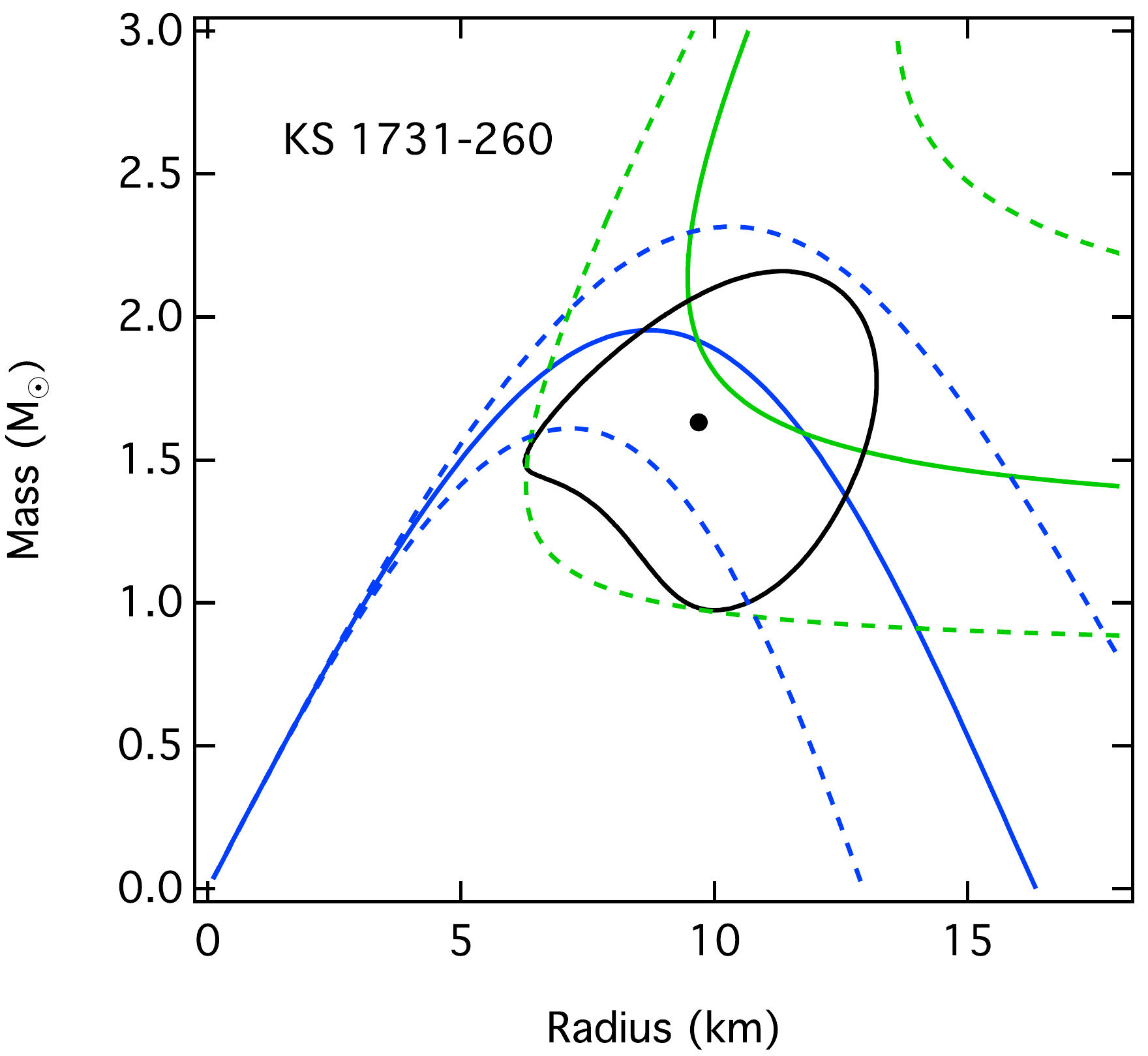}
   \includegraphics[scale=0.4]{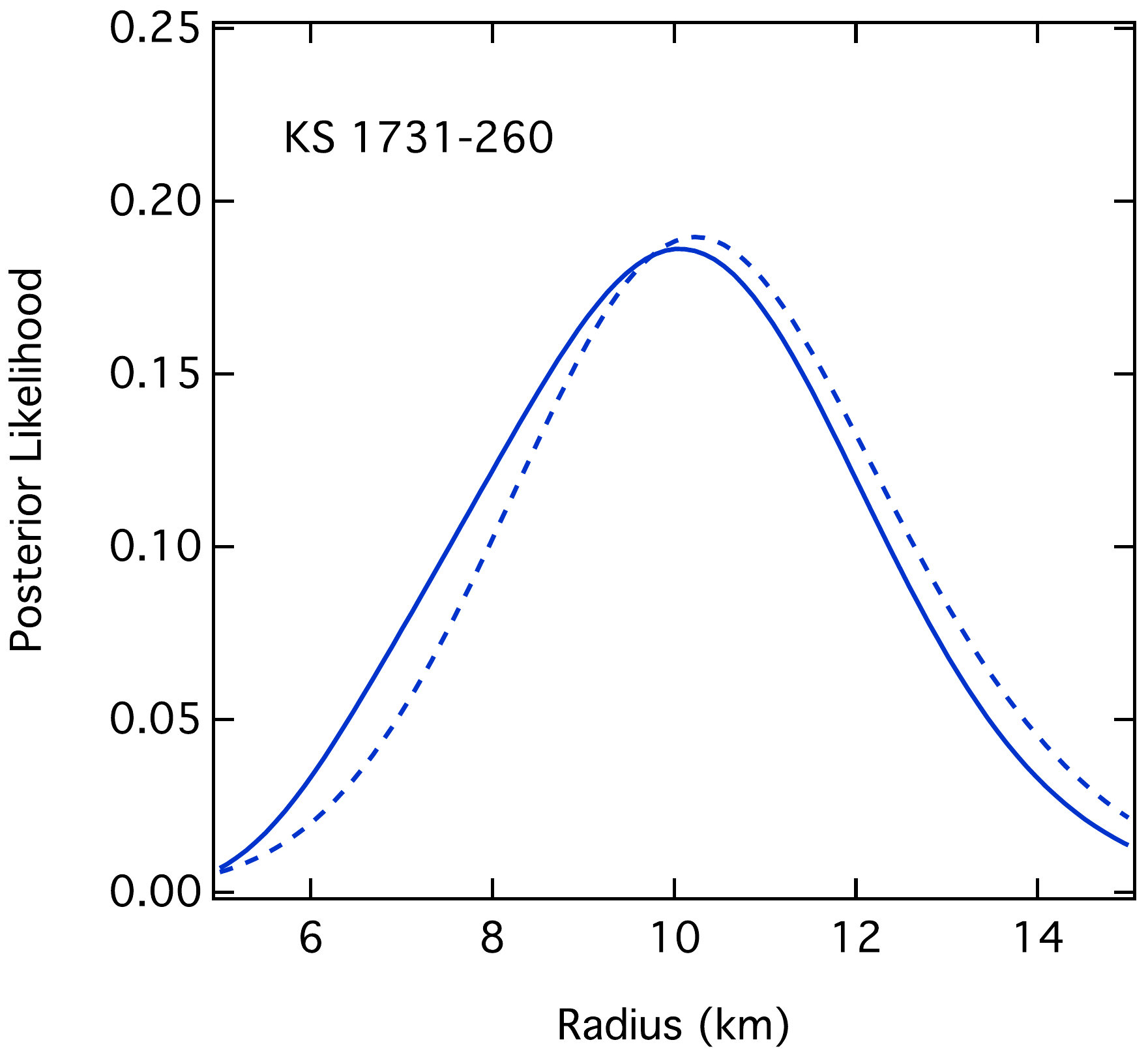}
\caption{Same as in Figure~\ref{fig:data_1820} but for \thirtyone.}
\label{fig:data_1731} 
\end{figure}

\subsubsection{\fortyfive}

\fortyfive\ is located in Terzan 5, one of the most metal-rich
globular clusters in the Galaxy. The distance to Terzan~5 was obtained
using HST/NICMOS data (Ortolani et al.\ 2007). The sources of
uncertainty in the distance measurement were discussed in detail in
\"Ozel et al.\ (2009). We adopt here the same flat likelihood over
distance centered at 6.3~kpc with a width of 0.63~kpc.

No burst oscillations or persistent pulsations have ever been observed
from \fortyfive. As before, we adopt a flat prior over its spin
frequency between 250 and 650~Hz when calculating the spin corrections
to the apparent angular size. The nature of the companion of
\fortyfive\ is ambiguous (Heinke et al.\ 2003). While the empirical
comparison of its spectrum to those of ultracompact sources suggested
an ultracompact binary with a hydrogen-poor companion, the
identification of a possible infrared counterpart leaves open the
possibility of a hydrogen-rich donor. To account for both
possibilities, we take a flat prior over the hydrogen mass fraction in
the range $X=0-0.7$.

The top panels of Figure~\ref{fig:data_1745} show the evolution of the
flux and temperature during the cooling tails of two bursts (left) and
the 68\% and 95\% confidence contours in the measurement of the
blackbody normalization vs. temperature during the touchdown phases of
these two Eddington-limited bursts (right). 

The lower panels of Figure~\ref{fig:data_1745} show, as before, the
68\% confidence contour in mass and radius (left) derived in the
Bayesian framework from the measurements of the apparent angular size,
touchdown flux, and the distance, as well as the posterior likelihood
marginalized over mass (right).

\subsubsection{\thirtyone}

\thirtyone\ is a binary in the Galactic bulge, lying in the direction
of Baade's window. \"Ozel et al.\ (2012a) derived a distance prior to
this source based on the stellar density along the line of sight. We
use the same numerical prior in the current study, which places
\thirtyone\ at a distance of approximately 7$-$9~kpc.

\thirtyone\ has a spin frequency of 524~Hz based on the detection of
burst oscillations (Smith et al.\ 1997). Its optical counterpart has
been identified (Zurita et al. 2010) and the duration and the
energetics of some of its X-ray bursts point to accreted fuel that is
hydrogen-rich. Nevertheless, because there is no conclusive evidence
on the hydrogen content of the bursts we analyze here, we allow for a
flat distribution in the hydrogen mass fraction $X$ between 0 and 0.7.

We show in the top panels of Figure~\ref{fig:data_1731} the flux
vs. temperature observed during the cooling tails of twenty four X-ray
bursts used for the measurement of the apparent angular size and the
68\% and 95\% confidence contours in the blackbody normalization and
temperature measured during the touchdown phases of two Eddington
limited bursts.

The lower left panel of Figure~\ref{fig:data_1731} shows the 68\%
confidence contours over the mass and radius of \thirtyone\ inferred
within the Bayesian framework, along with the contours of constant
apparent angular size (blue) and touchdown flux (green) obtained for
this source. The lower right panel shows the likelihood over the
radius when we marginalized the two-dimensional likelihood over mass.

\begin{figure}
\centering
   \includegraphics[scale=0.4]{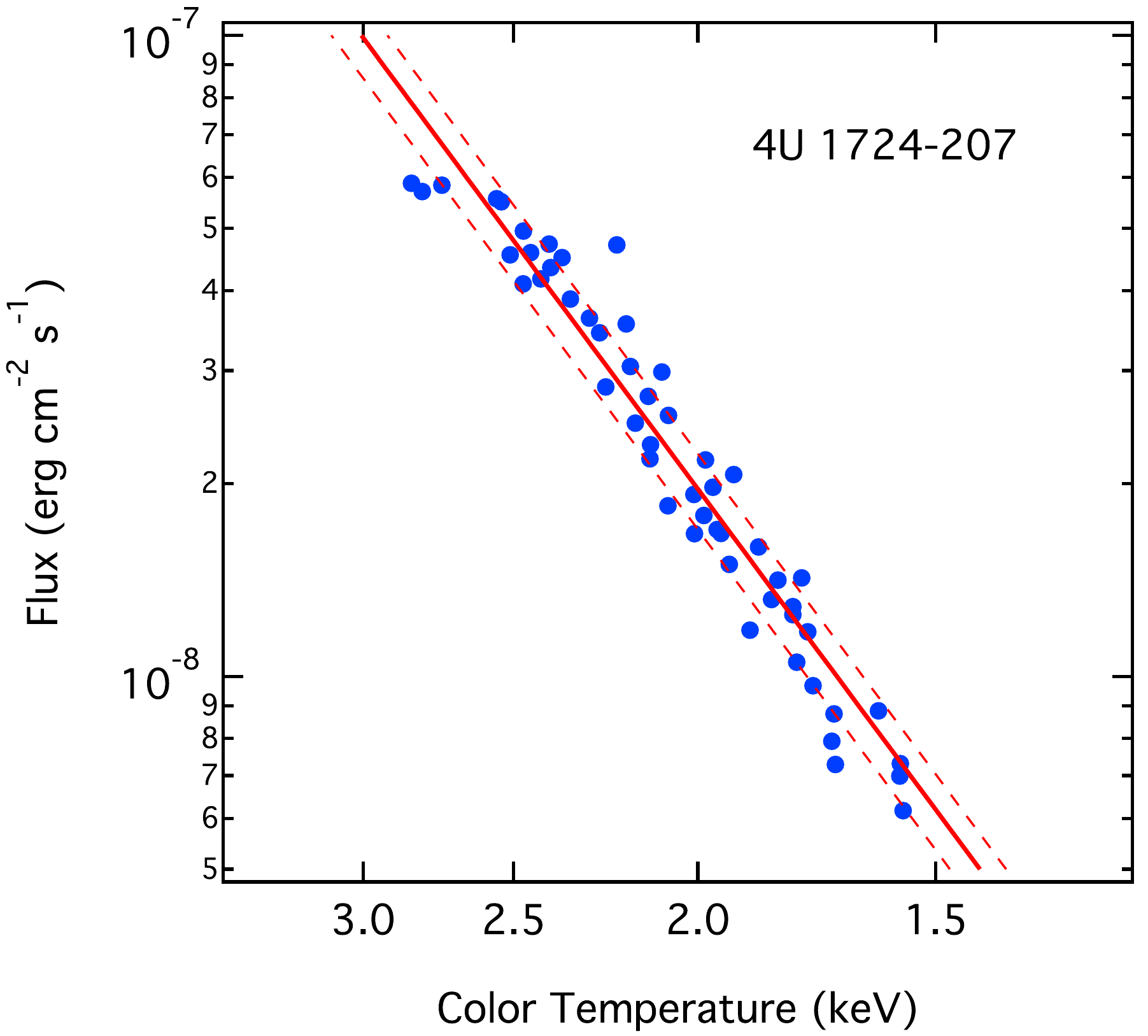}
   \includegraphics[scale=0.4]{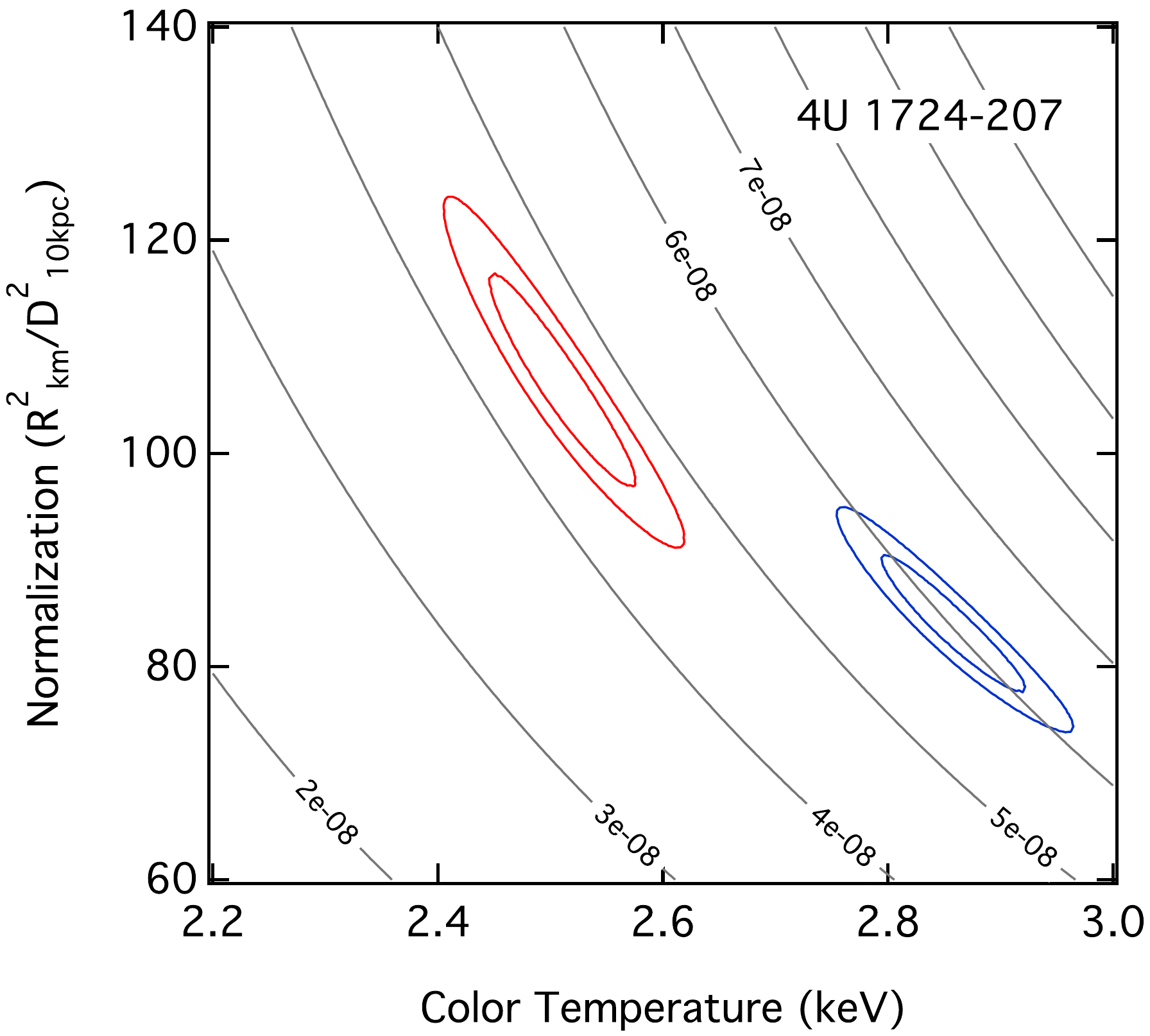}
   \includegraphics[scale=0.4]{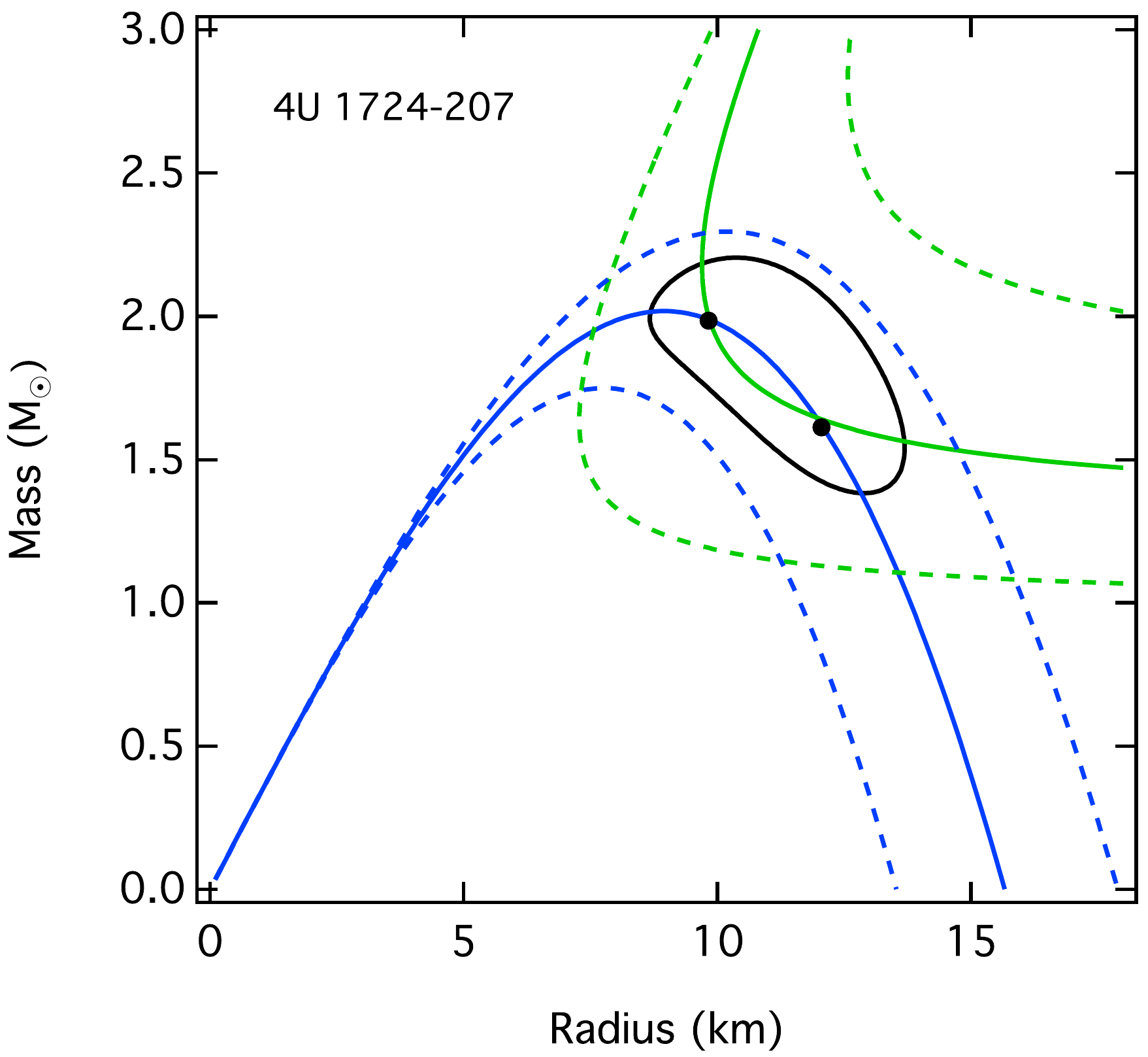}
   \includegraphics[scale=0.4]{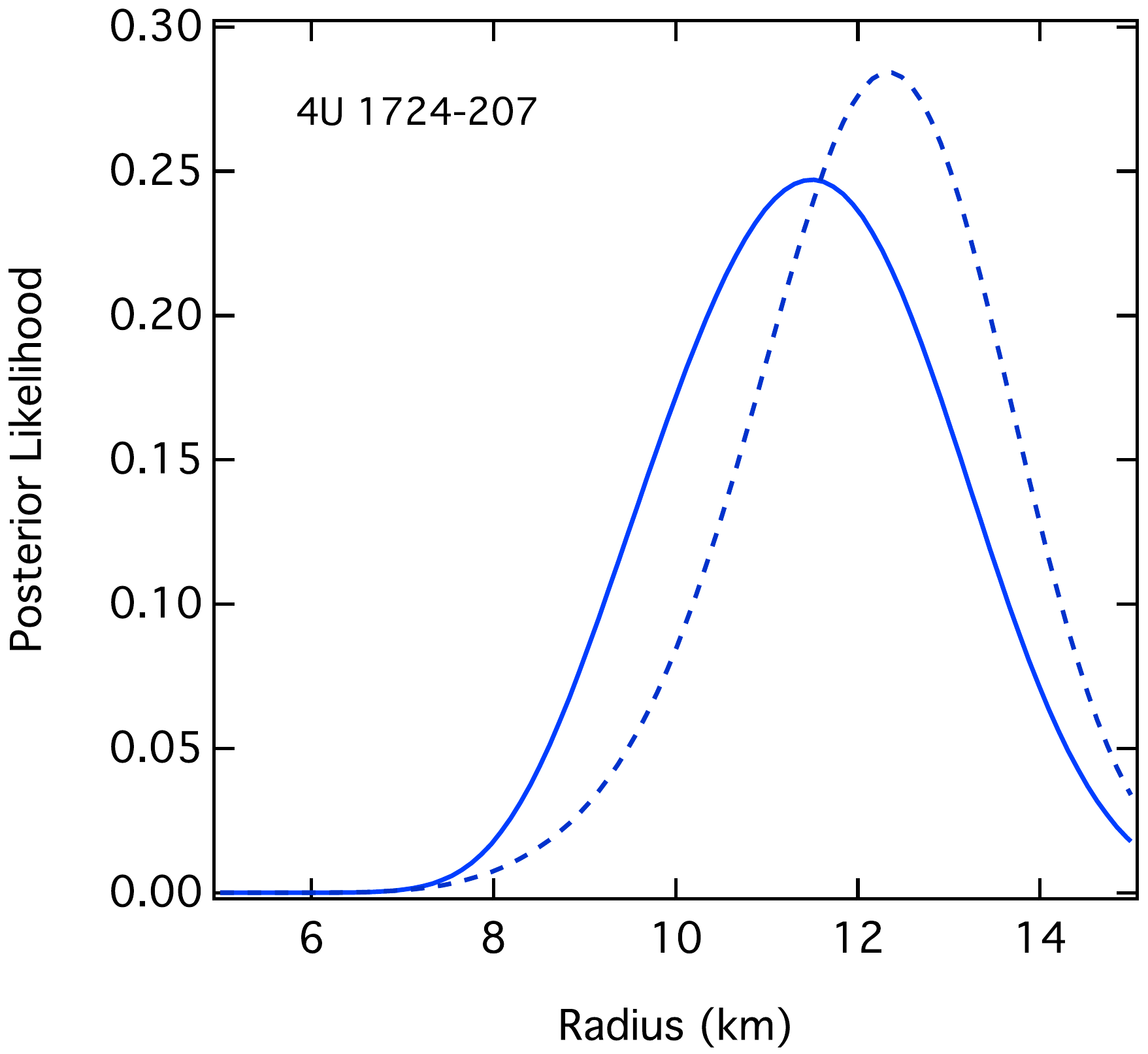}
\caption{Same as in Figure~\ref{fig:data_1820} but for \twentyfour.}
\label{fig:data_1724} 
\end{figure}

\subsubsection{\twentyfour}

\twentyfour\ lies in the globular cluster Terzan~2. Early studies of
the distance to this cluster by Ortolani et al.\ (1997) obtained a
distance of 5.3 or 7.7~kpc, depending on whether the selective
extinction, R, was set to 3.1 or 3.6.  The most recent study by
Valenti et al.\ (2012) used near-IR observations of red giant branch
stars and led to a distance of 7.4~kpc. Independent of the reddening
or color-magnitude measurements, one can statistically argue that the
distance of Terzan~2 should be the same as the distance to the
Galactic center (Racine \& Harris 1989). This is because the whole
system of globular clusters is centrally concentrated around the
Galactic center and, given the fact that the direction of Terzan 2 is
within the Galactic bulge region, it is likely that its distance is
close to 8.0~kpc (Reid 1993). Based on these arguments and in order to
avoid using a measurement that depends strongly on the assumed
extinction, we adopt the recent measurement of Valenti et al. (2012)
of 7.4$\pm$0.5~kpc. The error primarily reflects the systematic
uncertainty in the measurements of the distances to globular clusters,
as estimated by using 47~Tuc as a reference.

There have been no studies on the composition of the companion to
\twentyfour\ and no detected burst oscillations or persistent
pulsations from this source. For this reason, when inferring its
radius, we use a flat distribution in hydrogen abundance between
X=0 and X=0.7 and a flat distribution in spin frequency between 250
and 650~Hz.

In the top left panel of Figure~\ref{fig:data_1724}, we show the flux
vs.\ temperature diagram during the cooling tails of the bursts for
which the blackbody model provides an acceptable fit to the data (see
discussion in G\"uver et al.\ 2010b). In the top right panel, we show
the 68\% and 95\% confidence contours in the measured blackbody
normalization vs.\ temperature during the touchdown phases of two
Eddington-limited bursts. Finally, in the lower two panels of the same
figure, we show {\em (left)\/} the 65\% confidence contours in the
inferred mass and radius of \twentyfour\ and {\em (right)\/} the
posterior likelihood over radius, after we marginalize over the mass
of the neutron star.

\begin{figure}
\centering
   \includegraphics[scale=0.4]{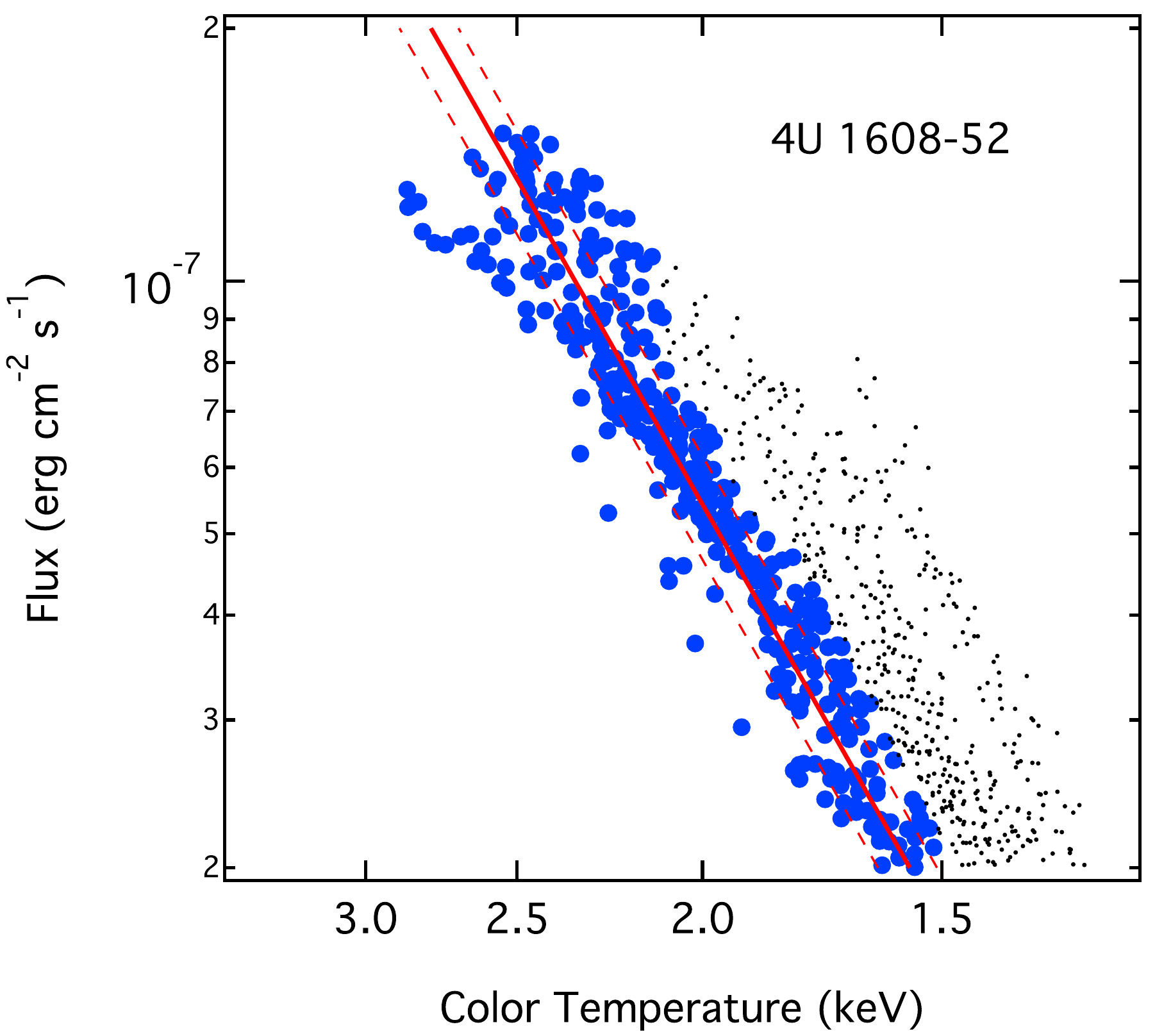}
   \includegraphics[scale=0.4]{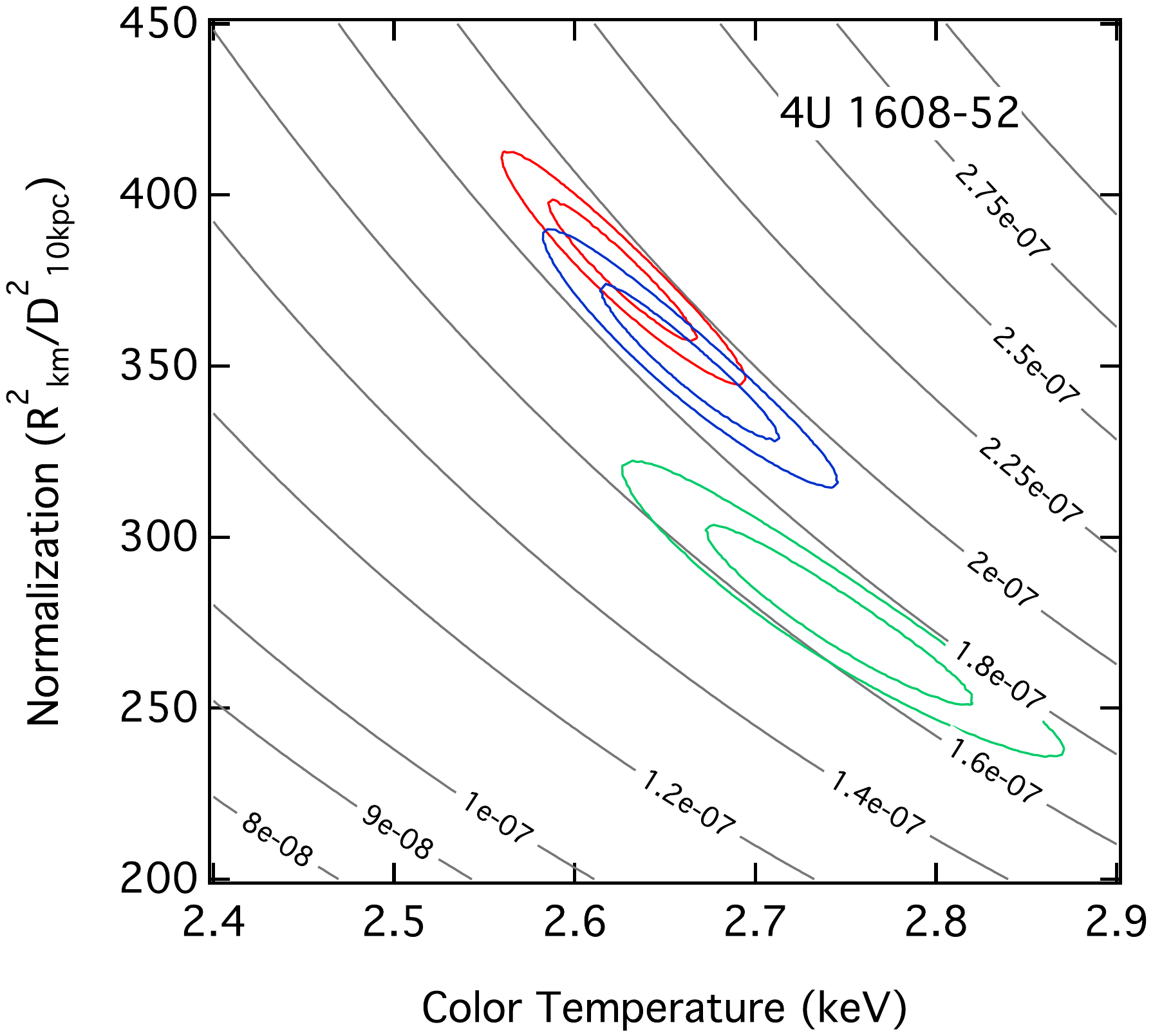}
   \includegraphics[scale=0.4]{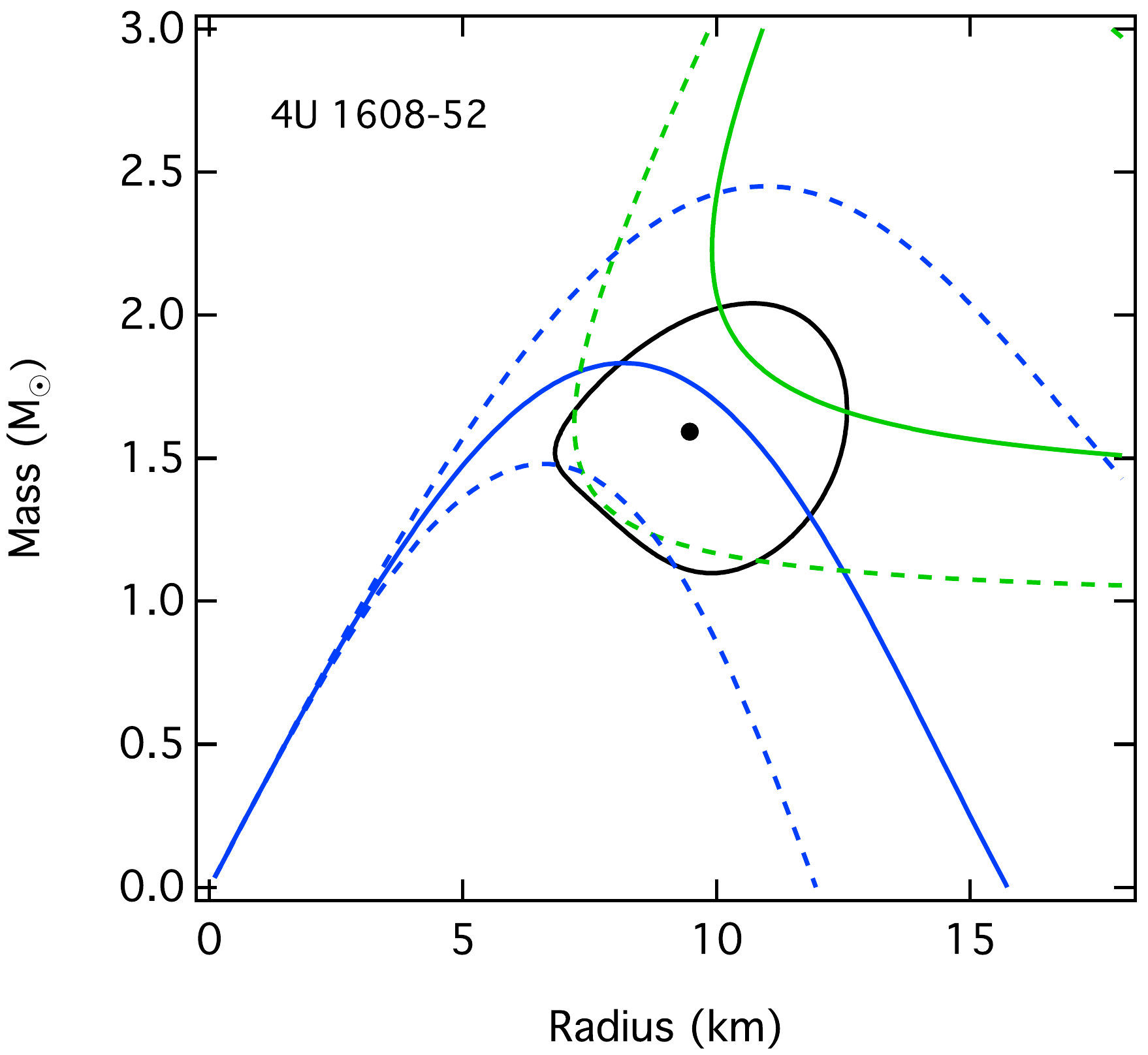}
   \includegraphics[scale=0.4]{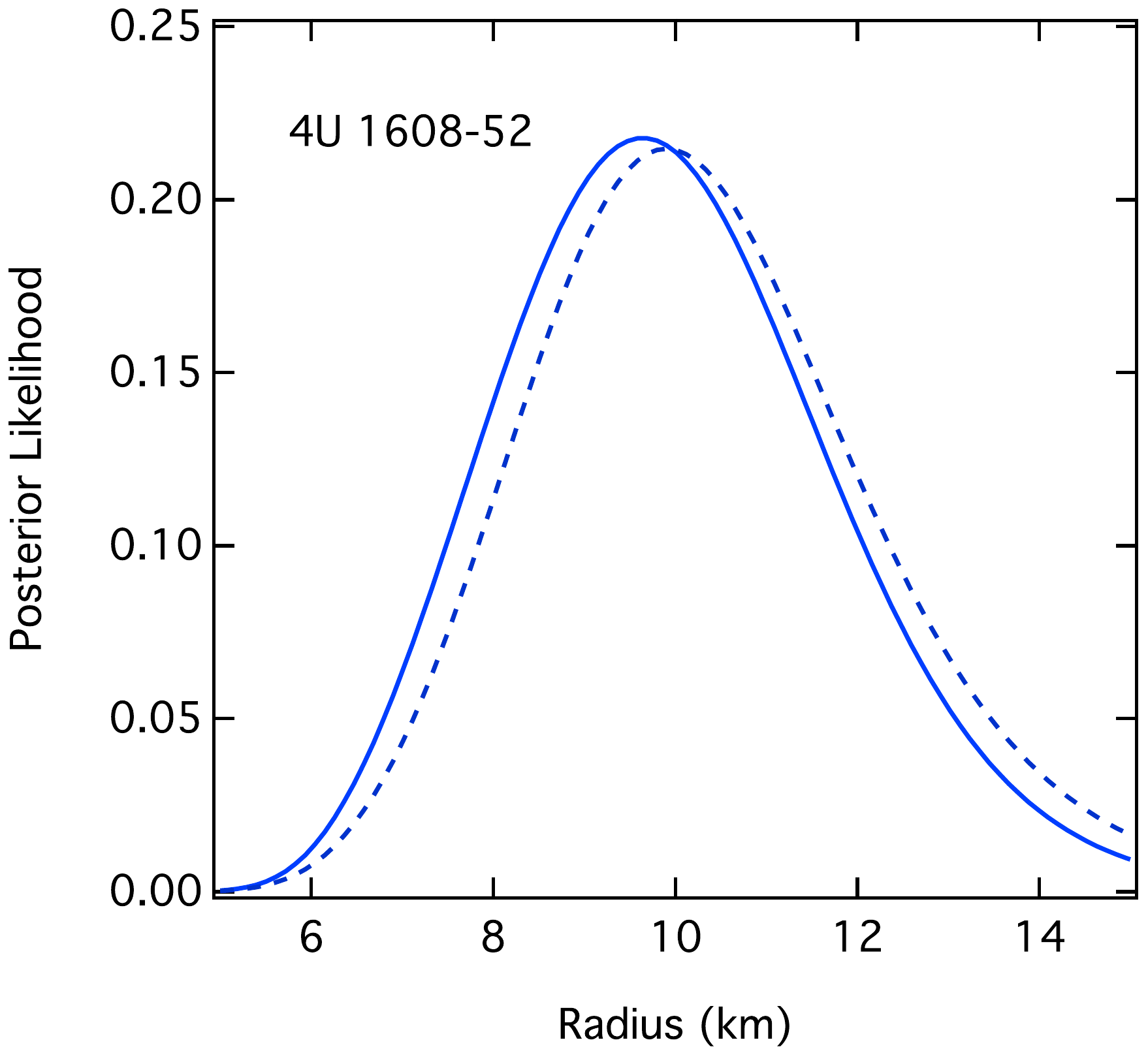}
\caption{Same as in Figure~\ref{fig:data_1820} but for \oeight. In the
  upper left panel, the blue points correspond to the main cooling
  tail and the black points denote the outliers, as discussed in the
  Appendix. }
\label{fig:data_1608} 
\end{figure}

\subsubsection{\oeight}

\oeight\ lies in the Galactic disk. The distance to this source was
measured in G\"uver et al.\ (2010a) by comparing the extinction
obtained from the red clump stars along the line of sight to the
extinction to the binary inferred from the high energy-resolution
X-ray observations. We repeat this analysis in the Appendix, utilizing
a new Chandra observation and the latest relation between the optical
extinction $A_{\rm V}$ and the hydrogen column density $N_{\rm H}$
obtained in a new study (Foight et al.\ 2015). The new results place a
lower limit of 3~kpc on the distance and give the highest likelihood
at $\sim 4$~kpc.

\oeight\ is the fastest spinning source in the current sample, with a
spin frequency of 620~Hz (Hartman et al.\ 2003). As with the other
sources whose companions do not have known compositions, we take a
boxcar prior for the hydrogen mass fraction between $X=0$ and $X=0.7$. 

In G\"uver et al.\ (2012b), we developed a Bayesian Gaussian-mixture
method for outlier detection and measuring the systematic
uncertainties in the apparent angular size during the cooling tails of
the bursts. \oeight\ is the only source in the present sample for
which such outliers were detected in the observed bursts. We discuss
these in the Appendix. The top left panel of
Figure~\ref{fig:data_1608} uses different color symbols to distinguish
the main sequence of the cooling tail from these outliers. 

The upper right panel in the same figure shows the confidence contours
in the blackbody normalization and temperature in the touchdown
moments of the three Eddington limited bursts, which includes the
newly detected burst during the simultaneous RXTE and Chandra
observations (see Appendix A2). Finally, the lower panels of
Figure~\ref{fig:data_1608} show the 68\% confidence contour in mass
and radius (left) as well as the posterior likelihood marginalized
over mass (right).

As with the other sources, these Eddington-limited bursts were
selected using the robust photospheric radius expansion (PRE) criteria
outlined in G\"uver et al.\ (2012a) and form a much smaller sample
than those initially identified as potential PRE events by Galloway et
al.\ (2008a). The earlier selection criteria of Galloway et
al.\ (2008) admitted a large number of non-PRE bursts into the sample,
because they were based primarily on the non-monotonic evolution of
the inferred apparent radii after the peak of each burst. As discussed
in G\"uver et al.\ (2012a), a careful scrutiny of these bursts clearly
demonstrates that the inferred touchdown fluxes (had they been PRE
events) are much smaller than the peak fluxes seen in the brightest
(true) PRE bursts, in a way that cannot be accounted for by the change
in the general relativistic redshift\footnote[2]{This is the same
  argument used in Galloway et al.\ (2008b) to reject bursts from high
  inclination sources.}. In addition, the inferred photospheric radii
during these misidentified PRE events are comparable to the asymptotic
radii of the same bursts. For these reasons, they are not PRE events
and do not pass the criteria of G\"uver et al.\ (2012a).

\begin{figure}
\centering
   \includegraphics[scale=0.3]{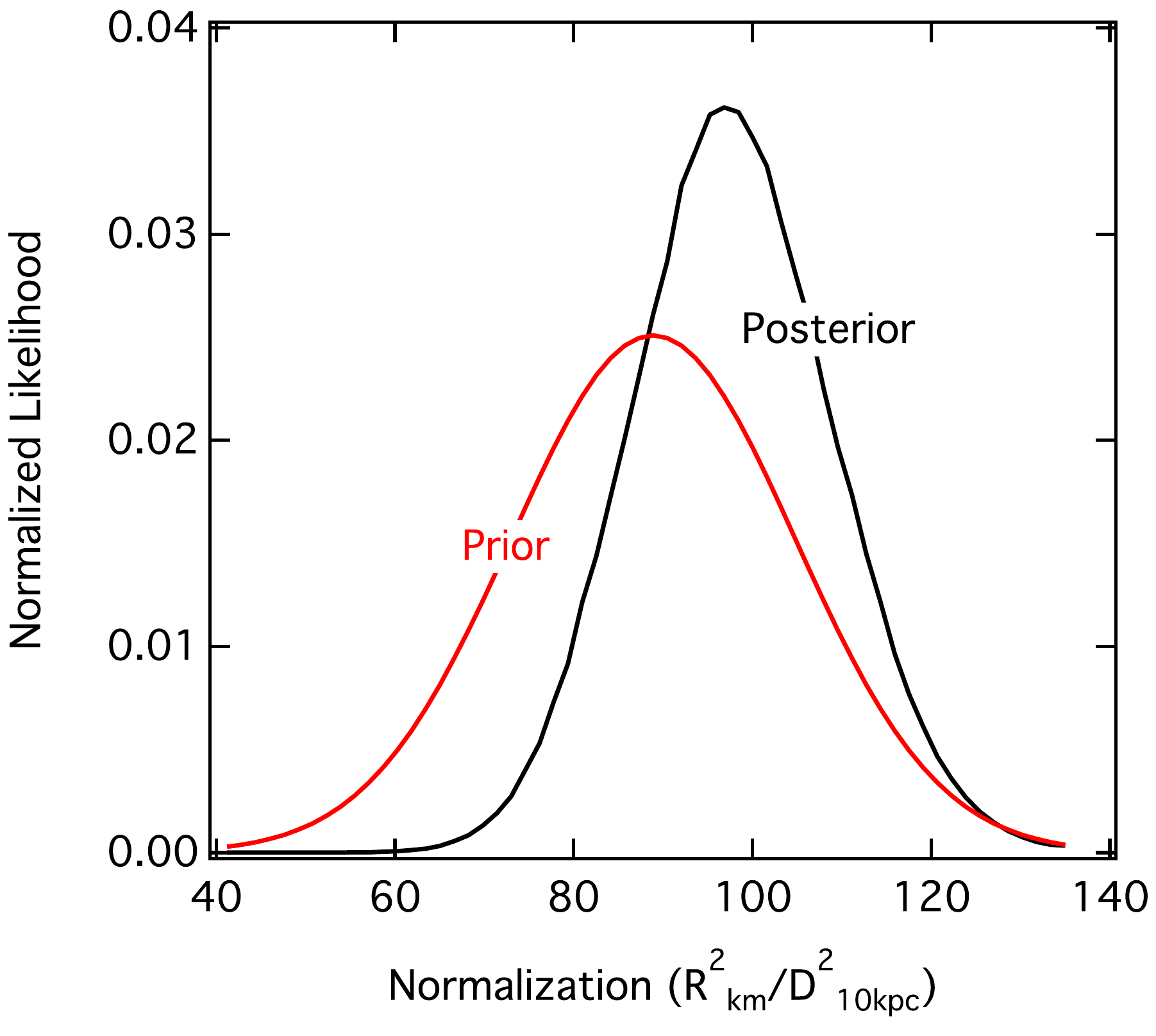}
   \includegraphics[scale=0.3]{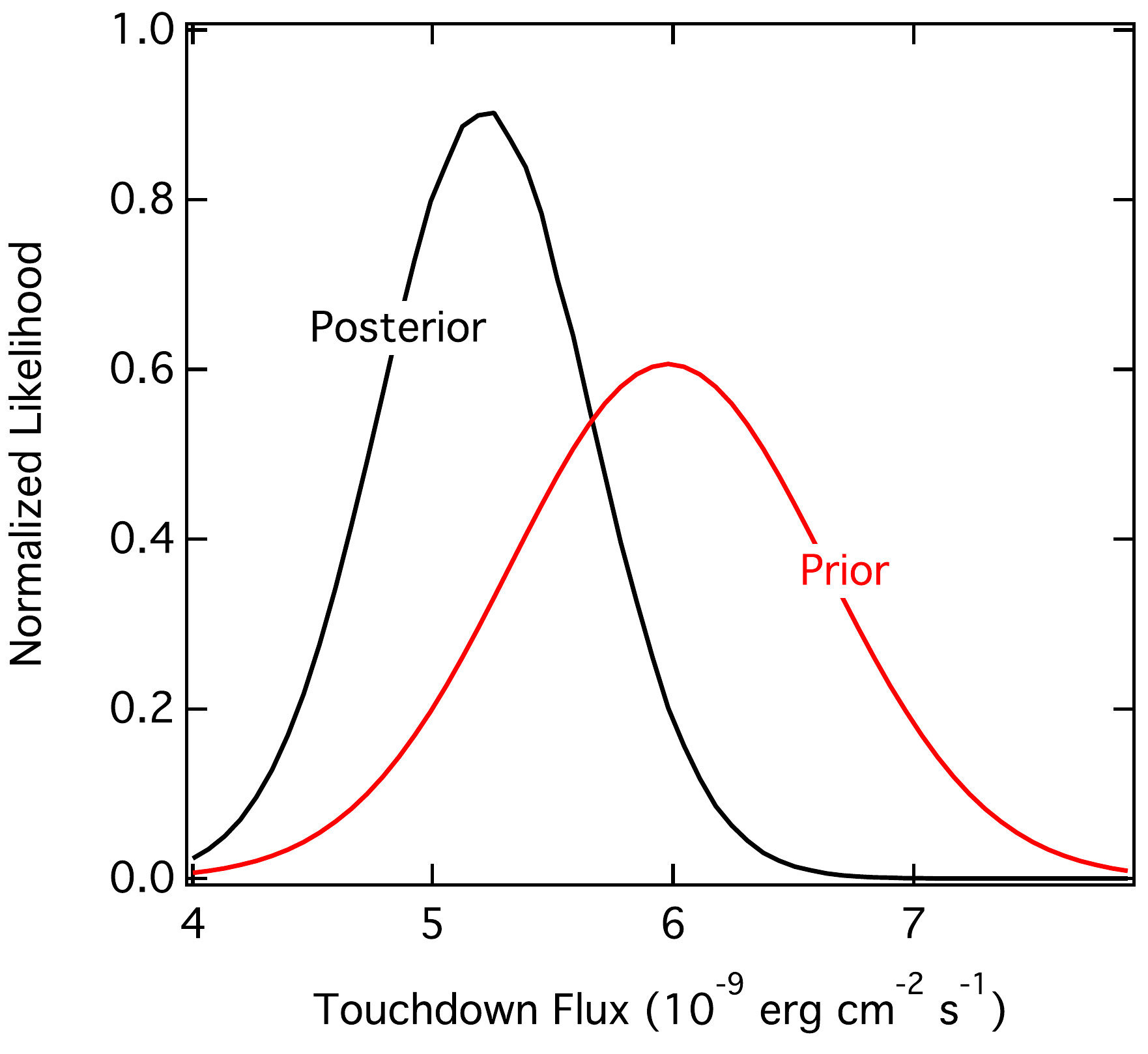}
   \includegraphics[scale=0.3]{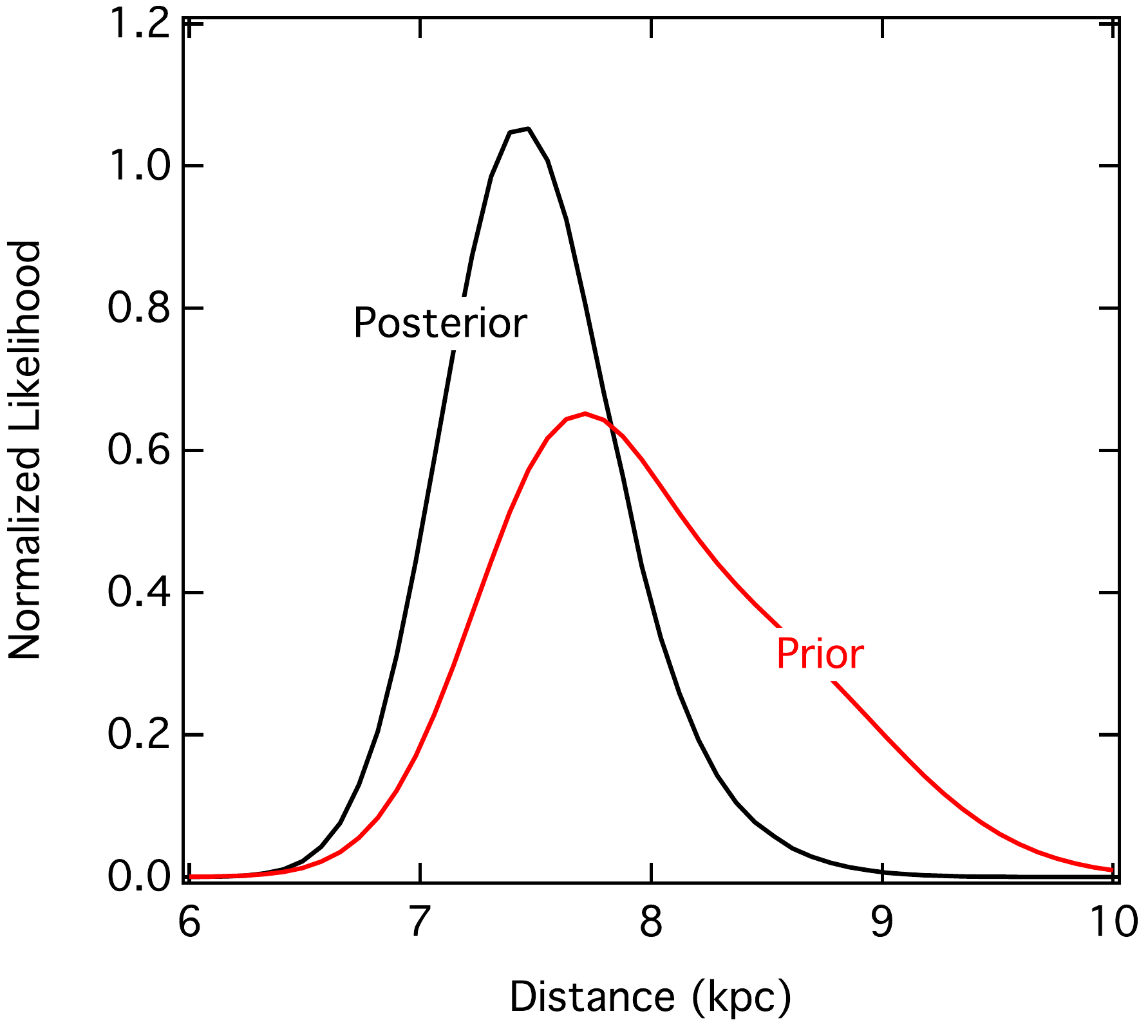}
\caption{A comparison between the prior and the posterior likelihoods
  over the blackbody normalization, the touchdown flux, and the
  distance to \twenty. When the measured systematic uncertainties in
  the two spectroscopic measurements as well as the full prior
  likelihood over the distance are taken into account, there is no
  evidence for inconcistencies between the observables.}
\label{fig:1820_post} 
\end{figure}

\subsubsection{Comparison with Previous Work}
  There are some differences in the mass-radius contours presented
  here compared to our earlier studies of the same sources. The
  primary reasons for these differences were discussed at the end of
  section 3.1 and include applying appropriate deadtime corrections to
  the observed countrates, incorporating the measured intrinsic
  scatter in the measurements (beyond the statistical uncertainties),
  applying spin and temperature corrections to the apparent angular
  sizes and touchdown fluxes, and using a Bayesian method to infer the
  masses and radii from the observables that does not suffer from the
  biases of the earlier frequentist approach. All of these
  improvements in the analysis methods lead to most likely values for
  the radii that are $\lesssim 1~km$ larger than before (compare with
  \"Ozel et al.\ 2010; \"Ozel et al.\ 2012a; G\"uver \& \"Ozel 2013)
  but with 68\% contours that encompass the most likely radii of the
  earlier studies.

  Our analysis and results are different from those of Steiner et
  al.\ (2010, 2013) and especially on the upper range of likely values
  of radii. Steiner et al.\ (2010) used the measurements reported in
  \"Ozel et al.\ (2009) and G\"uver et al.\ (2010a,b) but explored a
  number of different possibilities, including varying the location of
  the photosphere at the point we identify as the touchdown in a PRE
  burst. Their analysis favored the assumption that the photospheric
  radius at that point is much larger than the neutron star radius
  such that the general relativistic redshift is negligible. They
  followed this approach because they argued that the two
  spectroscopic measurements from each source are otherwise
  inconsistent with each other. In \"Ozel \& Psaltis (2015), we
  demonstrated that this potential inconsistency is alleviated when
  the true systematic scatter in the measurements is taken into
  account. Moreover, we showed in Figure~\ref{fig:rot_corr} the role
  that the rotational correction to the angular size and the
  temperature correction to the Eddington flux play in determining the
  consistency of observables. When these corrections are not taken
  into account, two highly accurate measurements of these quantities
  will appear to be inconsistent with each other and will not lead to
  a solution for the neutron star mass and radius.

  As the lower left panels of Figures~3-8 show, taking these effects
  into account makes the two spectroscopic observables in all sources
  consistent with each other at the 68\% level even when only the most
  likely value of the distance and the central value of the color
  correction factor are considered. This is further illustrated in
  Figure~\ref{fig:1820_post}, which compares the prior and posterior
  likelihoods of the blackbody normalization, the touchdown flux, and
  the distance for \twenty; this is the source for which Steiner et
  al. (2010) made the argument that the solutions were the least
  consistent. As is evident from this figure, combining the two
  observables lead to posterior likelihoods that are well within the
  prior likelihoods, indicating a high level of consistency. We report
  in Table~\ref{tb:posterior} of the Appendix the posterior
  likelihoods over each of the three measured quantities for all of
  the six thermonuclear burst sources used in this study. In all of
  the cases, the central values of the posterior likelihoods are
  within the 68\% range of the prior likelihoods shown in Table 1,
  pointing again to a high degree of consistency between the
  measurements that are used to infer the neutron star radii.
  
  Because the new analysis eliminates the concern over the consistency
  of solutions, it does not force us into the astrophysically
  unreasonable assumption of Steiner et al.\ (2010) that the
  photosphere at what we identified as the touchdown point is much
  larger than the neutron star radius. This was problematic for two
  reasons. First, in order for the blackbody normalization to remain
  small at that point while the photospheric radius is still extremely
  large, the color correction factor needs to be unphysically large;
  i.e, larger by factors of three or more than what the atmosphere
  models predict. Second, within 1-2 time bins, as the photosphere
  settles onto the neutron star, the color correction factor would
  need to evolve in such a way that it exactly cancels out the change
  in the photospheric radius, keeping a constant blackbody
  normalization. We do not need to make these implicit assumptions in
  the present study.

  Our radii are significantly smaller than those reported by
  Suleimanov et al.\ (2011) and Poutanen et al.\ (2014) who selected
  bursts and obtained radius measurements using the evolution of the
  blackbody normalization during the cooling tails of \twentyfour\ and
  \oeight, respectively. In the case of Suleimanov et al.\ (2011), the
  selection criteria identified one burst. Unfortunately, as shown in
  G\"uver et al.\ (2012b) and discussed earlier, the spectra from
  these bursts are inconsistent with the atmosphere models, leading to
  reduced $\chi^2$ values in the $2-8$ range, rendering them
  unsuitable for radius measurements.

  Poutanen et al.\ (2014) selected bursts from \oeight\ by requiring
  the bursts to follow the trends expected from the bursting neutron
  star atmosphere models of Suleimanov et al.\ (2012) at
  near-Eddington fluxes. In \"Ozel et al.\ (2015), we showed that this
  criterion is not useful for burst selection from RXTE data for three
  reasons. First, the spectral evolution at the end of a photospheric
  radius expansion episode occurs too rapidly to be resolved with the
  current data, because over the typical 0.25~s time bin used to
  extract spectral parameters, the flux evolves by $\sim 10\%$. This
  is exactly the range of fluxes near the Eddington limit that one
  needs to resolve in order to see the expected evolution of the color
  correction factor. Second, the scatter in the blackbody
  normalization due to even a mild change in the emitting area (due
  to, e.g., uneven burning or an evolving photosphere) masks the
  theoretical trends. Finally, the correlated measurement
  uncertainties between the blackbody normalization and temperature
  further smear any trends. By not taking these data limitations into
  account, Poutanen et al.\ (2014) selected a set of bursts that are
  not actual PRE bursts, contrary to the implicit assumption in their
  method.

  As discussed above, in both the Suleimanov et al.\ (2011) and
  Poutanen et al.\ (2014) studies, applying these theoretically
  motivated criteria led to selection of bursts that are inconsistent
  with the framework of the method: in the former by selecting spectra
  that are clearly not described by their atmopsheric models and, in
  the latter, by comparing models of the color correction factor
  evolution near the Eddington limit to bursts that have not reached
  it. Kajava et al.\ (2014) tried to generalize this selection
  procedure to several more sources (without reporting any additional
  radius measurements) but also did not consider the limitations of
  the data. We conclude that with the present data, the application of
  this procedure motivated by the spectral models leads neither to
  unbiased data selection, nor to reliable radius measurements.

\subsection{Quiescent Low-Mass X-ray Binaries}

The second group of sources on which radius measurements have been
performed are the accreting neutron stars in low-mass X-ray binaries
during their quiescent epochs (qLMXBs). It is thought that, in
quiescence, neutron stars reradiate the heat stored in the deep crust
during the accretion phases through a light element atmosphere (Brown
et al.\ 1998). This allows interpreting the observed thermal spectra
as surface emission from atmospheres in radiative equilibrium, while
allowing for the presence of a weak power-law spectral component at
higher energies due to residual accretion.  Because of the very short
settling time of heavier elements in a neutron-star atmosphere, the
photospheres of such neutron stars in quiescence are expected to be
composed of hydrogen, unless the companion star is hydrogen poor. In
that case, they will be composed of helium.

A number of qLMXBs in globular clusters has been observed with {\it
  Chandra} and XMM-{\it Newton}.  Because they are very faint and are
located in crowded fields, the high angular resolution and low
background of these instruments were crucial for obtaining
spectroscopic constraints of their apparent angular sizes (e.g.,
Heinke et al.\ 2006; Webb \& Barret 2007; Guillot et al.\ 2011).

Guillot et al.\ (2013) and Guillot \& Rutledge (2014) performed a
uniform analysis of six sources in this category, which are the
neutron stars located in the globular clusters M13, M28, M30,
$\omega$~Cen, NGC~6397, and NGC~6397.  These observations include
those summarized in Table~1 of Guillot et al.\ (2013) as well as the
{\it Chandra} observations of the qLMXB in M30 (ObsID 2679; Lugger et
al. 2007) and of the qLMXB in $\omega$~Cen (ObsIDs 13726 and 13727),
as described in Guillot \& Rutledge (2014). These two studies fit the
extracted spectra with hydrogen atmosphere models to measure the
apparent angular sizes for these neutron stars.  They explored the
dependence of the results on different hydrogen model atmosphere
spectra used. They also allowed for a Gaussian distribution of errors
in distances (albeit narrower than the uncertainties we assigned above
to bursters in globular clusters) when fitting all of the sources
simultaneously.


\begin{deluxetable}{cccccc}
\tabletypesize{\scriptsize}
\tablewidth{500pt}
\tablenum{2}
\tablecaption{Properties of Quiescent LMXBs}
\tablehead{
 \colhead{Source} &
 \colhead{$N_{\rm H}$\tablenotemark{a}} &
 \colhead{$kT_{\rm eff}$} &
 \colhead{P.L. Norm. \tablenotemark{b}} &
 \colhead{Distance\tablenotemark{c}} &
 \colhead{Radius\tablenotemark{d}} 
\cr
 \colhead{} &
 \colhead{($10^{22}~{\rm cm}^{-2}$)} &
 \colhead{(eV)} &
  \colhead{($10^{-7} {\rm keV}^{-1} {\rm s}^{-1} {\rm cm}^{-2}$)} &   
 \colhead{(kpc)} &
 \colhead{(km)}
}
\startdata  
M13           & $0.02^{+0.04}_{-0.02p}$ & $81^{+27}_{-12}$  & $4.2^{+3.6}_{-3.1p}$ & $7.1\pm0.4$\tablenotemark{1} & 10.9 $\pm$ 2.3 \\
M28           & $0.30^{+0.03}_{-0.03}$  & $128^{+35}_{-13}$ & $8.3^{+4.9}_{-4.7p}$ & $5.5\pm0.3$\tablenotemark{2} & 8.5 $\pm$ 1.3 \\
M30           & $0.02^{+0.03}_{-0.02p}$ & $96^{+30}_{-13}$  & $9.3^{+5.4}_{-5.3p}$ & $9.0\pm0.5$\tablenotemark{3,4} & 11.6 $\pm$ 2.1 \\
$\omega$Cen   & $0.15^{+0.04}_{-0.04}$  & $80^{+24}_{-10}$  & $0.8^{+1.3}_{-0.7p}$ & $4.59\pm0.08$\tablenotemark{5,6} & 9.4 $\pm$ 1.8  \\
NGC~6304      & $0.49^{+0.15}_{-0.13}$  & $100^{+33}_{-17}$ & $2.4^{+2.7}_{-1.9p}$ & $6.22\pm0.26$\tablenotemark{7} & 10.7 $\pm$ 3.1\\
NGC~6397      & $0.14^{+0.02}_{-0.02}$  & $66^{+17}_{-7}$   & $3.3^{+1.8}_{-1.8}$  & $2.51\pm0.07$\tablenotemark{8} & 9.2 $\pm$ 1.8 
\enddata
\scriptsize
\tablenotetext{a} {NGC6397 was fitted with a Helium atmosphere model ({\tt nsx} in XSPEC).}
\tablenotetext{b} {“p” indicates that the posterior distribution did
  not converge to zero probability within the hard limit of the
  model.}
\tablenotetext{c} {References:
  1.~Harris et al.\ (1996, 2010 revision);
  2.~Servillat et al.\ (2012);
  3.~Carretta et al.\ (2000);
  4.~Lugger et al.\ (2007);
  5.~Watkins et al.\ (2013);
  6.~see also the discussion in Heinke et al.\ (2014);
  7.~Guillot et al.\ (2013) and references therein;
  8.~Heinke et al.\ (2014)}
\tablenotetext{d} {The radius and its 68\% uncertainty obtained by marginalizing 
the mass-radius likelihood of each source over the observed mass distribution, as in 
Figure~\ref{fig:combined_R}.}
\mbox{} 
\end{deluxetable}

There are several additional sources of systematic uncertainties that
can affect the radius measurements that have been addressed to various
degrees: the composition of the atmosphere, the composition and
modeling of the interstellar medium that gives rise to the low-energy
extinction, and the modeling of the power-law spectral component that
is due to residual accretion. The majority of qLMXBs for which optical
spectra have been obtained show evidence for H$\alpha$ emission
(Heinke et al.\ 2014), indicating a hydrogen rich companion.  Most of
these qLMXBs are in the field and not in globular clusters as the ones
we are using here. Assuming that sources in both types of environments
have similar companions supports, in general, the use of hydrogen
atmospheres when modeling quiescent spectra. The one source among
those used here for which there is evidence to the contrary is the
qLMXB in NGC~6397.  Heinke et al.\ (2014) obtained only an upper limit
on the H$\alpha$ emission using HST observations and, because of this,
they applied a helium atmosphere model to the {\it Chandra}/XMM-{\it
  Newton} data sets described above.

\begin{figure}
\centering
   \includegraphics[scale=0.45]{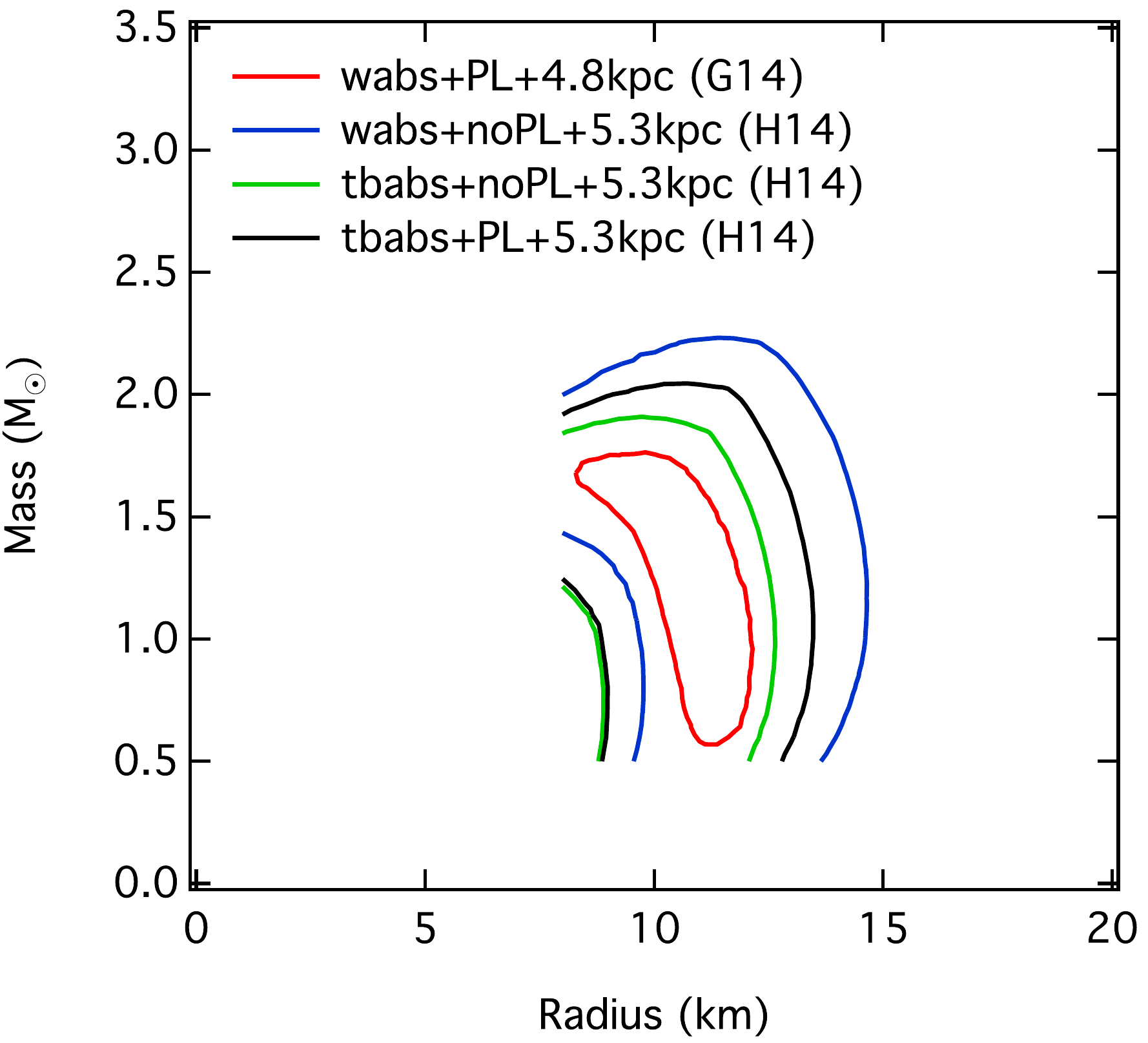}
\caption{The 68\% confidence contours in mass and radius for the
  quiescent neutron star in $\omega$~Cen, inferred by Heinke et
  al.\ (2014; H14) and by Guillot \& Rutledge (2014; G14) using
  different assumptions regarding the interstellar extinction (wabs:
  Morrison \& McCammon 1983; tbabs: Wilms et al.\ 2000), the presence
  of a power-law spectral component, and for different distances to
  the globular cluster (4.8~kpc vs.\ 5.3~kpc)}.
\label{fig:omegaCen} 
\end{figure}

Heinke et al.\ (2014) also explored the effect of assuming different
spectral indices in modeling the power-law component. Even though the
low counts preclude an accurate measurement of this parameter, the
specific value has a small effect on the radius measurement, which can
be folded in as a sytematic uncertainty. Finally, because of the low
temperature of the surface emission from qLMXBs, the spectral modeling
is affected significantly by the assumed model of the interstellar
medium to account for the low-energy extinction. Heinke et al.\ (2014)
explored different models for the interstellar extinction in their
analysis of the qLMXBs in $\omega$~Cen and NGC~6397 and found
statistically consistent results, with small differences in the
central values but larger differences in the uncertainties. In the
left panel of Figure~\ref{fig:omegaCen}, we show the effect of
different assumptions on the power-law index, the distance, and the
interstellar extinction model on the inference of the mass and radius
of the neutron star in $\omega$Cen. In particular, one of the larger
effects arises from the use of the two common interstellar extinction
models they consider (the earlier Morrison \& McCammon 1983 model with
solar abundances, referred to as wabs in the spectral fitting software
XSPEC, and the more recent Wilms et al.\ 2000 model, with ISM
abundances from the same paper, referred to as tbabs with {\it wilms}
in XSPEC). The wabs model (employed by Guillot et al.\ 2013) leads to
somewhat larger radii for the same distance.

\begin{figure}
\centering
   \includegraphics[scale=0.45]{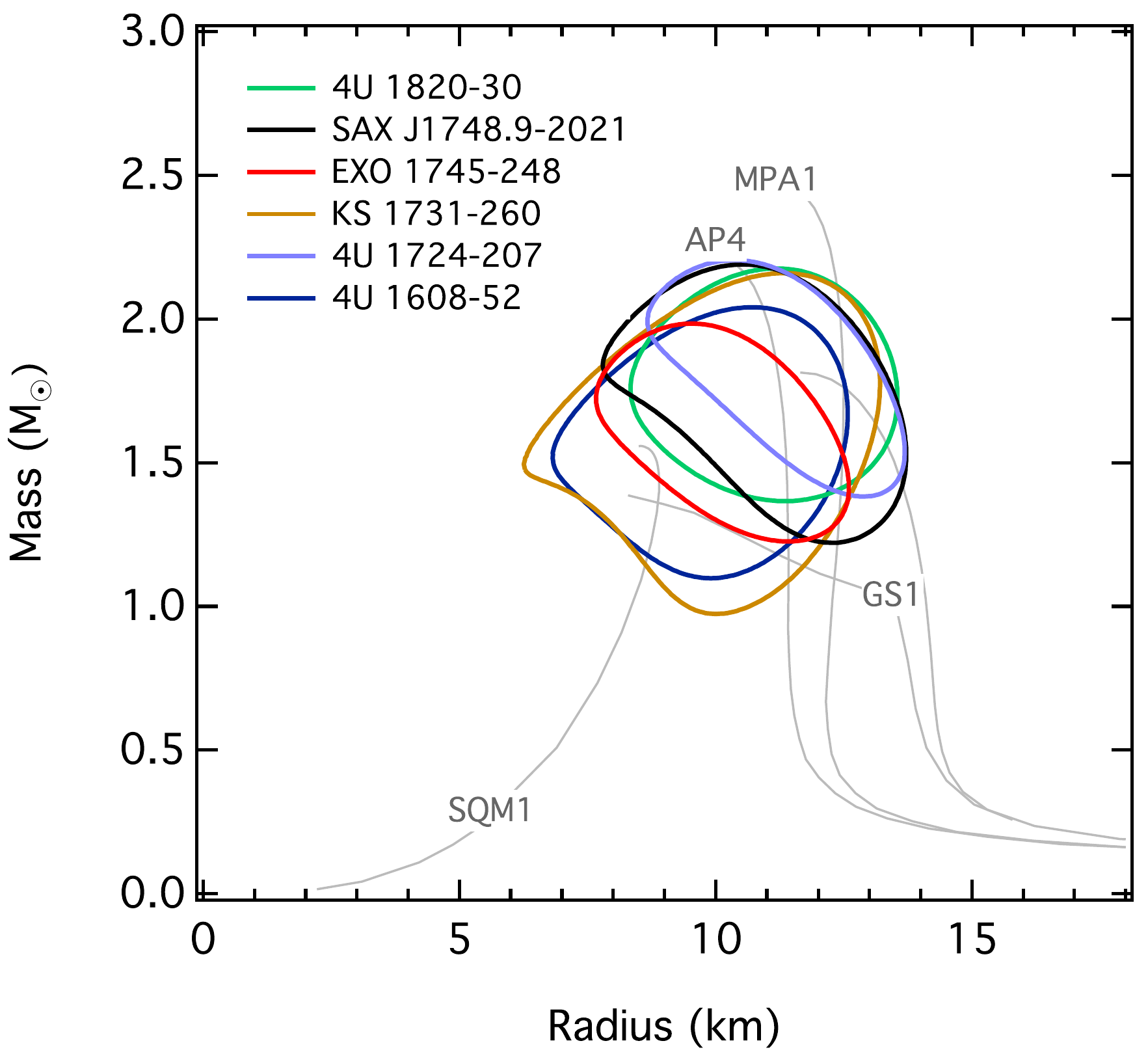}
   \includegraphics[scale=0.45]{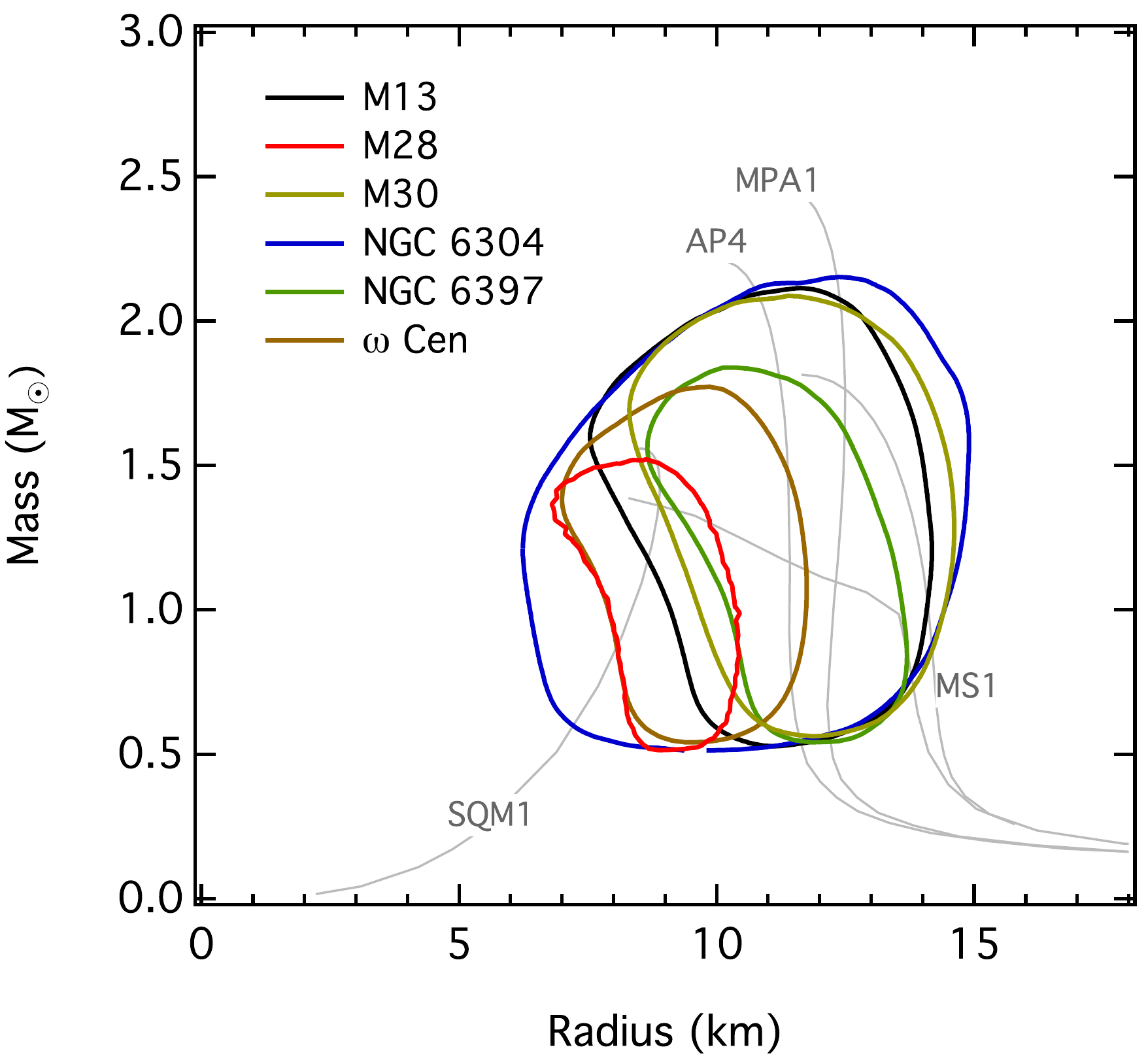}
\caption{The combined constraints at the 68\% confidence level over
  the neutron star mass and radius obtained from (Left) all neutron
  stars with thermonuclear bursts (Right) all neutron stars in
  low-mass X-ray binaries during quiescence.}
\label{fig:combined_MR} 
\end{figure}

In the present study, we repeat the analysis of Guillot et al.\ (2013)
individually for all the sources in M13, M28, NGC~6304, NGC~6397, M30,
and $\omega$Cen. (Note that for the last two sources, the observations
were reported in Guillot \& Rutledge 2014). In all of the spectral
fits, we allow for a power-law component with a fixed photon index
$\Gamma =1$ but a free normalization. We use the Wilms et al.\ (2001)
ISM abundances in all of the analyses for a uniform treatment of all
qLMXBs. We leave the hydrogen column density as a free parameter in
the fits and when calculating the posterior likelihoods over mass and
radius. We use a hydrogen atmosphere model for all of the sources
except the one in NGC~6397, for which we use a helium atmosphere model
(see also Heinke et al.\ 2014). The best-fit spectral parameters for
each source are shown in Table~2.  We also fold in distance
uncertainties using a Gaussian likelihood for the distance to each
source with a mean and standard deviation given in Table~2.

We show the resulting posterior likelihoods over the mass and radius
for all of the qLMXBs in the right panel of
Figure~\ref{fig:combined_MR} and compare them to the combined
constraints from the X-ray bursters discussed earlier.  There is a
high level of agreement between all of these measurements.  Note that
the larger widths of the 68\% confidence contours of each source
compared to those presented in Guillot et al.\ (2013, their figures
$3-7$) are due to the fact that in the present work, we leave the
hydrogen column density as a free parameter.\footnote[3]{While the
  simultaneous ``Constant Rns'' fits of Guillot et al.\ (2013) were
  performed by leaving the column density $N_{\rm H}$ as a free
  parameter, Figures~$3-7$ in that study display results from fits
  performed with fixed $N_{\rm H}$. Furthermore, note that the present
  study also includes additional X-ray data for $\omega$~Cen, which
  refined the M-R contours.}


\section{The Constraints on the Neutron Star Radius}

Having obtained posterior likelihoods over the mass and radius for a
number of neutron stars, we can follow one of the inversion techniques
developed earlier to infer the equation of state of neutron star
matter. We defer this analysis to the following section and first
carry out a simple exercise to illustrate how tight constraints on the
equation of state can be obtained when a large number of measurements
with relatively large uncertainties are used.

For this purpose, we consider a mono-parametric equation of state in
which all neutron stars have the same radius independent of
mass. (Note that this is the same in spirit as the CstR$_{\rm NS}$
model of Guillot et al.\ 2013. It is indeed a meaningful assumption
for nearly all nucleonic equations of state, which predict
approximately constant radii for the astrophysically relevant mass
range). We also assume that all the neutron stars in our sample are
drawn from the observationally inferred Gaussian distribution of
masses (see \"Ozel et al.\ 2012). Specifically, we write
\begin{equation}
P(R \; \vert \; {\rm data})=C\prod_{i=1}^N\int P_i(R,M \; \vert \; {\rm data}) P_p(M) dM
\end{equation}
where $C$ is an appropriate normalization constant, $P_i(R,M \; \vert
\; {\rm data})$ is the two-dimensional posterior likelihood over mass
and radius for each of the $N$ sources (as given, e.g., in
equation~\ref{eq:MR_burst} for the bursters), and $P_p(M)$ is the
Gaussian likelihood with a mean of 1.46~M$_\odot$ and a dispersion of
0.21~M$_\odot$ for the mass distribution inferred by \"Ozel et
al.\ (2012) for the descendants of these systems.

\begin{figure}
\centering
   \includegraphics[scale=0.45]{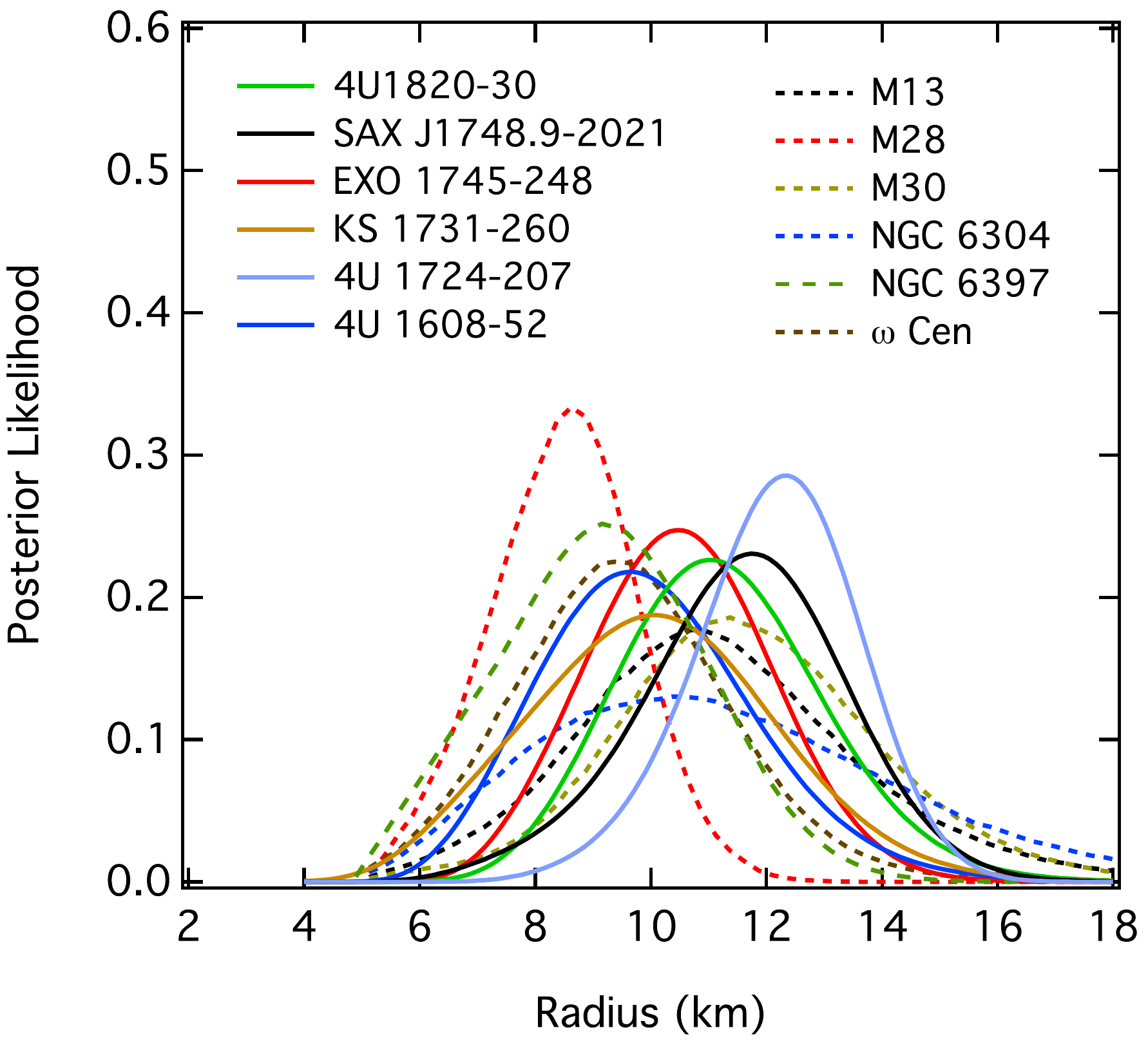}
  \includegraphics[scale=0.45]{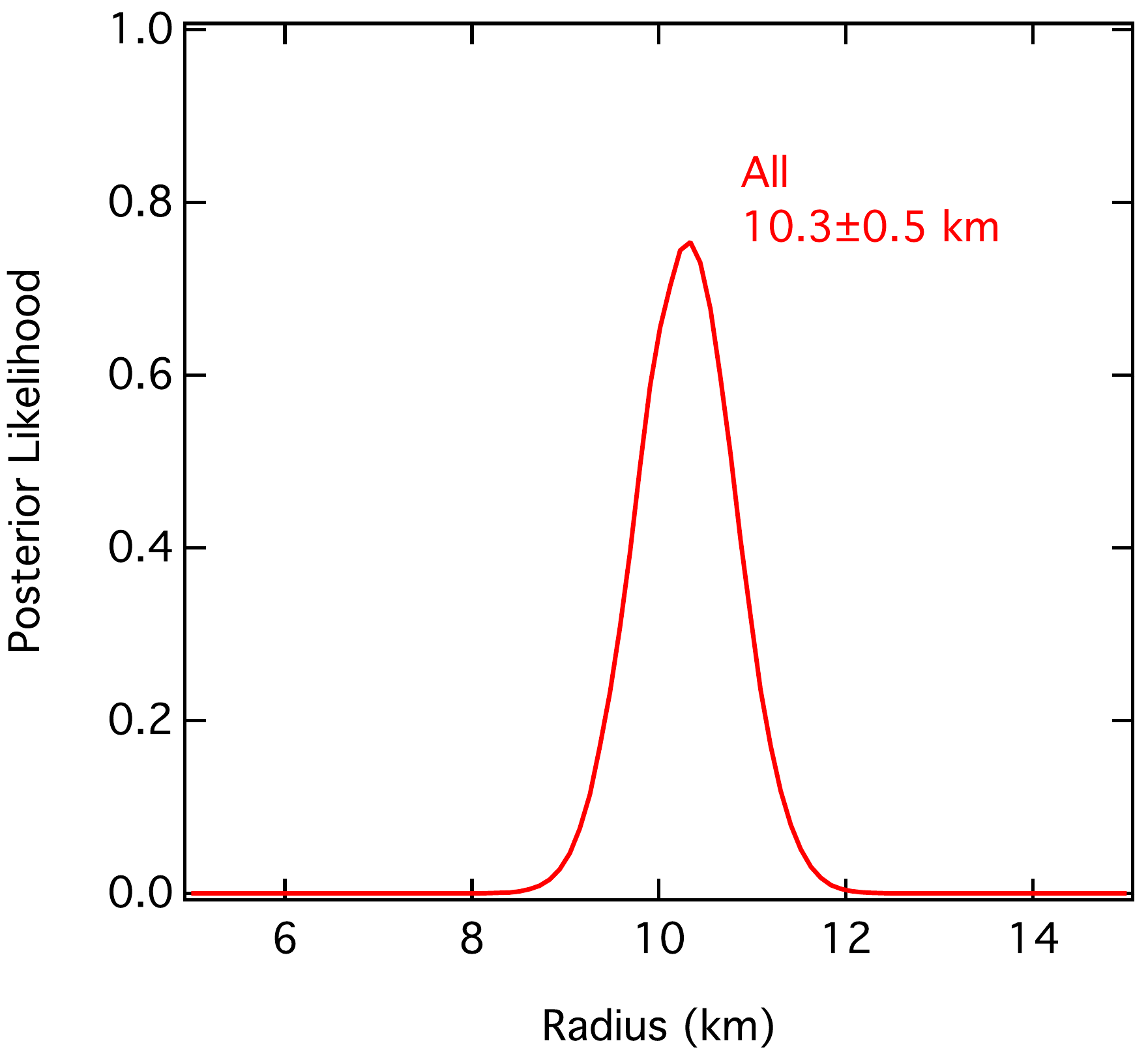}
\caption{{\em (Left)\/} The posterior likelihood over the radius
  obtained by marginalizing the two dimensional likelihoods over the
  neutron star mass, with a prior equal to the observationally
  inferred distribution of recycled pulsar masses, for all twelve
  sources in our sample. The peak probabilities are highly clustered
  in the 9-12~km range. {\em (Right)\/} The combined posterior
  likelihood assuming that all sources in our sample have the same
  radius and masses drawn from the observationally inferred
  distribution of recycled pulsar masses. We use this inferrence only
  as an illusration of the fact that using radius measurements for
  twelve sources leads to a highly accurate constraint on the
  neutron-star equation of state. \\}
\label{fig:combined_R} 
\end{figure}

The left panel of Figure~\ref{fig:combined_R} shows the individual
terms of the product in the equation above; i.e., the posterior
likelihoods over radius for each of the twelve sources. They are all
well approximated by Gaussian distributions that peak between 9-12~km
and typical uncertainties $\sim$2~km. The right panel of
Figure~\ref{fig:combined_R} shows the posterior likelihood over the
single radius in this mono-parametric equation of state, which is
peaked at a radius of 10.3~km with an uncertainty of 0.5~km. As
expected, given that all radii are statistically consistent with each
other, combining the data of twelve sources led to a reduction in the
uncertainty by a factor $\sqrt{12}\simeq 3.5$. The result is a level
of uncertainty that is comparable to what is required to severely
constrain the neutron star equation of state, as we will show in
detail in the next section.

\section{The Neutron Star Equation of State From Radii and Low-Energy Experiments}

We now make use of the one-to-one mapping between the neutron star
mass-radius relation and the pressure-density relation of cold dense
matter to put direct constraints on the neutron-star equation of
state. In this procedure, we take the most general approach and do not
assume that neutron stars have a constant radius or make assumptions
about their masses based on the observationally determined mass
distribution.

The structure equations for relativistic stars provide, for each
equation of state, a unique mass-radius curve, with no adjustable
parameters. Because of this, a large number of radius measurements
across a range of neutron star masses allow us to invert the
measurements formally and obtain the unique equation of state that
gave rise to the measured mass-radius pairs (Lindblom 1992).
Moreover, more recent parametric forms of the equation of state that
do not rely on particular nuclear physics models (e.g., Read et
al.\ 2009; \"Ozel \& Psaltis 2009; Steiner et al.\ 2010; Lattimer \&
Steiner 2014b) allow us to put this inversion into practice even
without sampling the entire mass-radius curve. This is because the
radii of astrophysically relevant neutron stars turn out to be
sensitive to the equation of state in a fairly narrow range of
densities between 1.8 and 7.4 times the nuclear saturation density
($\rho_{\rm ns}$) and this, in turn, enables a discretization of the
equation of state over this narrow range of densities using only three
sampling points that are connected by piecewise polytropes (Read et
al.\ 2009; \"Ozel \& Psaltis 2009). Using the SLy equation of state
(Douchin \& Haensel 2001) up to a density of
$\rho_0=10^{14}$~g~cm$^{-3}$ and taking the fiducial densities at
1.85~$\rho_{\rm ns}$, 3.7~$\rho_{\rm ns}$, and 7.4~$\rho_{\rm ns}$,
which we denote as $\rho_1$, $\rho_2$, and $\rho_3$, respectively,
these studies have shown that the discretized equations of state
generate mass-radius relations that faithfully reproduce the
mass-radius relations of the continuous $P(\rho$) functions for a
large number of proposed equations of state. Of these three pressures,
the pressure at 1.85~$\rho_{\rm ns}$ sets the overall radii of neutron
stars, the pressure at 3.7~$\rho_{\rm ns}$ determines the slope of the
predicted $M-R$ relation, and the pressure at 7.4~$\rho_{\rm ns}$ sets
the maximum mass.

In \"Ozel et al.\ (2010), we used the framework devised in \"Ozel \&
Psaltis (2009) to convert the mass-radius measurements of three
sources to posterior likelihoods over the pressures at these three
fiducial densities. In order to incorporate the mass-radius
measurements of the twelve sources presented in Section~3, we will
follow here the Bayesian approach outlined below. (See Steiner et
al.\ 2010 for a similar Bayesian inference approach.)

To calculate the posterior likelihood over the pressures 
$P_1(\rho_1)$, $P_2(\rho_2)$, and $P_3(\rho_3)$ using the likelihoods 
$P_i(M,R)$ for twelve sources, we write 
\begin{equation}
P(P_1, P_2, P_3\; \vert {\rm data}) = C P( {\rm data} \; \vert \; P_1, P_2, P_3) 
P_p(P1) P_p(P_2) P_p(P_3), 
\end{equation}
where $P_p(P1)$, $P_p(P_2)$, and $P_p(P_3)$ are the priors over the three 
pressures and 
\begin{equation}
P( {\rm data} \; \vert \; P_1, P_2, P_3)  = 
\prod_{i=1}^N P_i(M_i, R_i \; \vert \; P_1, P_2, P_3)
\end{equation}
To obtain $(M_i, R_i)$ from the pressures $P_1, P_2, P_3$, we also 
need to specify, and marginalize over, the central density of the star $\rho_c$, 
i.e., 
\begin{equation}
P_i(M_i, R_i \; \vert \; P_1, P_2, P_3) = C_1 \int_0^\infty P_i(M_i, R_i \; \vert \; P_1, P_2, P_3, \rho_c)
P_p(\rho_c) d\rho_c.
\end{equation}
Because there is a one-to-one correspondence between the central density $\rho_c$ 
and mass, we can write the integral over the mass instead as
\begin{equation}
P_i(M_i, R_i \; \vert \; P_1, P_2, P_3) = C_2 \int_{M_{\rm min}}^{M_{\rm max}} 
P_i(M, R(M) \; \vert \; P_1, P_2, P_3) P_p(M) dM,
\end{equation}
where we take $M_{\rm min}$ to be $0.1 M_\odot$ and $M_{\rm max}$ to
be the maximum mass for the equation of state specified by that $P_1,
P_2, P_3$ triplet. Here, $P_p(M)$ is the prior likelihood over the
mass of each neutron star, which we take to be constant.

We use a variety of physical and observational constraints to define the 
priors on $P_1, P_2$, and $P_3$. \\
{\bf (i)} We require that the equation of state be microscopically stable, 
i.e., $P_3 \ge P_2 \ge P_1$, and that $P_1$ be greater than or equal to the 
pressure of matter at $\rho_0=10^{14}$~g~cm$^{-3}$ that is specified by 
the SLy equation of state (see \"Ozel \& Psaltis 2009). \\
{\bf (ii)} We impose the physically plausible condition of causality that 
\begin{equation}
c_s^2=\frac{\partial P}{\partial \epsilon} \le c^2
\end{equation}
when evaluated at all three fiducial densities; here, $c_s$ is the
sound speed and $\epsilon$ is the energy density. \\ 
{\bf (iii)} We require that the maximum stable mass for each equation 
of state corresponding to a $P_1, P_2, P_3$ triplet exceeds 
$1.97 M_\odot$, consistent with the heaviest neutron stars observed to
date. Specifically, this corresponds to the central value of the
measurement by Demorest et al.\ (2010) and is within the $1\sigma$
lower limit of the measurement by Antoniadis et al.\ (2013). \\ 
{\bf (iv)} We impose a lower limit on $P_1 = 7.56$~MeV~fm$^{-3}$ such
that the equation of state is consistent with laboratory experiments
and low density calculations, as we will describe in detail below.
This value is quantitatively consistent with the APR equation of 
state for pure neutron matter (Akmal et al.\ 1998). \\ 
{\bf (v)} Finally, in order to explore the dependence of 
our results on the prior distributions, we consider two sets: one 
that is flat in $\log P_1$, $\log P_2$, and $\log P_3$ and one 
that is flat in $P_1$, $P_2$, and $P_3$.

The first two constraints are required on microphysical grounds (but
see Ruderman \& Bludman 1968 and Ellis et al.\ 2007 for caveats on the
causality argument). In the next two constraints, we fold in
information about the equation of state inferred from other
astrophysical observations or nuclear experiments.  This ensures that
the equation of state derived from neutron star radii is consistent
with these other results. In addition, combining these several
different avenues of information allows us to achieve the highest
precision in the resulting constraints.

\begin{figure}
\centering
\includegraphics[scale=0.55]{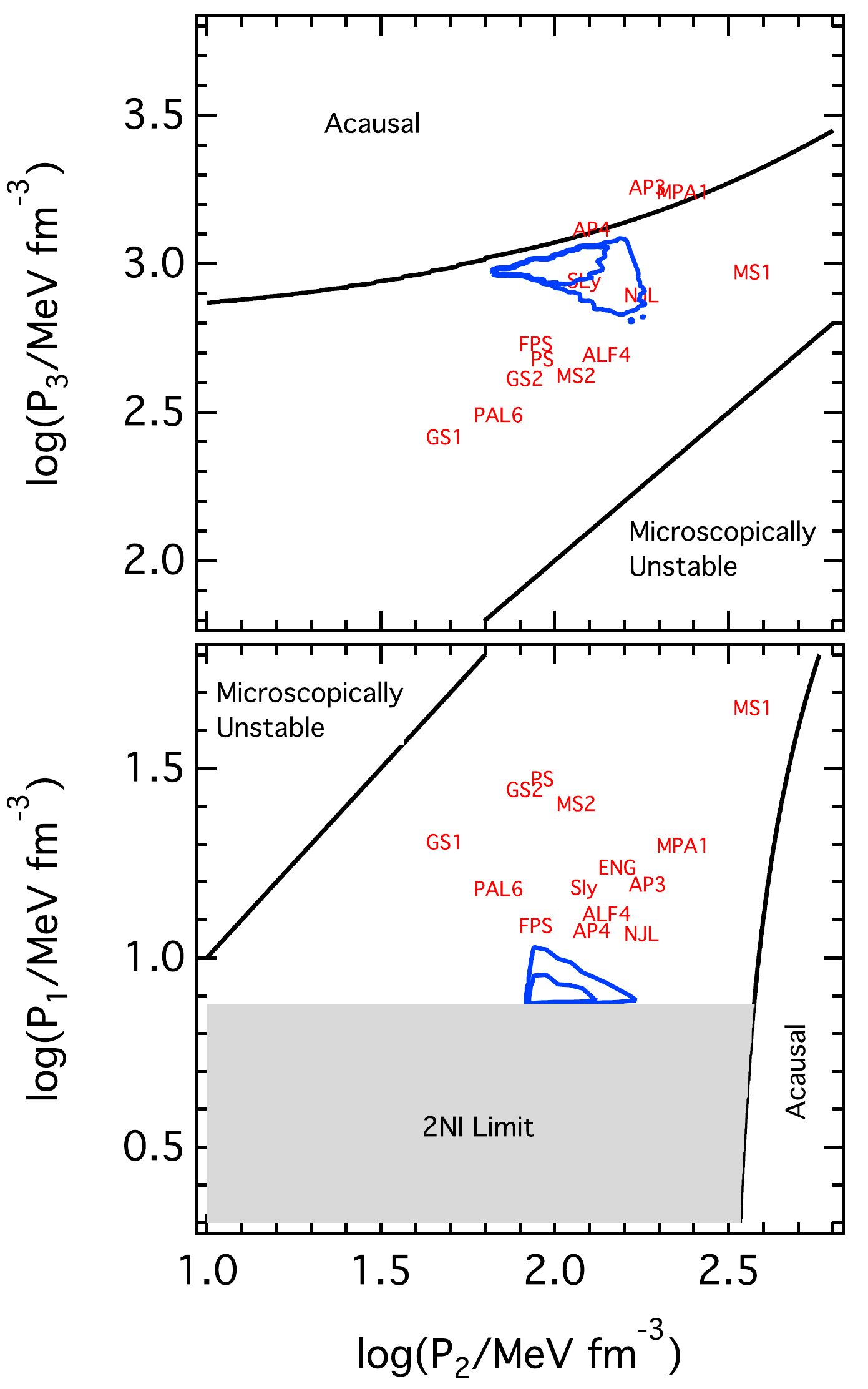}
\includegraphics[scale=0.55]{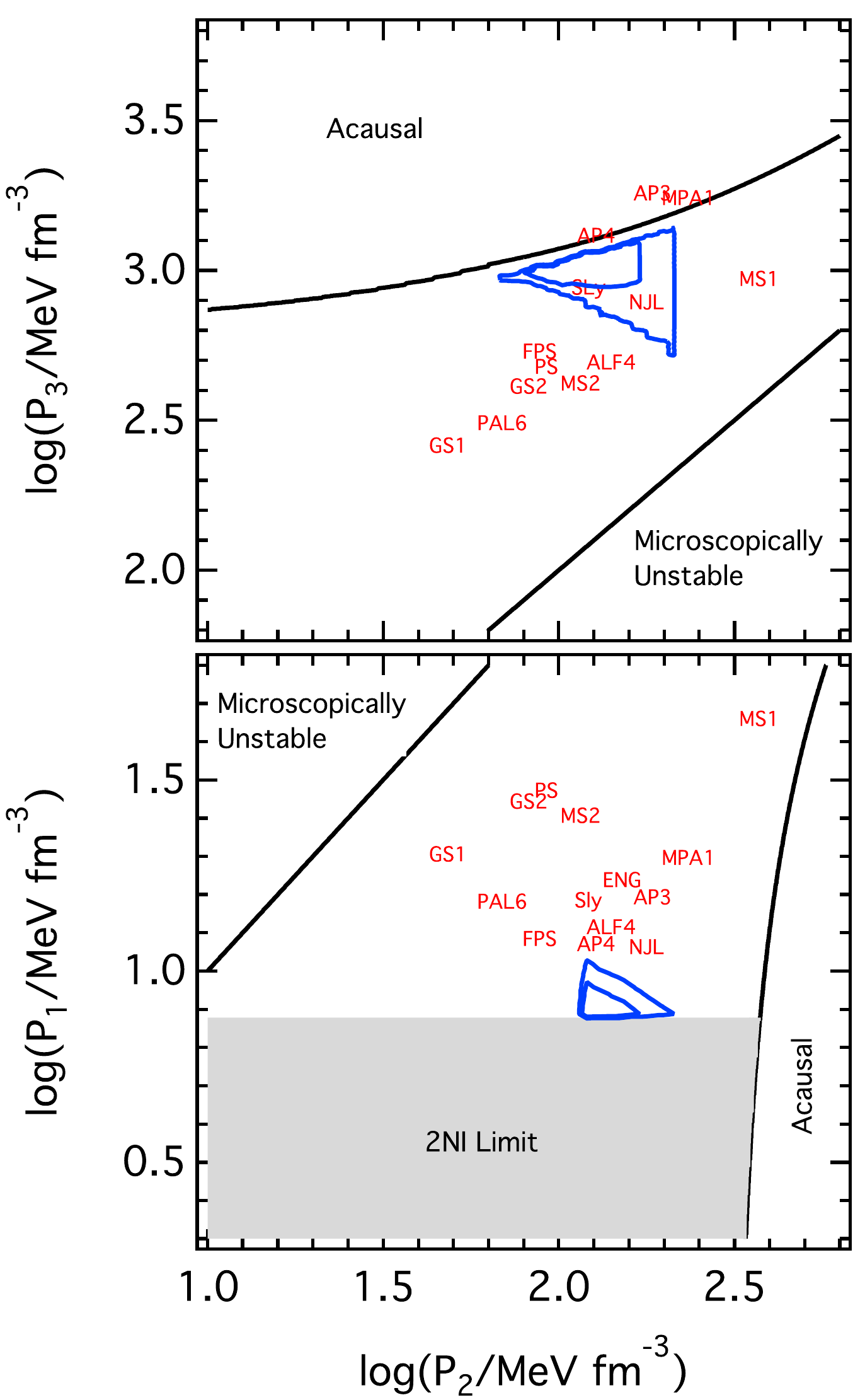}
\caption{The highest likelihood regions in the pressure of neutron
  star matter at 1.85~$\rho_{\rm ns}$ ($P_1$), 3.7~$\rho_{\rm ns}$
  ($P_2$), and 7.4~$\rho_{\rm ns}$ ($P_3$) obtained by a parametric
  inversion of all the neutron star radius measurements. To visualize
  the structure of the three-dimensional posterior likelihood
  function, the contours outline the regions in which the posterior
  likelihoods drop down to $e^{-1/2}$ and $e^{-1}$ of the highest
  value. The allowed regions of the parameter space are consistent
  with the constraints from calculations based on low-energy
  scattering experiments, are microscopically stable, and ensure that
  the equation of state remains causal. The left panel shows the
    result for flat priors in the logarithms of $P_1$, $P_2$, and
    $P_3$, while the right panel shows the result for flat priors in
    $P_1$, $P_2$, and $P_3$ within the physically allowed ranges of
    these parameters. }
\label{fig:P1P2P3} 
\end{figure}

Our understanding of the equation of state in the vicinity of the
nuclear saturation density is firmly founded on nucleon-nucleon
scattering experiments below 350~MeV and on the properties of light
nuclei. An approach that makes use of these data most directly is
based on describing the interactions between particles via static two-
and three-body potentials at this density (Akmal et al.\ 1998). As we
noted in \"Ozel et al.\ (2010), beyond a few times nuclear saturation
density, the interaction between particles can no longer be described
by static few-body potentials and, at even higher densities, a
well-defined expansion in terms of two-, three-, or many-body forces
no longer exists. For this reason, the most model-independent
constraint that we can impose is on $P_1$, i.e., at $\rho_1=1.85
\rho_{\rm ns}$. Following Gandolfi et al.\ (2012), we consider the
expansion of the interaction in terms of two- and three-body
potentials and use the contribution of the two-body potential (Argonne
AV8) to determine a lower bound to the pressure at 1.85~$\rho_{\rm
  ns}$.  We obtain $P_1 = 7.56$~MeV~fm$^{-3}$ using AV8, but
essentially the same result is found with the AV18 potential.

Using the two nucleon interaction pressure as an absolute lower bound
is justified by the fact that the three-body interactions in pure
neutron matter are always repulsive (J. Carlson, private
communication); including their contributions serves to increase the
pressure (see, e.g., Pieper et al. 2001 and Gandolfi et al. 2012, for
pure neutron matter; see also Figure~3 of Akmal et al.\ 1998).  The
$\delta v$ relativistic boost correction, which we do not include,
also gives a positive contribution to the pressure when calculated
with the two body interaction alone.  Note though that the
contributions of the three body forces and the relativistic boost
corrections are not simply additive.

A few words of caution are in order.  In symmetric nuclear matter at
nuclear matter density, the three nucleon interaction is quite
attractive at nuclear matter density, as it must be to achieve
sufficient binding energy for light nuclei; the three nucleon
interaction turns repulsive only for densities roughly above
1.5~$\rho_{\rm ns}$.  Furthermore, the three nucleon interaction
including the relativistic boost correction, softens the equation of
state of symmetric nuclear matter above nuclear matter density.  While
the three nucleon interaction as implemented in the APR equation of
state is indeed repulsive in pure neutron rich matter, there remain
theoretical uncertainties in the interaction itself.  One should, in
addition, take into account the modification of the effects of the
three body interaction when imposing beta equilibrium and hence
allowing for a finite proton fraction; as can be seen from Figure~16 of
Akmal et al.\ (1998), the three body interaction again increases the
pressure.  A further complication is the onset of a neutron
pion-condensed phase in neutron rich matter at a density $\sim
0.2~{\rm fm}^{-3}$, which lowers the pressure (as can be seen for pure
neutron matter in Fig. 5 of Akmal et al.\ 1998).

\begin{figure}
\centering
\includegraphics[scale=0.45]{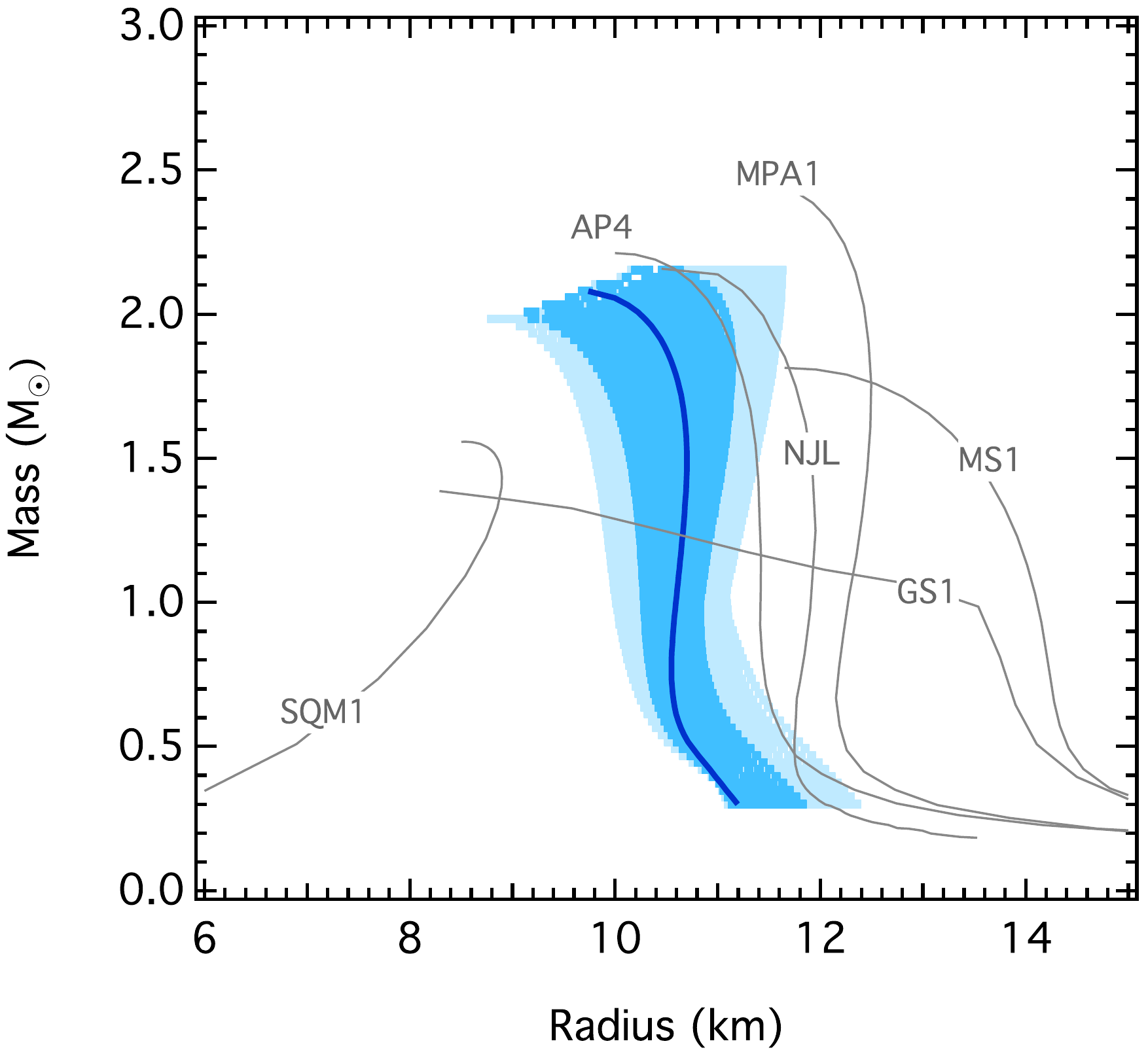}
\includegraphics[scale=0.45]{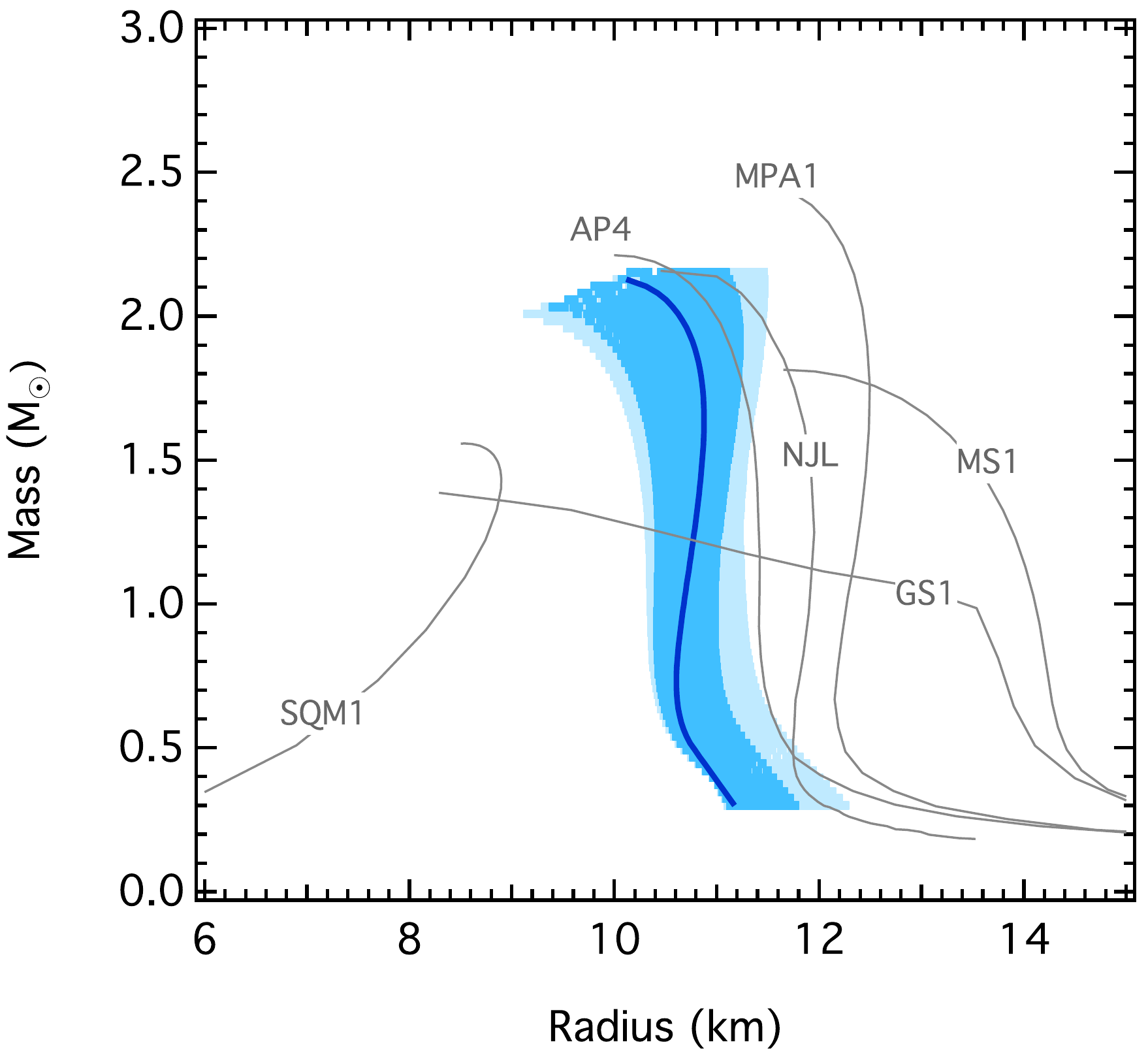}
\caption{The mass-radius relation (solid blue curve) corresponding to
  the most likely triplet of pressures that agrees with all of the
  neutron star radius and low energy nucleon-nucleon scattering data
  and allows for a $M>1.97~M_\odot$ neutron star mass.  The ranges of
  mass-radius relations corresponding to the regions of the $(P_1,
  P_2, P_3)$ parameter space in which the likelihood is within
  $e^{-1/2}$ and $e^{-1}$ of its highest value are shown in dark and
  light blue bands, respectively. The left panel shows the result for
  flat priors in the logarithms of $P_1$, $P_2$, and $P_3$, while the
  right panel shows the result for flat priors in $P_1$, $P_2$, and
  $P_3$ within the physically allowed ranges of these parameters.}
\label{fig:eos_mr} 
\end{figure}

In Figure~\ref{fig:P1P2P3}, we show the posterior likelihoods over the
pressures at the three fiducial densities, as well as the microscopic
and experimental bounds on these pressures. We also plot the pressures
of a number of proposed equations of state with widely differing
assumptions and calculation techniques. Because pressure $P_1$ has the
largest effect on the stellar radius, it is significantly constrained
by the radius data from above. The lower limit on $P_1$ coming from
the two-body interaction potential obtained at low densities excludes
the gray region labeled 2NI. The most likely value, as well as the
entire region within the highest posterior likelihood, are, in fact,
lower than the pressure predicted by most equations of state at that
density, as shown in the lower panel (see Read et al.\ 2009 for the
acronyms and the references for the various equations of state).  We
also include in this figure the recent equation of state labeled NJL
(Kojo et al.\ 2015), based on a smooth interpolation in pressure
vs. baryon chemical potential of a nucleonic equation of state (APR)
at densities below $\sim \rho_{\rm ns}$ with a quark matter equation
of state at densities above $\sim 5-7 \rho_{\rm ns}$.

The combination of $P_2$ and $P_3$, on the other hand, is constrained
by the maximum mass requirement: a lower value of $P_2$ pushes $P_3$
to be as high as possible within the causality limit, whereas for
moderate to high values of $P_2$, which already lead to M-R relations
that allow high mass stars and are consistent with the radius
measurements, the allowed range of $P_3$ extends to lower values. The
combination of $P_2$ and $P_3$ excludes to high confidence the stiff
equations of state such as MPA1 and MS1, which produce radii that are
too large (see also their inconsistency with $P_1$ in the lower
panel). This combination also excludes equations of state with
condensates, such as GS1, with pressures that are too low to be
consistent with the maximum mass requirement. 

Figure~\ref{fig:P1P2P3} shows that the combination of the radius
measurements with the low density experimental data and the
requirement of a $\sim 2\;M_\odot$ maximum mass pins down the
parameters of the equation of state extremely well across a wide range
of supranuclear densities and points to a preferred equation of state
that is somewhat softer than the nuclear equation of state AP4 (a
version of the APR equation of state). To see this on the mass-radius
plane, we also show in Figure~\ref{fig:eos_mr} the mass-radius
relation corresponding to the most likely triplet of pressures as well
as the range of mass-radius relations for the region of the $(P_1,
P_2, P_3)$ parameter space with the highest likelihood.  We limit the
range of masses in this figure to $\le 2.2\;M_\odot$ because of the
absence of any data to constrain the relation at higher masses. As
expected from the above discussion, the preferred mass-radius relation
lies to the left of most model predictions and is closest to AP4,
especially at low masses, where the main uncertainty in AP4 is in the
strength of the three-nucleon interactions. It also rises along a
nearly constant radius in order to reach the $\sim 2~M\odot$ limit.
Depending on the choice of the prior (i.e., flat prior on $P$ or $\log
P$), the predicted radius for a $1.5 \; M_\odot$ neutron star is
between 10.1 and 11.1~km.

Because of the relatively narrow range of the statistically acceptable
values of the three pressures, the effect of changing the prior
distributions is only marginal. It leads to a small shift in the most
likely values of the pressures and a change in the predicted radii
that is $\sim 0.4$~km. This is similar in magnitude to the result
obtained by Steiner et al.\ (2013), even when they considered extreme
scenarios (see their Fig.~3).

When compared to the earlier inference of \"Ozel et al.\ (2010) of the
pressures at three fiducial densities, the current measurements point
to much more constrained values of $P_1$ and to larger values of $P_2$
and $P_3$ by $\sim 0.3 $~dex. There are three reasons for this change.
First, in the present study, we include the radius measurements from
qLMXBs. Second, as discussed in Section 3.1.7, the improved analysis
methods lead to radii from bursters that are larger by $\lesssim
1$~km. Third, we incorporate the fact that, since that study, two
$\sim 2 M_\odot$ neutron stars have been discovered. This effectively
places a lower limit on the slope of the mass-radius relation,
controlled by $P_2$, and its turnover point, controlled by $P_3$ (see
\"Ozel \& Psaltis 2009; Lattimer \& Prakash 2010). 

Guillot et al.\ (2013) used the qLMXBs and a constant radius model to
infer typical neutron star radii of $9.4 \pm 1.2$~km, which also
indicate low values of $P_1$. Even though this inference is not
statistically inconsistent with our measurement, the small differences
can be understood in terms of the different distances and absorption
model used here, compared to those of Guillot et al.\ (2013). We also
incorporate the radius measurements from thermonuclear bursts as well
as the mass limits from the heaviest known neutron stars.

Our results point to smaller neutron star radii by $\sim 1$~km and to
lower pressures than the analyses of Steiner et al.\ (2010, 2013) and
Lattimer \& Steiner (2014a). As explained in Section 3.1.7, in their
analysis of the burst data, these authors obtained larger radii
because of their assumption regarding the location of the photosphere
at touchdown. In their analysis of the data from several qLMXBs,
Lattimer \& Steiner (2014a) also obtained larger radii by using a
range of estimated values of $N_{\rm H}$ to alter the $M-R$ contours
from Guillot et al.\ (2013), instead of reanalyzing these
data. Because of the strong correlation between $N_{\rm H}$ and the
apparent radius at infinity, this method underestimates the width of
$M-R$ contours reported in that study. Finally, they assumed different
distances to some of the globular clusters as well as helium
atmospheres for any qLMXB other than $\omega$Cen, all of which push
the inferred radii to larger values.

\section{Conclusions}

We performed a comprehensive study of spectroscopic radius
measurements of neutron stars using thermonuclear bursters and
quiescent low-mass X-ray binaries. We included a number of corrections
to the mass-radius inference that have recently been calculated,
incorporated systematic uncertainties in the measurements, and
employed Bayesian statistical tools to map the observed
quantities to neutron star masses and radii and the latter to the
neutron star equation of state.

Using a total of twelve sources allows us to place strong and
quantitative constraints on the properties of the equation of state
between $\approx 2-8$ times the nuclear saturation density, even
though the individual measurements themselves do not have the
precision to lead to tight constraints. We find that around
$M=1.5~M_\odot$, the preferred equation of state predicts a radius of
$10.1-11.1$~km. When interpreting the constraints on the pressure at
1.85~$\rho_{\rm ns}$ in the context of an expansion in terms of
few-body potentials (see, e.g., Akmal et al.\ 1998; Pieper et
al.\ 2001; Hebeler et al.\ 2010; Gandolfi et al.\ 2012), our results
suggest a relatively weak contribution of the three-body interaction
potential. In the framework of quark matter equations of state, the
inferred lower pressure at $1.85\; \rho_{\rm ns}$ is consistent with
an increased effective pairing interaction in the interpolated
equation of state (the NJL parameter H) at densities $\gtrsim 2
\rho_{\rm ns}$. While one can see this effect in model dependent
calculations (Kojo et al.\ 2015), such a lower pressure is physically
reasonable, independent of any particular model, since quark matter at
lower densities, en route with decreasing density to the strong three
quark correlations that eventually become well defined nucleons, is
expected to have greater pairing correlations than at higher
density. Accounting for all sources of the inferred pressure decrease
remains a theoretical challenge.

Even though we have taken into account, in the present work, a large
number of systematic effects in both the observations and in the
theoretical framework, our conclusion still relies on the validity of
the astrophysical interpretation of two phenomena, namely,
Eddington-limited thermonuclear bursts and the emission from neutron
stars in quiescence. We rely on the assumptions that the entire
neutron star surface is visible during the cooling tails of bursts and
during quiescence, that the Eddington limit is reached in photospheric
radius expansion bursts, and that the surface compositions in
quiescence is dominated by hydrogen due to gravitational settling,
unless there is evidence to the contrary.  The fact that the two sets
of measurements are in agreement with each other strongly argues
against an overall systematic bias. Nevertheless, radius measurements
obtained by non-spectroscopic techniques with potential biases that
are different than spectroscopic ones will be necessary to confirm the
results of our study. Pulse profile modeling with observations with
NICER and LOFT will offer such an opportunity in the near future.

\acknowledgments

We thank Elena Valenti and Bill Harris for useful discussions on the
distances to globular clusters and Toru Kojo for very helpful input on
the equations of state. We thank Harvey Tananbaum and the CXC team for
approving and executing the Chandra DDT observation for \oeight.  We
thank the participants of the ``The Neutron Star Radius'' conference
in Montreal, and especially Jim Lattimer and Cole Miller, for helpful
discussions. F\"O acknowledges support from NSF grant AST~1108753. DP
acknowledges support from NASA ADAP grant NNX12AE10G.  TG thanks the
University of Arizona for their hospitality and acknowledges support
from Istanbul University Project numbers 49429 and 48285. GB was
supported in part by NSF Grant PHY-1305891.  COH acknowledges support
through an NSERC Discovery Grant and an Alexander von Humboldt
Fellowship. SG is a FONDECYT Fellow, and acknowledges support through
the FONDECYT Post-doctoral grant \#3150428.


\appendix 
\setcounter{table}{0}
\renewcommand{\thetable}{A\arabic{table}}

\section{A1. Distance to 4U\,1608$-$52}

In G\"uver et al. (2010), we measured the distance to the neutron star
in the X-ray binary 4U\,1608$-$52 using a technique that relies on
comparing the hydrogen column density measurement to the equivalent
infrared extinction obtained from red clump stars along the line of
sight. In order to obtain the hydrogen column density $N_{\rm H}$ in a
way that does not depend on the assumed continuum model, we made use
of X-ray grating data and measured the absorption edges of individual
elements in the spectra caused by the attenuation in the ISM. At that
time, the only suitable X-ray dataset was a short XMM-{\it Newton} RGS
data obtained in 2003.

On March 1 2010, RXTE-ASM countrate of \oeight\ started a systematic
increase as expected from an outburst. Based on this, we triggered a
{\it Chandra} DDT observation, which was performed on March 15. We
obtained high resolution MEG spectra with these observations, with a
total number of counts that was comparable to the archival XMM-Newton
observation. Using these grating spectra, we determined the column
density of Mg and Ne elements along the line of sight, which we
discuss below. We also present the new resulting source distance
obtained with these new data.

{\it Chandra} observed 4U\,1608$-$52 on March 15 2010 with a net
exposure time of 23.08~ks using ACIS HETG in CC mode
(OBSID:12127). The calibration and the extraction of the 1st order
grating spectra was performed using
TGCAT\footnote[3]{\href{http://space.mit.edu/cxc/analysis/tgcat/index.html}{http://space.mit.edu/cxc/analysis/tgcat/index.html}}
scripts (Huenemoerder et al.\ 2011) with CIAO v4.3 and CALDB v.4.1.5.

\begin{figure*}
\centering
\includegraphics[scale=0.4]{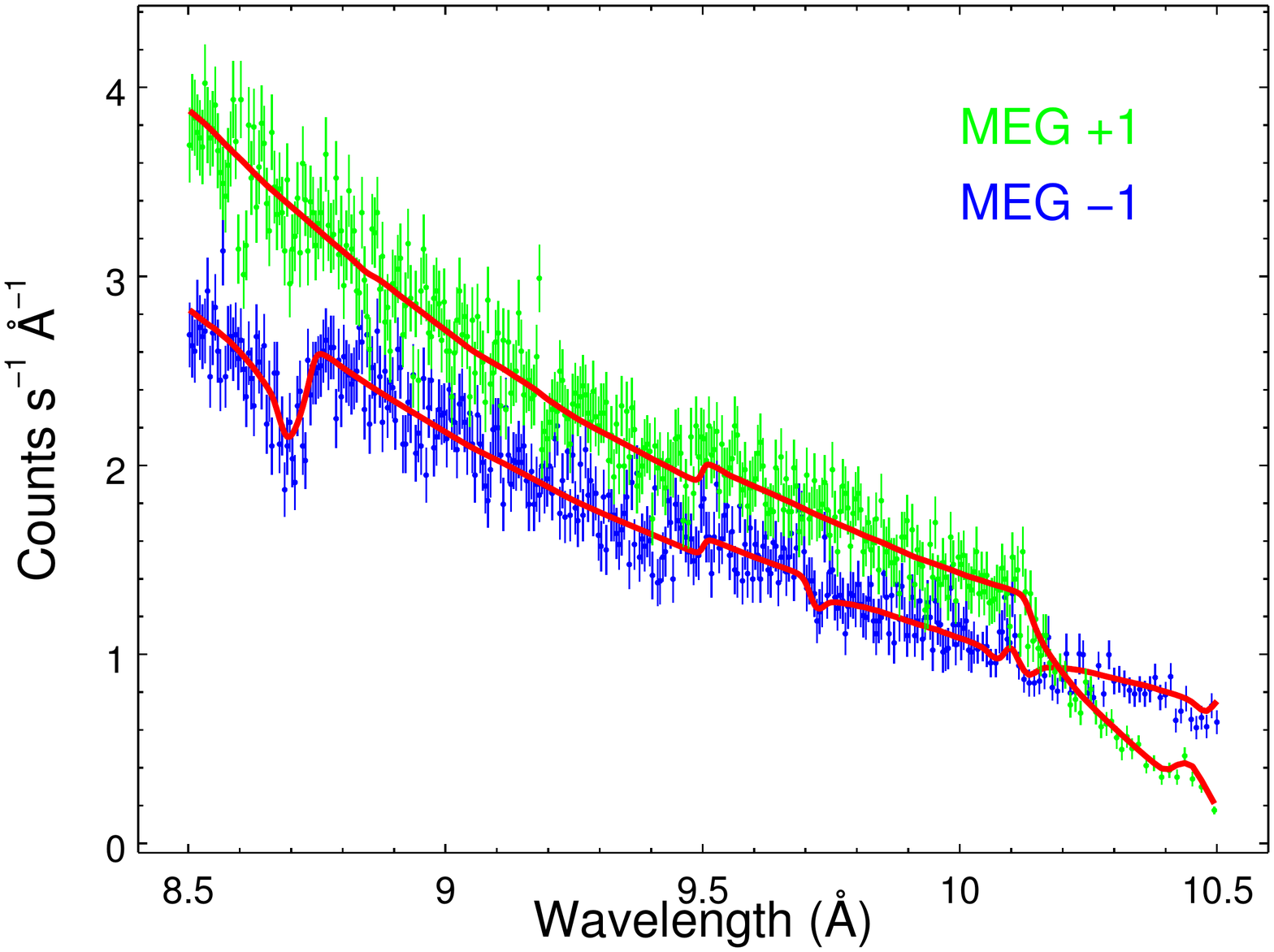}
\includegraphics[scale=0.4]{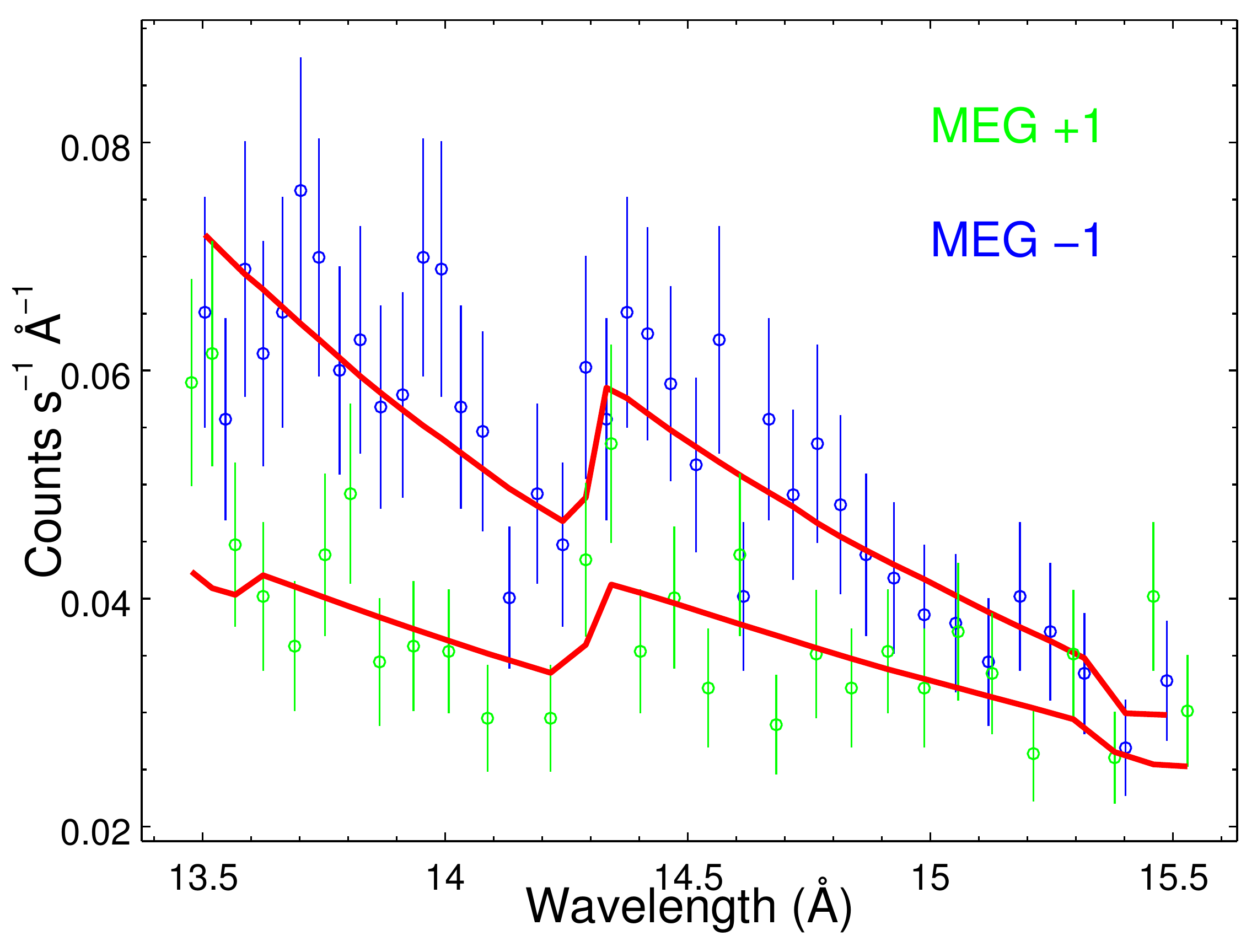}
\caption{Chandra Medium Energy Grating data and the best fit model
     around the Mg edge at $\lambda = 9.5~\AA$ (top panel) and the Ne
     edge at $\lambda = 14.3~\AA$ (bottom panel).}
\label{fig:edges}
\end{figure*}

Following the analysis detailed in G\"uver et al. (2010), we fit the
Mg and Ne edges using the MEG +/- 1st order data using {\it Sherpa}
software (Freeman et al. 2001; Doe et al. 2007). Due to its low number
of counts, we did not use the HEG data in this analysis. We binned the
spectra to have at least 50 counts in each spectral channel. We show
the data and the best-fit continuum model in Figure~\ref{fig:edges}
and the best fit values for both edges in Table~\ref{tbedges}. The
total hydrogen column density inferred from the {\it Chandra}
observation is $(1.09\pm0.16)\times10^{22}$~cm$^{-2}$ assuming ISM
abundances (Wilms et al.\ 2000), which is in excellent agreement with
the value $(1.08\pm0.16)\times10^{22}$~cm$^{-2}$ that was found by
G\"uver et al.\ (2010). Combining these two measurements, we found a
hydrogen column density of $(1.085\pm0.113)\times 10^{22}$~cm$^{-2}$
to 4U\,1608$-$52.

The good agreement between measurements separated by 7 years indicates
that the column density is dominated by absorption in the interstellar
medium as opposed to the absorption that is intrinsic to the binary,
which is variable (see Miller et al.\ 2009). Therefore, the distance
measurement to \oeight\ using this technique is not significantly
affected from possible intrinsic absorption in the system.

\begin{deluxetable}{cccccc}
  \tablecolumns{6} 
\tablewidth{400pt} 
\centering
\tablecaption{Best-fit Values for the Mg and Ne Edges}
  \tablehead{Edge & Grating & Absorption Coefficient & $N$ & $N_{H}$ & $\chi^{2}/$d.o.f. \\
    & & & (10$^{17}$cm$^{-2}$) & (10$^{22}$cm$^{-2}$) & } 
\startdata
Mg & MEG$-$1 & 0.067$\pm$0.017 & 3.06$\pm$0.75 & 1.22$\pm$0.30 & \\
      & MEG$+$1& 0.067$\pm$0.015 & 3.06$\pm$0.66 & 1.22$\pm$0.26 & \\
     & Combined & 0.067$\pm$0.011 & 3.06$\pm$0.50 & 1.22$\pm$0.20 & 0.919 \\
Ne  & MEG$-$1 & 0.28$\pm$0.10 & 7.95$\pm$2.83 & 0.91$\pm$0.33 & \\
      & MEG$+$1 & 0.26$\pm$0.12 & 7.40$\pm$3.41 & 0.85$\pm$0.40 & \\
      & Combined & 0.27$\pm$0.08 & 7.66$\pm$2.27 & 0.88$\pm$0.26 & 0.844
\enddata
\label{tbedges}
\end{deluxetable}

To convert the equivalent hydrogen column density to an optical
extinction, we used the most recent results in the relation between
these two quantities. Foight et al.\ (2015) analysed the archival {\it
  Chandra} observations of all the supernova remnants used in the
G\"uver \& \"Ozel (2009) study to improve the relation between the
optical extinction A$_{V}$ and the hydrogen column density N$_{\rm
  H}$. Performing a uniform analysis of the data set and taking into
account the abundances of the interstellar matter (Wilms et
al.\ 2000), Foight et al.\ (2015) determined the relation between
these parameters as $N_{\rm H} = (2.87\pm0.12) \times 10^{21} A_{\rm
  V}$. Note that the higher coefficient found by Foight et al.\ (2015)
is primarily because of using different abundances for the ISM.  We
used this improved relation to convert the hydrogen column density
found from the X-ray spectral analysis to optical extinction, and
following the same methods detailed in G\"uver et al.\ (2010a), we
found that the optical extinction $A_{\rm V}$ to the X-ray binary is
$3.78\pm0.42$~mag, which corresponds to a near-IR extinction of
$A_{\rm K} = 0.42 \pm 0.11$ using the relation of Cardelli et
al.\ (1989). The errors here reflect those arising from the
uncertainties in the hydrogen column density as well as those arising
from the uncertainties in the conversions.

\begin{figure*}
\centering
\includegraphics[scale=0.50]{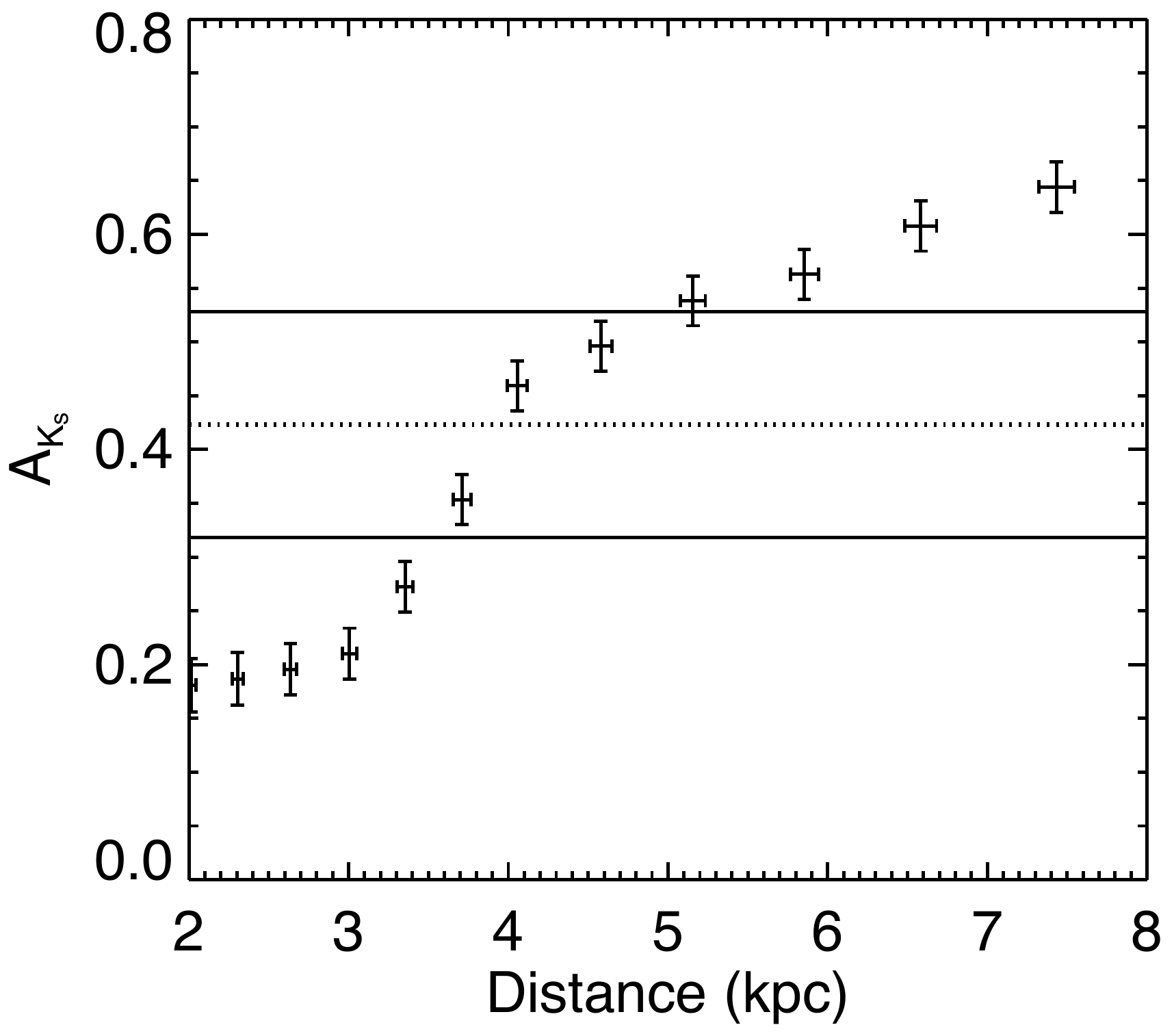}
\caption{The data points show the evolution of the extinction with
  distance along the line of sight to 4U~1608$-$52 as presented by
  G\"uver et al.\ (2010). The lines show the best-fit value of A$_K$
  of the source as derived from high resolution X-ray spectroscopy.}
\label{fig:dist_ext}
\end{figure*}

\begin{figure*}
\centering
\includegraphics[scale=0.50]{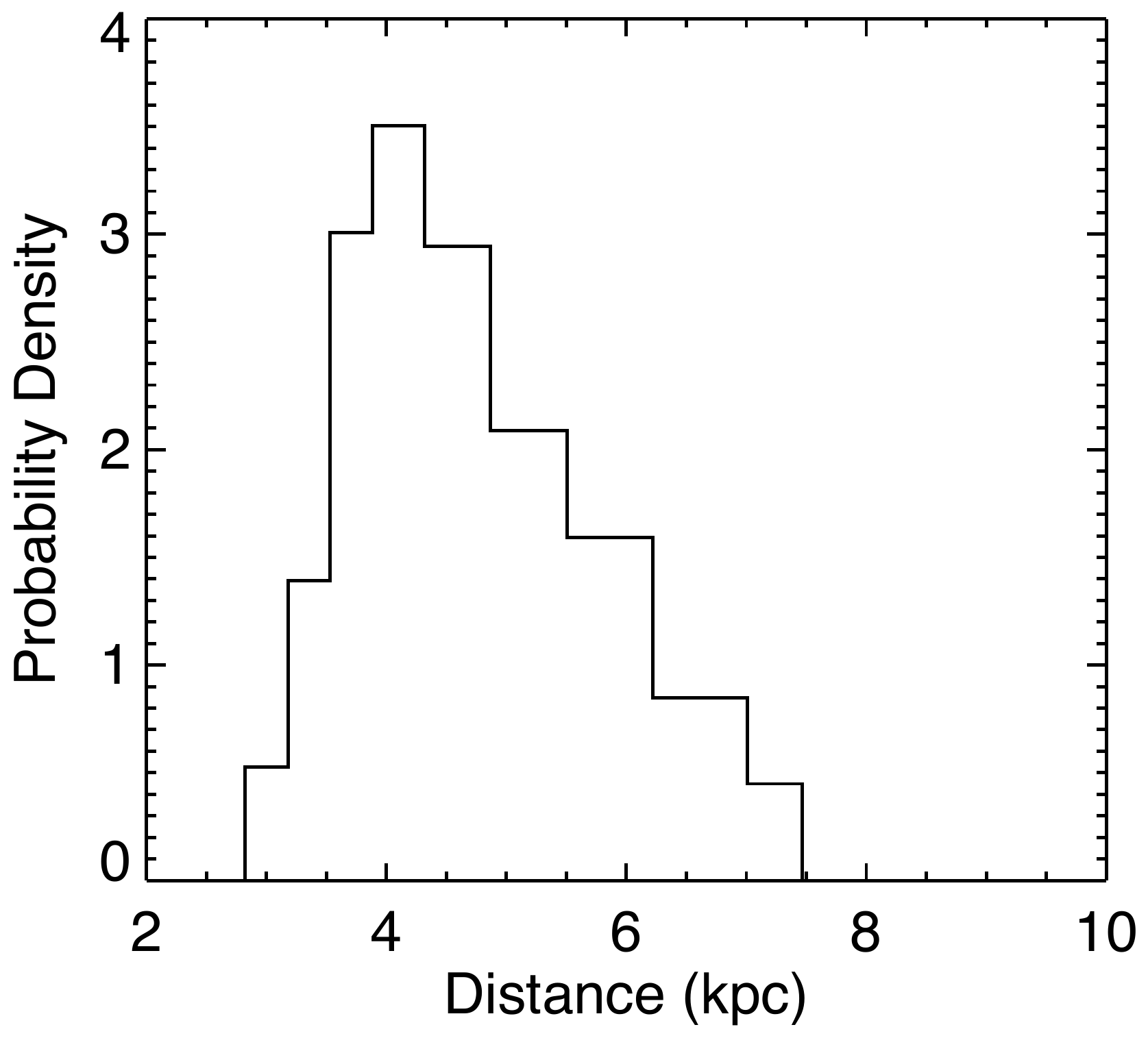}
\caption{The likelihood over the distance to 4U~1608$-$52, which
  indicates that it lies at a distance $D>3$~kpc and has the highest
  likelihood at $D\approx4$~kpc.}
\label{fig:dist_prob}
\end{figure*}

We compare in Figure~\ref{fig:dist_ext} the extinction values derived
above with the extinction curve derived using the red clump giants in
the field of view of \oeight\ (see G\"uver et al.\ 2010a). Using this,
we show in Figure~\ref{fig:dist_prob} the derived likelihood over the
distance to \oeight, which indicates a source distance $D>3$~kpc and
the most likely distance at 4~kpc.

\section{A2. The Photospheric Radius Expansion and Cooling Tails of \oeight\ Bursts}

RXTE carried out an observation of \oeight\ that was simultaneous with
the {\it Chandra} DDT observation. During the coincident observing
period, we detected a very bright X-ray burst $\approx 2800$~s after
the {\it Chandra} observation started, which corresponds to MJD
58070.18852974 (ObsID: 95334$-$01$-$03$-$08).

\begin{figure*}
\centering
\includegraphics[scale=0.60]{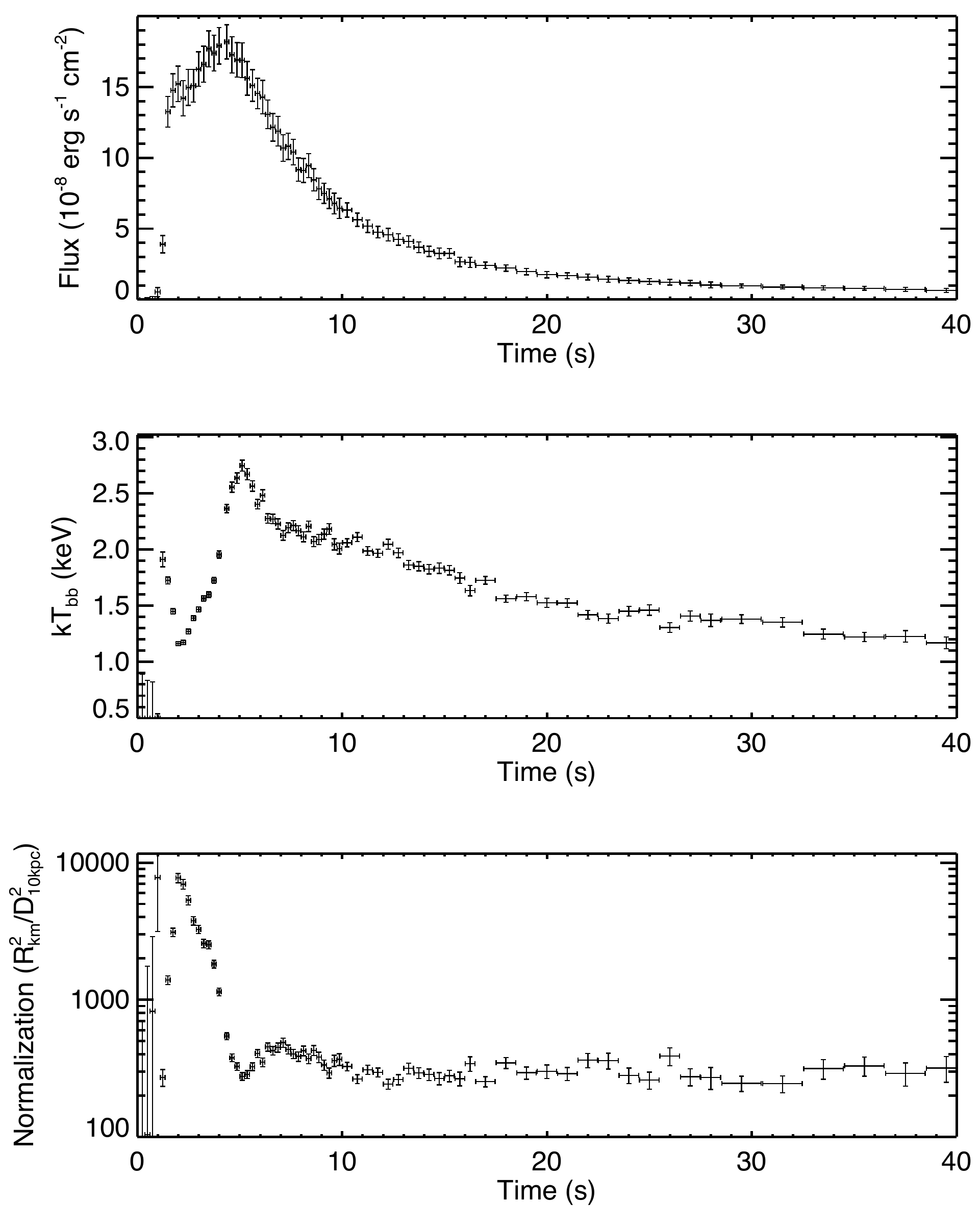}
\caption{Evolution of the bolometric flux (top), color temperature (middle), and the
  blackbody normalization (bottom) of burst 32 obtained from the spectral fits.}
\label{fig:ev_burst}
\end{figure*}

Following the methods outlined in the previous papers (see, e.g.,
G\"uver et al.  2012a,b) we performed time resolved X-ray spectroscopy
on the RXTE PCA data of the burst and measured the $2-10$~keV flux,
the spectral temperature obtained by fitting the spectrum with a
blackbody, and the blackbody normalization (i.e., the apparent angular
size). The evolution of the spectral parameters during the burst is
shown in Figure~\ref{fig:ev_burst}.  With a peak blackbody
normalization of $\approx 6400\; (R^{2}_{\rm km} / D^{2}_{\rm 10kpc}$),
which is the highest ever reached by this source, the X-ray burst
shows clear evidence of a photospheric radius expansion
event. Extrapolating from the Galloway et al.\ (2008a) catalog, we use
the ID number 32 for this burst and incorporate it into the touchdown
flux and the apparent angular size measurement of \oeight.

\begin{figure*}
\centering
   \includegraphics[scale=0.35]{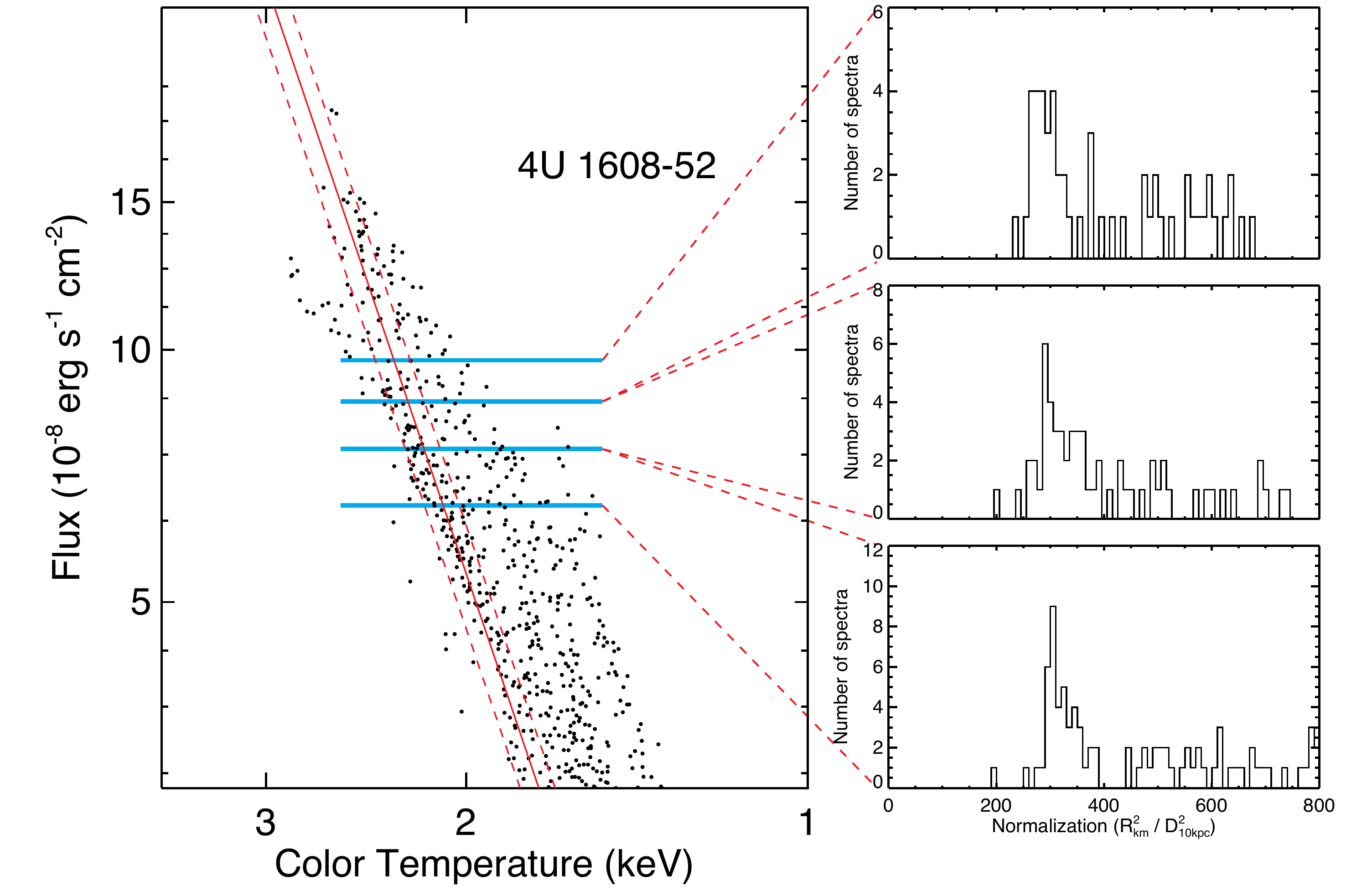}
\caption{The evolution of the flux and temperature measured for all
  the spectra in the cooling tails of bursts from \oeight. The
  diagonal lines show the best-fit blackbody normalization and its
  1$\sigma$ uncertainty. The histograms at three different flux levels
  visually identify the main apparent angular size track and the
  outliers.}
\label{fig:1608_hist}
\end{figure*}

To determine the apparent angular size that we use in the radius
measurements, we combined all of the data in the cooling tails of the
X-ray bursts from \oeight, including burst 32, using the methods
described in G\"uver et al.\ (2012b). We used the Bayesian
Gaussian-mixture model to determine the signal and the outliers and
Bayesian technique to infer the width of the distribution. We show in
Figure~\ref{fig:1608_hist} the data points along with three
representative histograms of the blackbody normalization at three
different flux bins. Despite the scatter in the outliers, the main
cooling track is clearly seen in these plots. We finally assigned the
width that we found from the Bayesian analysis as the systematic
uncertainty in the apparent angular size, which we found to be $314\pm
44.3$~(km/kpc)$^2$ (see Table~1).

Because it is a photospheric radius expansion event, we also included
burst 32 in the determination of the touchdown flux and the touchdown
temperature for \oeight\ (see upper right panel of
Figure~\ref{fig:data_1608} and Table~1).

\section{A3. Bursts Used in the Analysis}


\begin{deluxetable}{ccc}
  \tablecolumns{3} 
\tablewidth{400pt} 
\centering
\tablecaption{Burst IDs used in the Analysis}
  \tablehead{Source Name & Burst ID\tablenotemark{a} & Burst ID \\
    & for Touchdown Flux & for App. Angular Size } 
\startdata
4U~1820$-$30 & 1, 2, 3, 4, 5 & 1,2,3,4,5 \\
SAX~J1748.9$-$2021 & 1, 2 & 1,2,15,16 \\
EXO~1745$-$248 & 21, 22 & 21,22 \\
KS~1731$-$260 & 8, 9 & 3,4,5,6,7,8,9,10,11,12,13,14,15,16,17,18,19,20,21,22,23,24,25,27  \\
4U~1608$-$52 & 5, 23, 32 & 1,2,4,5,8,9,10,12,13,14,15,16,17,21,22,23,24,25,26,27,28,30,31,32\\
\enddata
\scriptsize
\tablenotetext{a} {Burst ID numbers follow the Galloway et al.\ (2008a) numbering system.}
\label{tb:burstid}
\end{deluxetable}

In Table~\ref{tb:burstid}, we list the IDs of the bursts we used in the determination
of the touchdown flux and the apparent angular size for each source.

\section{A4. The Posterior Likelihoods over the Observed Parameters of Burst Sources}

We report in Table~\ref{tb:posterior} the posterior likelihoods over
each of the three measured quantities for the six sources that show
thermonuclear bursts. The central values of the posterior likelihoods
are always within the 68\% range of the prior likelihoods shown in
Table~1, indicating a high degree of consistency between the
measurements that are used to infer the neutron star radii.


\begin{deluxetable}{cccc}
\tabletypesize{\scriptsize}
\tablewidth{400pt}
\tablecaption{Posterior Likelihoods over the Observed Parameters of the Burst Sources}
\tablehead{
 \colhead{Source} &
 \colhead{App. Angular Size} &
 \colhead{Touchdown Flux} &
 \colhead{Distance} 
\cr
 \colhead{} &
 \colhead{(km/10~kpc)$^2$} &
 \colhead{($10^{-8}$~erg~s$^{-1}$~cm$^{-2}$)} &
 \colhead{(kpc)} 
}
\startdata
\twenty\     & 97.6$\pm$11.2   & 5.22$\pm$0.44 & $7.47 \pm 0.38$ \\
\fortyeight\ & 91.8$\pm$8.4    & 3.74$\pm$0.46 & $8.01 \pm 0.47$ \\
\fortyfive\  & 129.8$\pm$15.0  & 6.21$\pm$0.65 & $6.08 \pm 0.40$  \\
\thirtyone\  & 97.2$\pm$7.5    & 4.48$\pm$0.47 & $6.77 \pm 0.75$ \\
\twentyfour\ & 121.1$\pm$11.9  & 4.95$\pm$0.50 & $7.27 \pm 0.37$ \\
\oeight\     & 333.4$\pm$36.8  & 16.9$\pm$1.78 & $3.63 \pm 0.29$  
\enddata
\scriptsize
\label{tb:posterior}
\end{deluxetable}

\end{document}